\documentclass[11pt,english,onecolumn]{IEEEtran}
\usepackage[T1]{fontenc}
\usepackage[latin9]{inputenc}
\usepackage{units}
\usepackage{amsthm}
\usepackage{amsmath}
\usepackage{amssymb}
\usepackage{mathdots}
\usepackage{graphicx}
\usepackage{setspace}
\PassOptionsToPackage{version=3}{mhchem}
\usepackage{mhchem}
\usepackage{esint}
\PassOptionsToPackage{normalem}{ulem}
\usepackage{ulem}
\makeatletter
\theoremstyle{plain}
\newtheorem{thm}{\protect\theoremname}
\theoremstyle{plain}
\newtheorem{lem}[thm]{\protect\lemmaname}
\theoremstyle{plain}
\newtheorem{prop}[thm]{\protect\propositionname}
\theoremstyle{remark}
\newtheorem{rem}[thm]{\protect\remarkname}

\usepackage{hyperref}
\usepackage{breqn}
  
 \DeclareMathOperator*{\argmin}{arg\,min}
 
\DeclareMathOperator{\interior}{int} 
\allowdisplaybreaks

\global\long\def\s[#1]{\textnormal{\scriptsize #1}}
\global\long\def\st[#1]{\textnormal{\tiny #1}}

\global\long\def\P{\mathbb{P}}
\global\long\def\E{\mathbb{E}}
\global\long\def\I{\mathbb{I}}

\global\long\def\RRc{\mathsf{R}_{\st[L]}}
\global\long\def\EEc{\mathsf{E}_{\st[L]}}
\global\long\def\RR{\mathsf{R}}

\global\long\def\DDc{\mathsf{D}_{\st[L]}}
\global\long\def\DD{\mathsf{D}_{\st[E]}}
\global\long\def\DDbar{\mathsf{\overline{D}}_{\st[E]}}

\newcommand{\dotleq}{\overset{\cdot} {\leq}}
\global\long\def\teq{\triangleq}

\global\long\def\trre[#1,#2]{\overset{{\scriptstyle (#2)}}{#1}} 
\global\long\def\bin[#1,#2]{\mathbb{B}[#1;#2]} 
\global\long\def\dec[#1]{\mathbb{D}_{#1}} 

\author{
\authorblockN{Nir Weinberger and Neri Merhav}

\authorblockA{Dept. of Electrical Engineering\\
    	    Technion - Israel Institute of Technology\\
Technion City, Haifa 3200004, Israel
} \\
\authorblockA{\{nirwein@tx, merhav@ee\}.technion.ac.il}
}
\makeatother

\usepackage{babel}
\providecommand{\lemmaname}{Lemma}
\providecommand{\propositionname}{Proposition}
\providecommand{\remarkname}{Remark}
\providecommand{\theoremname}{Theorem}

\begin{document}

\title{A Large Deviations Approach to Secure Lossy Compression%
\thanks{This work was supported by the Israel Science Foundation (ISF), grant
no. 412/12.%
}}

\maketitle
\renewcommand\[{\begin{equation}}
\renewcommand\]{\end{equation}}
\renewenvironment{align*}{\align}{\endalign}
\thispagestyle{empty}
\begin{abstract}
We consider a Shannon cipher system for memoryless sources, in which
distortion is allowed at the legitimate decoder. The source is compressed
using a rate distortion code secured by a shared key, which satisfies
a constraint on the compression rate, as well as a constraint on the
exponential rate of the excess-distortion probability at the legitimate
decoder. Secrecy is measured by the exponential rate of the exiguous-distortion
probability at the eavesdropper, rather than by the traditional measure
of equivocation. We define the perfect secrecy exponent as the maximal
exiguous-distortion exponent achievable when the key rate is unlimited.
Under limited key rate, we prove that the maximal achievable exiguous-distortion
exponent is equal to the minimum between the average key rate and
the perfect secrecy exponent, for a fairly general class of variable
key rate codes. \end{abstract}

\section{Introduction}

In his seminal paper \cite{shannon1949communication}, Shannon has
introduced a mathematical framework for secret communication. The
cipher system is considered \emph{perfectly secure} if the cryptogram
and the message are statistically independent, and so, an eavesdropper
does not gain any information when he observes the cryptogram. To
achieve secrecy, the sender and the legitimate recipient share a secret
key, which is used to encipher and decipher the message. It is rather
apparent from ordinary compression \cite{Shannon} that a necessary
and sufficient condition for perfect secrecy is that the available
key rate is larger than the information rate required to compress
the source (the entropy or rate-distortion function of the source
in case of lossless or lossy compression, respectively). Usually,
the supply of key bits is a limited resource, as they need to be transferred
to the intended recipient via a completely secure channel. When the
key rate is less than the information rate, secrecy is traditionally
measured in terms of \emph{equivocation,} that is, the conditional
entropy of the message given the cryptogram.  The use of equivocation
as a secrecy measure was advocated by other models of secrecy systems,
which do not assume a shared key. Instead, secrecy is achieved by
the fact that the message intercepted by the eavesdropper is of lower
quality than the one received by the legitimate receiver. For example,
in the ubiquitous wire-tap model \cite{wire-tap,CK_confidential},
the channel of the wiretapper is degraded (or more noisy) with respect
to (w.r.t.) the channel of the legitimate receiver. In the model of
\cite{Gunduz_ITW2008,Gunduz_ISIT2008,merhav2008shannon} the legitimate
recipient has better quality of side information than the eavesdropper. 

The equivocation is indeed an unambiguous measure for statistical
dependence when it is equal to either its minimal value of zero (the
random variables are deterministic functions of each other), or its
maximal value of the unconditional entropy (the two random variables
are independent). Nonetheless, for \emph{partial secrecy}, i.e., when
the equivocation takes values strictly between these two extremes,
its operational meaning is disputable. Thus, in \cite{Hellman}, it
was proposed to measure partial secrecy by the expected number of
spurious messages that explain the given cryptogram (which is somewhat
equivalent to the probability of correctly decrypting the message).
Later, in \cite{Yamamoto}, it was proposed to measure partial secrecy
by the minimum average distortion that an eavesdropper can attain
(this was also considered previously, to some extent, in \cite{Lu}).
In addition, in \cite{Yamamoto} the possibility that the legitimate
recipient can tolerate a certain distortion level was also incorporated
into the system model. In \cite[Theorems 2 and 3]{Yamamoto}, inner
and outer bounds were obtained on the achievable trade-off between
the coding rate, the key rate, and distortion levels at the legitimate
recipient and eavesdropper. However, in \cite{Schieler_ISIT2012},
it was revealed that this trade-off is, in fact, degenerated. It was
demonstrated there that in some cases, a negligible key rate can cause
maximum distortion at the eavesdropper. The following simple example
(from \cite[Section I.A]{Schieler_rate_distortion_secrecy}) demonstrates
this: Consider an memoryless source $\mathbf{X}=\left(X_{1},\ldots,X_{n}\right)\in\{0,1\}^{n}$
where $\P(X_{i}=1)=\frac{1}{2}$ for $i=1,\ldots n$, and a \emph{single}
key bit $U$, shared by the two legitimate parties, where $\P(U=1)=\frac{1}{2}$.
Suppose that the distortion measure at the eavesdropper side is the
Hamming distortion measure. Then, if the encrypted message is $\mathbf{Y}=\left(Y_{1},\ldots,Y_{n}\right)$,
where $Y_{i}=X_{i}\oplus U$, then the distortion at the eavesdropper
attains its maximal possible value of $\frac{1}{2}$, regardless of
the estimate of the eavesdropper. Nonetheless, such a secrecy is severely
insecure. If the eavesdropper becomes aware of just a single bit of
the source, then it can decrypt the entire message. It was therefore
proposed to consider models which are more robust to assumptions concerning
the eavesdropper. These models indeed lead to a non-degenerated trade-off,
that requires a positive key rate. In \cite{Schieler_rate_distortion_secrecy,Foil_Allerton2010}
it was assumed that the eavesdropper's estimation is performed sequentially,
and at the time it estimates the $i$-th symbol, it has noiseless/noisy
estimates of all the previous message symbols and the previous reproduced
symbols (at the legitimate recipient), in addition to the public cryptogram.
This model was termed \emph{causal disclosure}. It was justified by
the scenario in which the sender and legitimate recipient attempt
to coordinate actions in a distributed system in order to maximize
a certain payoff, and the eavesdropper acts in order to minimize the
payoff. In a different line of work \cite{Henchman_ISIT}, the eavesdropper
produces a fixed-size list (of exponential cardinality in the block-length),
and the distortion is measured w.r.t. the reproduction word in the
list which attains the minimal distortion. 

However, the fact that the trade-off in \cite{Yamamoto} is degenerated
can be attributed to the way that the distortion is measured, rather
than to the weakness of the eavesdropper. For a given strategy of
the eavesdropper, the average distortion, as assumed in \cite{Yamamoto,Schieler_rate_distortion_secrecy,Henchman_ISIT},
may be large due to message and key-bit combinations that lead to
a very large distortion, albeit with small probability. A more refined
figure of merit would include the probability that the distortion
is less than some level, rather than the average distortion. Such
a performance criterion is customary in ordinary rate-distortion theory
(e.g. the $\epsilon$-fidelity criterion in \cite[Chapter 7]{csiszar2011information}).
Indeed, in the above single key-bit example, the eavesdropper can
estimate the message exactly with probability $\frac{1}{2}$, irrespective
of its length. Thus, for any positive distortion level, the probability
of an exiguous-distortion event is $\frac{1}{2}$, which is clearly
unacceptable for most applications. 

For most source models, good estimation of the message at the eavesdropper
should be a rare event, and finding its exact probability is difficult.
Instead, an asymptotic analysis can be carried in order to find the
exponential decrease rate (i.e. the \emph{exponent}) of the correct
decryption probability. The results of \cite{Lu} can be considered
as a special case of this line of thought, for the restricted class
of instantaneous encoders. In \cite{Lu}, the exponent of decrypting
the message by the eavesdropper was found as a function of the exponent
of exiguous-distortion of the estimation by the eavesdropper. For
the same model, the exponent of the minimal probability of correct
decryption by the eavesdropper was found in \cite{ahlswede1982bad}.
Later, in \cite{secrecy_large_dev} secrecy was defined in a large-deviations
sense: A system is considered secure if the exponent of the probability
of the eavesdropper \emph{correctly} decrypting the message is the
same with and without the cryptogram. This, in turn, required the
analysis of the correct decryption probability. In \cite{Lu,ahlswede1982bad,secrecy_large_dev},
it was assumed that the legitimate recipient must reproduce the message
exactly (i.e., with zero distortion).

\begin{figure}
\begin{centering}
\includegraphics{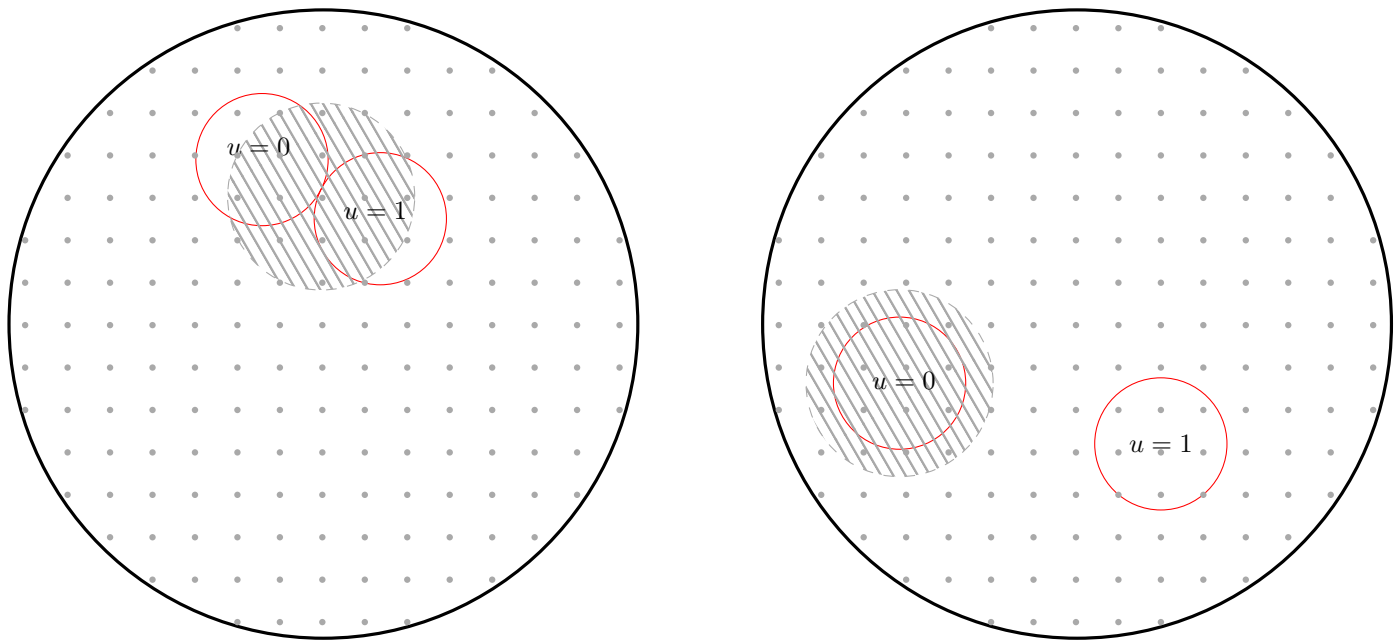}
\par\end{centering}

\protect\caption{Two cases of ambiguity for the eavesdropper, for a single key bit
code. Left side: Assume for simplicity that the source is distributed
uniformly over the dots encapsulated by the outermost circle. The
two small solid line circles represent two reproduction cells, which
are mapped to the same cryptogram by the two possible values of the
key bit $u$. The dashed larger circle represents all the source block
for which the distortion between the source block and the best estimate
of the eavesdropper is less than $\DD$. As can be seen, there is
a large exiguous-distortion probability. Right side: Under the same
assumptions, in this case the two reproduction cells are far apart.
The best estimate of the eavesdropper can `cover' at most one of the
reproduction cells, and the exiguous-distortion probability is $\frac{1}{2}$.
\label{fig:Two-cases-of-ambiguity}}

\end{figure}

In this paper, we adopt a similar large-deviations approach to measuring
secrecy, using a distortion measure, and generalize the results of
\cite{secrecy_large_dev}. For a memoryless source, we allow an imperfect
reproduction at the legitimate recipient, and measure distortion both
at the legitimate recipient and at the eavesdropper using a large-deviations
measure. Specifically, we will define two exponents. First, for a
given distortion level $\DDc$, the \emph{excess-distortion exponent
}is defined in the usual way \cite[Chapter 9]{csiszar2011information},
as the exponent of the probability that the distortion between the
legitimate recipient reproduction and the source sequence is larger
than $\DDc$. Second, for a given distortion level $\DD$, we define
the \emph{exiguous-distortion exponent }as the exponent of the probability
that the distortion between the eavesdropper estimate and the source
sequence is \emph{less} than $\DD$. We will derive the \emph{perfect
secrecy exponent }function $E_{e}^{*}(\DD)$, which is the exiguous-distortion
exponent of the eavesdropper when it estimates the message blindly,
without the cryptogram (alternatively, for codes with unlimited key
rate). It will be assumed that the secrecy system has a limited coding
rate $\RRc$, and that for a given distortion level $\DDc$, the excess-distortion
exponent must be larger than $\EEc$. Our main result is that under
mild conditions on the \emph{compression constraints} $(\RRc,\DDc,\EEc)$,
the maximal achievable exiguous-distortion exponent is equal to the
minimum between the key rate $\RR$, and $E_{e}^{*}(\DD)$, calculated
at distortion level required by the eavesdropper $\DD$. Since this
maximal exiguous-distortion exponent does not depend on $(\RRc,\DDc,\EEc)$
(in the interesting domain of these parameters), such a result implies
that as far as performance trade-offs are concerned, the compression
and secrecy problems are essentially decoupled: The fact that the
message is required to be kept secret does not affect the compression
performance. It should be stressed, however, that this result does
not imply a separation theorem from the operational point of view.
The rate-distortion code should be designed in a certain manner in
order to provide secrecy, in contrast to, e.g., \cite{Yamamoto,merhav2008shannon,merhav2006shannon}.
A concatenation of an arbitrary good rate-distortion code, followed
by encryption using the available key bits, does not necessarily achieve
a good exiguous-distortion exponent. For intuition, consider an ordinary
rate-distortion code, assume that one key bit is available, and that
the distortion measures of the legitimate decoder and eavesdropper
are the same. The eavesdropper, in this case, knows that the reproduction
of the legitimate decoder is one of two possible reproductions (of
equal probability). If these two reproductions are close, then it
can approximate them using a single reproduction, and achieve a distortion
which may be only slightly larger than the distortion of the legitimate
decoder. If, however, the rate-distortion code is designed in such
a way that these two reproductions are sufficiently far apart, then
the eavesdropper will have a poor compromise between them, and will
achieve high distortion. This is illustrated in Figure \ref{fig:Two-cases-of-ambiguity}.
More generally, unlike ordinary rate-distortion codes, in which the
performance is determined only by the reproduction cells, and the
way in which the reproduction cells are mapped to transmitted bits
is immaterial, here, the latter will be crucial for the security performance.

To show this result, we will prove both achievability (lower bound
on the exiguous-distortion exponent) and a matching converse (upper
bound). In the achievability part, we will demonstrate the existence
of a secrecy system in which the compression constraints are satisfied,
and it has a fixed key rate $\RR$. For this secrecy system, the best
strategy of the eavesdropper will be either to (1) guess the secret
key and reproduce the message as a legitimate recipient (using the
cryptogram), or (2) blindly estimate the message. The secrecy system
constructed will also be \emph{universal }in the following two senses.
First, it does not require the knowledge of the source statistics,
as long it is a memoryless source. Second, it is not designed for
a specific value of $\DD$, yet the exiguous-distortion exponent $\min\{\RR,E_{e}^{*}(\DD)\}$
will be achieved for any value of $\DD$, \emph{by the same sequence
of codes}, as long as $\DD\geq\DDc$. As a converse, we will show
that even if \emph{variable }key rate is allowed, yet with average
key rate less than $\RR$, then the exiguous-distortion exponent cannot
be larger than $\min\{\RR,E_{e}^{*}(\DD)\}$. The results of \cite{secrecy_large_dev}
are essentially recovered from our results, as a special case with
$\DDc=\DD=0$. We also remark that in our model, the distortion measures
of the legitimate recipient and the eavesdropper can be different,
as long as they satisfy a certain relationship.

Finally, we briefly mention a related work in which large-deviations
aspects were also incorporated. In \cite{Haroutunian_ISIT2002}, the
\emph{guessing} model of \cite{Neri_guessing,Neri_guessing2} was
relaxed to allow, after a maximum of possible guesses has passed,
a small probability of large distortion for the eavesdropper. To analyze
the asymptotic limits of the system, the excess-distortion\emph{ }exponent
of the \emph{eavesdropper} was restricted, and the maximal normalized
logarithm of the number of guesses was found%
\footnote{Reference \cite{Haroutunian_ISIT2002} is a one page abstract, and
contains only a description of the problem. The results were not published,
but a detailed version of \cite{Haroutunian_ISIT2002} can be found
in \cite{Haroutunian_arxiv}. However, we believe that the achievability
results provided in \cite{Haroutunian_arxiv} are not actually proven.
Specifically, in the achievability proof, no system is actually constructed,
and the claims about the expected number of guesses of the eavesdropper
are made on \emph{any} given secrecy system. Obviously, there are,
particularly bad, secrecy systems, in which a single guess suffices
to find the message exactly.%
}. However, in our model, no testing mechanism is assumed to be available
to the eavesdropper, which allows it to validate its estimate.

The outline of the rest of the paper is as follows. In Section \ref{sec:Notation-Conventions},
we establish notation conventions, and in Section \ref{sec:Problem Statement},
we formulate the problem. In Section \ref{sec:Main-Result}, we present
our main theorem, and discuss its implications. In Section \ref{sec:Outline-of-the proof},
we provide the outline and the main ideas of the proof.  The proof
of the main theorem appears in Section \ref{sec:Proof-of-Main}.

\section{Notation Conventions\label{sec:Notation-Conventions}}

Throughout the paper, random variables will be denoted by capital
letters, specific values they may take will be denoted by the corresponding
lower case letters, and their alphabets will be denoted by calligraphic
letters. Random vectors and their realizations will be denoted, respectively,
by capital letters and the corresponding lower case letters, both
in the bold face font. Their alphabets will be superscripted by their
dimensions. For example, the random vector $\mathbf{X}=(X_{1},\ldots,X_{n})$
($n$ positive integer), may take a specific vector value $\mathbf{x}=(x_{1},\ldots,x_{n})$
in ${\cal X}^{n}$, the $n$th order Cartesian power of ${\cal X}$,
which is the alphabet of each component of this vector. For any given
vector $\mathbf{x}$, we will also denote $\mathbf{x}_{i}^{j}=\left(x_{i},\ldots,x_{j}\right)$
for $1\leq i\leq j\leq n$, and use the shorthand $\mathbf{x}_{1}^{j}=\mathbf{x}^{j}$. 

We will follow the standard notation conventions for probability distributions,
e.g., $P_{X}(x)$ will denote the probability of the letter $x\in{\cal X}$
under the distribution $P_{X}$. The arguments will be omitted when
we address the entire distribution, e.g., $P_{X}$. Similarly, generic
distributions will be denoted by $Q$, $Q^{*}$, and in other forms,
subscripted by the relevant random variables/vectors/conditionings,
e.g. $Q_{XZ}$, $Q_{X|Z}$. Whenever clear from context, these subscripts
will be omitted.  An exceptional case will be the `hat' notation.
For this notation, $\hat{Q}_{\mathbf{x}}$ will denote the empirical
distribution of a vector $\mathbf{x}\in{\cal {\cal X}}^{n}$, i.e.,
the vector of relative frequencies $\hat{Q}_{\mathbf{x}}(x)$ of each
symbol $x\in{\cal X}$ in $\mathbf{x}$. The type class of $\mathbf{x}\in{\cal X}^{n}$,
which will be denoted by ${\cal T}_{n}(\hat{Q}_{\mathbf{x}})$, is
the set of all vectors $\mathbf{x}'$ with $\hat{Q}_{\mathbf{x}'}=\hat{Q}_{\mathbf{x}}$.
The set of all type classes of vectors of length $n$ over ${\cal X}$
will be denoted by ${\cal P}_{n}({\cal X})$, and the set of all possible
types over ${\cal X}$ will be denoted by ${\cal P}({\cal X})\teq\bigcup_{n=1}^{\infty}{\cal P}_{n}({\cal X})$.
Similar notation for type classes will also be used for generic types
$Q_{X}\in{\cal P}({\cal X})$, i.e., ${\cal T}_{n}(Q_{X})$ will denote
the set of all vectors $\mathbf{x}$ with $\hat{Q}_{\mathbf{x}}=Q_{X}$.
In the same manner, the empirical distribution of a pair of vectors
$(\mathbf{x},\mathbf{z})$ will be denoted by $\hat{Q}_{\mathbf{x}\mathbf{z}}$
and the joint type class will be denoted by ${\cal T}_{n}(\hat{Q}_{\mathbf{x}\mathbf{z}})$.
The joint type classes over the Cartesian product alphabet ${\cal X}\times{\cal Z}$
will be denoted by ${\cal P}_{n}({\cal X}\times{\cal Z})$, and ${\cal P}({\cal X}\times{\cal Z})\teq\bigcup_{n=1}^{\infty}{\cal P}_{n}({\cal X}\times{\cal Y})$.
For a joint type $Q_{XZ}\in{\cal P}({\cal X}\times{\cal Z})$, ${\cal T}_{n}(Q_{XZ})$
will denote the set of all pairs of vectors $(\mathbf{x},\mathbf{z})$
with $\hat{Q}_{\mathbf{x}\mathbf{z}}=Q_{XZ}$. The conditional type
class, namely, the set $\{\mathbf{x}':\hat{Q}_{\mathbf{x}'\mathbf{z}}=\hat{Q}_{\mathbf{x}\mathbf{z}}\}$,
will be denoted by ${\cal T}_{n}(\hat{Q}_{\mathbf{x}|\mathbf{z}},\mathbf{z})$,
or more generally ${\cal T}_{n}(Q_{X|Z},\mathbf{z})$ for a generic
empirical conditional probability distribution $Q_{X|Z}$. The probability
simplex for ${\cal X}$ will be denoted by ${\cal Q}({\cal X})$,
and the simplex for the alphabet ${\cal X}\times{\cal Z}$ will be
denoted by ${\cal Q}({\cal X}\times{\cal Z})$. Similar notations
will be used for triplets of random variables. 

For two distributions $P_{X},Q_{X}$ over the same finite alphabet
${\cal X}$, we will denote the variational distance (${\cal L}_{1}$
norm) by
\[
||P_{X}-Q_{X}||\teq\sum_{x\in{\cal X}}|P_{X}(x)-Q_{X}(x)|.
\]
When optimizing a function of a distribution $Q_{X}$ over the entire
probability simplex ${\cal Q}({\cal X})$, the explicit display of
the constraint will be omitted. For example, for a function $f(Q)$,
we will write $\min_{Q}f(Q)$ instead of $\min_{Q\in{\cal Q}({\cal X})}f(Q)$.
The same will hold for optimization of a function of a distribution
$Q_{XZ}$ over the probability simplex ${\cal Q}({\cal X}\times{\cal Z})$,
and for similar optimizations. 

The expectation operator w.r.t. a given distribution, e.g., $Q_{XZ}$,
will be denoted by $\E_{Q}[\cdot]$ where, the subscript $Q_{XZ}$
will be omitted if the underlying probability distribution is clear
from the context. In general, information-theoretic quantities will
be denoted by the standard notation \cite{Cover:2006:EIT:1146355},
with subscript indicating the distribution of the relevant random
variables, e.g. $H_{Q}(X|Z),I_{Q}(X;Z),I_{Q}(X;Z|W)$, under $Q=Q_{XZW}$.
For notational convenience, the entropy of $X$ under $Q$ will be
denoted both by $H_{Q}(X)$ and $H(Q_{X})$, depending on the context.
The binary entropy function will be denoted by $h_{\st[B]}(q)$ for
$0\leq q\leq1$. The information divergence between two distributions,
e.g. $P_{X}$ and $Q_{X}$, will be denoted by $D(P_{X}||Q_{X})$.
In all information measures above, the distribution may also be an
empirical distribution, for example, $H(\hat{Q}_{\mathbf{x}})$, $D(\hat{Q}_{\mathbf{x}}||P_{X})$
and so on. 

We will denote the Hamming distance between two vectors, $\mathbf{x}\in{\cal X}^{n}$
and $\mathbf{z}\in{\cal X}^{n}$, by $d_{\st[H]}(\mathbf{x},\mathbf{z})$.
The length of a string $b$ will be denoted by $|b|$, the concatenation
of strings $b_{1},b_{2},\ldots$ will be denoted by $(b_{1},b_{2},\ldots)$,
and the empty string will be denoted by $\phi$. We will denote the
complement of a set ${\cal A}$ by ${\cal A}^{c}$, and its interior
by $\interior({\cal A})$. For a finite set ${\cal A}$, we will denote
its cardinality by $|{\cal A}|$. The probability of the event ${\cal A}$
will be denoted by $\P({\cal A})$, and $\I({\cal A})$ will denote
its indicator function. 

For two positive sequences, $\{a_{n}\}$ and $\{b_{n}\}$ the notation
$a_{n}\doteq b_{n}$, will mean asymptotic equivalence in the exponential
scale, that is, $\lim_{n\to\infty}\frac{1}{n}\log(\frac{a_{n}}{b_{n}})=0$.
Similarly, $a_{n}\dotleq b_{n}$ will mean $\limsup_{n\to\infty}\frac{1}{n}\log(\frac{a_{n}}{b_{n}})\leq0$,
and so on. The ceiling function will be denoted by $\left\lceil \cdot\right\rceil $.
The notation $[t]_{+}$ will stand for $\max\{t,0\}$. For two integers,
$a,b$,  we denote by $a\bmod b$ the modulo of $a$ w.r.t. $b$.
Logarithms and exponents will be understood to be taken to the binary
base.

Throughout, we will ignore integer code length constraints for the
sake of simplicity, as they do not have any effect on the results.
For example, instead of $\left\lceil n\RR\right\rceil $ bits we will
write $n\RR$ bits. For a given finite ordered set, ${\cal A}=\{\mathbf{a}_{1},\ldots,\mathbf{a}_{|{\cal A}|}\}$,
we will denote by $\bin[\mathbf{a},\log|{\cal A}|]$ the binary representation
of the index of $\mathbf{a}$ in ${\cal A}$, i.e. $\bin[\mathbf{a},\log|{\cal A}|]=i$
if $\mathbf{a}=\mathbf{a}_{i}$, for $i=1,\ldots|A|$. 

In general, the subscript `L' will be used for quantities related
to the legitimate decoder, and the subscript `E' will be used for
eavesdropper-related quantities.

\section{Problem Statement\label{sec:Problem Statement}}

Let the source vector $\mathbf{X}=\left(X_{1},\ldots,X_{n}\right)$
be formed by $n$ independent copies of a random variable $X\in{\cal X}$,
where ${\cal X}$ is a finite alphabet, and $X_{i}$ is distributed
according to $P_{X}(x)=\P(X=x)$. Let ${\cal W}$ and ${\cal Z}$
be finite reproduction alphabets. In addition, let $\{U_{i}\}_{i=1}^{\infty}$
be a sequence of purely random bits (i.e. a Bernoulli process with
$\P(U_{i}=1)=\frac{1}{2}$), independent of the source $\mathbf{X}$. 

A \emph{secure rate-distortion} \emph{code} ${\cal S}_{n}=(f_{n},\varphi_{n})$
of block-length $n$ is defined by a \emph{key-length} function $k_{n}:{\cal X}^{n}\to\mathbb{Z}_{+}$,
which assigns a key length $k_{n}(\mathbf{x})$ to every $\mathbf{x}\in{\cal X}^{n}$,
an \emph{encoder} $f_{n}:{\cal X}^{n}\times\left\{ 0,1\right\} ^{*}\to{\cal Y}_{n}$,
which generates a cryptogram, $y=f_{n}(\mathbf{x},\mathbf{u})$, where
$\mathbf{u}=(u_{1},\ldots,u_{k_{n}(\mathbf{x})})$, and where ${\cal Y}_{n}$
is a finite alphabet%
\footnote{This alphabet need not be the $n$th order Cartesian power of some
alphabet ${\cal Y}$.%
}, and a\emph{ legitimate decoder} $\varphi_{n}:{\cal Y}_{n}\times\left\{ 0,1\right\} ^{*}\to{\cal W}^{n}$,
which generates a reproduction $\mathbf{w}=\varphi_{n}(y,\mathbf{u})$%
\footnote{It is implicit in the definition of the encoder and decoder that both
are aware of the key-length $k_{n}(\mathbf{x})$. Specifically, one
can define an \emph{inverse-key length }function $l_{n}:{\cal Y}_{n}\times\left\{ 0,1\right\} ^{*}\to\mathbb{Z}_{+},$
which reproduces the key-length at the decoder side, i.e. $k_{n}(\mathbf{x})=l_{n}(y,\{u_{i}\}_{i=1}^{\infty})$. %
}. A sequence of codes $\{{\cal S}_{n}\}_{n\geq1}$, indexed by the
block-length $n$, is denoted by ${\cal S}$. The performance of the
legitimate decoder is evaluated by a distortion measure $d_{\st[L]}:{\cal X}\times{\cal W}\to\mathbb{R}_{+}$,
where without loss of generality (w.l.o.g.), it is assumed that for
every $x\in{\cal X}$, there exists $w\in{\cal W}$ such that $d_{\st[L]}(x,w)=0$.
Also, with a slight abuse of notation, the distortion between $\mathbf{x}$
and $\mathbf{w}$ is defined as the average, 
\begin{equation}
d_{\st[L]}(\mathbf{x},\mathbf{w})\teq\frac{1}{n}\sum_{i=1}^{n}d_{\st[L]}(x_{i},w_{i}).\label{eq: distortion definition leg}
\end{equation}
 We say that ${\cal S}$ satisfies a \emph{compression constraint}
$(\RRc,\DDc,\EEc)$, if the \emph{coding rate }satisfies%
\footnote{This constraint can be weakened to a constraint on the normalized
entropy of the cryptogram. See discussion in Section \ref{sec:Main-Result}. %
} 
\begin{equation}
\limsup_{n\to\infty}\frac{1}{n}\log|{\cal Y}_{n}|\leq\RRc,\label{eq: code-rate constraint}
\end{equation}
and for any given $\{U_{i}\}_{i=1}^{\infty}=\{u_{i}\}_{i=1}^{\infty}$
the \emph{excess-distortion exponent}, at distortion level $\DDc$,
is larger than $\EEc$ for the legitimate decoder, i.e.%
\footnote{This constraint can be weakened to be only satisfied for an excess-distortion
probability averaged over $\left\{ U_{i}\right\} _{i=1}^{\infty}$.
See discussion in Section \ref{sec:Main-Result}. %
} 
\begin{equation}
\liminf_{n\to\infty}-\frac{1}{n}\P\left[d_{\st[L]}(\mathbf{X},\varphi_{n}(f_{n}(\mathbf{X},\mathbf{u}),\mathbf{u}))\geq\DDc\right]\geq\EEc.\label{eq: excess-distortion exponent constraint}
\end{equation}
Note that for a zero excess-distortion exponent $\EEc=0^{+}$, this
requirement implies that an \emph{average-distortion constraint}%
\footnote{Indeed, suppose that $\P\left(d_{\st[L]}(\mathbf{X},\varphi_{n}(f_{n}(\mathbf{X},\mathbf{u}),\mathbf{u}))\geq\DDc\right)$
decays to zero for all $\{u_{i}\}_{i=1}^{\infty}$ , but only sub-exponentially.
Assuming $\overline{d}_{\st[L]}\teq\min_{w\in{\cal W}}\max_{x\in{\cal X}}d_{\st[L]}(x,w)<\infty$,
for any $\delta>0$ and all $n$ sufficiently large 
\begin{alignat*}{1}
\E\left[d_{\st[L]}(\mathbf{X},\mathbf{W})\right] & \leq\DDc\cdot\P\left[d_{\st[L]}(\mathbf{X},\varphi_{n}(f_{n}(\mathbf{X},\mathbf{u}),\mathbf{u}))\leq\DDc\right]+\overline{d}_{\st[L]}\cdot\P\left[d_{\st[L]}(\mathbf{X},\varphi_{n}(f_{n}(\mathbf{X},\mathbf{u}),\mathbf{u}))\leq\DDc\right]\\
 & \leq\DDc+\overline{d}_{\st[L]}\cdot\P\left[d_{\st[L]}(\mathbf{X},\varphi_{n}(f_{n}(\mathbf{X},\mathbf{u}),\mathbf{u}))\leq\DDc\right]\\
 & \leq\DDc+\delta.
\end{alignat*}
} $\E\left[d_{\st[L]}(\mathbf{X},\mathbf{W})\right]\leq\DDc$ is also
satisfied. An \emph{eavesdropper} decoder is a function $\sigma_{n}:{\cal Y}_{n}\to{\cal Z}^{n}$,
where $\mathbf{z}=\sigma_{n}(y)$ is the \emph{estimate} of the eavesdropper.
It is assumed that the eavesdropper has full knowledge of all system
properties: The source statistics, the encoder $(f_{n},k_{n})$, and
the legitimate decoder $\varphi_{n}$. The set of all eavesdropper
decoders for a block-length $n$ is denoted by $\Sigma_{n}$. In what
follows, we also consider genie-aided eavesdropper decoders, which
are aware of the type class of the source block, i.e., $\tilde{\sigma}_{n}:{\cal Y}_{n}\times{\cal P}_{n}\to{\cal X}^{n}$,
and in this case, the estimate of the decoder is $\mathbf{z}=\tilde{\sigma}_{n}(y,\hat{Q}_{\mathbf{x}})$.
The set of all genie-aided eavesdropper decoders of block-length $n$
is denoted by $\tilde{\Sigma}_{n}$.

The performance of the eavesdropper is evaluated by a distortion measure
$d_{\st[E]}:{\cal X}\times{\cal Z}\to\mathbb{R}_{+}$, where again,
it is assumed that for every $x\in{\cal X}$, there exists $z\in{\cal Z}$
such that $d_{\st[E]}(x,z)=0$. As before, the distortion between
$\mathbf{x}$ and $\mathbf{z}$ is defined as 
\begin{equation}
d_{\st[E]}(\mathbf{x},\mathbf{z})\teq\frac{1}{n}\sum_{i=1}^{n}d_{\st[E]}(x_{i},z_{i}).\label{eq: distortion defintion eve}
\end{equation}
 For a given $\DD\geq0$, the \emph{exiguous-distortion probability},
for a given code ${\cal S}_{n}$, is denoted by 
\[
p_{d}({\cal S}_{n},\DD)\teq\max_{\sigma_{n}\in\Sigma_{n}}\P\left[d_{\st[E]}(\mathbf{X},\mathbf{Z})\leq\DD\right].
\]
The\emph{ limit inferior exiguous-distortion exponent}, achieved for
a sequence of codes ${\cal S}$, is defined as\textbf{
\begin{equation}
{\cal E}_{d}^{-}({\cal S},\DD)\triangleq\liminf_{n\to\infty}-\frac{1}{n}\log p_{d}({\cal S}_{n},\DD),\label{eq: error exponent definition given code}
\end{equation}
}and the \emph{limit superior exiguous-distortion exponent} achieved,
${\cal E}_{d}^{+}({\cal S},\DD)$, is defined analogously, with limit
superior replacing the limit inferior.\textbf{ }While, ${\cal E}_{d}^{-}({\cal S},\DD)\leq{\cal E}_{d}^{+}({\cal S},\DD)$,
it is guaranteed that $p_{d}(s_{n},\DD)\dot{\geq}\exp\left[-n{\cal E}_{d}^{-}({\cal S},\DD)\right]$
for all sufficiently large block-lengths, while $p_{d}(s_{n},\DD)\doteq\exp\left[-n{\cal E}_{d}^{+}({\cal S},\DD)\right]$
may hold only for some sub-sequence of block-lengths. Thus, ${\cal E}_{d}^{-}({\cal S},\DD)$
is less sensitive to the choice of the block-length. For a given $Q_{X}\in{\cal P}({\cal X})$,
let $n_{l}=n_{0}l$, $l=1,2,\ldots$, be the sub-sequence of block-lengths
such that ${\cal T}_{n}(Q_{X})$ is non-empty, where $n_{0}$ is the
minimal such block-length. We define, with a slight abuse of notation,
the \emph{conditional limit inferior exiguous-distortion exponent}
as 
\begin{equation}
{\cal E}_{d}^{-}({\cal S},\DD,Q_{X})\teq\liminf_{l\to\infty}-\frac{1}{n_{l}}\log\max_{\sigma_{n_{l}}\in\Sigma_{n_{l}}}\P\left[d_{\st[E]}(\mathbf{X},\mathbf{Z})\leq\DD|\mathbf{X}\in{\cal T}_{n_{l}}(Q_{X})\right],\label{eq: error exponent definition given code conditional}
\end{equation}
and ${\cal E}_{d}^{+}({\cal S},\DD,Q_{X})$ is defined analogously. 

The key rate of $\mathbf{x}\in{\cal X}^{n}$ is defined as $r_{n}(\mathbf{x})\triangleq\frac{1}{n}\left|k_{n}(\mathbf{x})\right|$.
A code is termed a \emph{fixed key rate} code of rate $\RR_{0}$ if
$r_{n}(\mathbf{x})=\RR_{0}$ for all $\mathbf{x}\in{\cal X}^{n}$,
otherwise, it is called a \emph{variable key rate} code\emph{,} and
it has an \emph{average key rate} $\E[r_{n}(\mathbf{X})]$. We define
the \emph{conditional key rate} of $Q_{X}\in{\cal P}({\cal X})$ as
\begin{equation}
\overline{R}({\cal S},Q_{X})\teq\lim_{l\to\infty}\E[r_{n_{l}}(\mathbf{X})|\mathbf{X}\in{\cal T}_{n_{l}}(Q_{X})]\label{eq: conditional rates def}
\end{equation}
whenever the limit exist. 

The rate-distortion function of a memoryless source $Q_{X}$, under
the distortion measure $d_{\st[L]}(\cdot,\cdot)$ is denoted by
\[
R_{\st[L]}(Q_{X},\DDc)\teq\min_{Q_{W|X}:\E_{Q}\left[d_{\st[L]}(X,W)\right]\leq\DDc}I_{Q}(X;W)
\]
and, similarly, the rate-distortion function of $Q_{X}$ under the
distortion measure $d_{\st[E]}(\cdot,\cdot)$ is denoted by $R_{\st[E]}(Q_{X},\DD)$. 

The main result of this paper, in Theorem \ref{thm:exiguous distortion expoent},
is a single-letter formula for the largest achievable exiguous-distortion
exponent for codes under a compression constraint $(\RRc,\DDc,\EEc)$
and limited key rate.

\section{Main Result\label{sec:Main-Result}}

The achievability part will be proved using fixed key rate codes,
but in the converse part, we will allow also variable key rate codes,
that satisfy the following assumptions:
\begin{enumerate}
\item \emph{\uline{Upper bound on the key rate:}} As $k_{n}(\mathbf{x})=n\log\left|{\cal X}\right|$
key-bits are always sufficient to perfectly encrypt the source, even
without distortion, it will be assumed that $k_{n}(\mathbf{x})\leq n\log\left|{\cal X}\right|$
for all $\mathbf{x}\in{\cal X}^{n}$.
\item \emph{\uline{Uniform convergence of the conditional key rate:}}
We assume that for every $Q_{X}\in{\cal P}({\cal X})$, conditioned
on $\mathbf{X}\in{\cal T}_{n}(Q_{X})$, the key rate $r_{n}(\mathbf{X})$
converges in probability to $\overline{R}({\cal S},Q_{X})$, and moreover,
this convergence is uniform over ${\cal P}({\cal X})$. Namely, for
any $\delta>0$ 
\begin{equation}
\max_{Q_{X}\in{\cal P}_{n}({\cal X})}\P\left[\left|r_{n}(\mathbf{X})-\overline{R}({\cal S},Q_{X})\right|>\delta|\mathbf{X}\in{\cal T}_{n}(Q_{X})\right]\xrightarrow[n\to\infty]{}0.\label{eq: uniform convergence in probability}
\end{equation}
It is easy to prove that since $0\leq r_{n}(\mathbf{X})\leq\log|{\cal X}|$
with probability $1$, then uniform convergence in the mean (${\cal L}_{1}$
norm) is also satisfied, and the limit in \textbf{\eqref{eq: conditional rates def}}
exists, uniformly over $Q_{X}\in{\cal P}({\cal X})$. 
\item \emph{\uline{Admissible encoders:}} An encoder $f_{n}$ will be
termed \emph{admissible}, if $\mathbf{u}\neq\mathbf{u}'$ implies
that $f_{n}(\mathbf{x},\mathbf{u})\neq f_{n}(\mathbf{x},\mathbf{u}')$
for all $\mathbf{x}\in{\cal X}^{n}$. We assume that $f_{n}$ is an
admissible encoder.

In addition, we make two more assumptions. These assumptions are inessential,
and are only made in order to simplify the exposition of our results.

\item \emph{\uline{Upper bound on the legitimate excess-distortion exponent:}}
It is well known \cite[Theorem 9.5]{csiszar2011information},\cite{marton1974},
that for a given $\DDc$, if 
\[
\liminf_{n\to\infty}\frac{1}{n}\log|{\cal Y}_{n}|\geq\RRc
\]
then there exist a sequence of codes ${\cal S}$ which satisfies the
compression constraint $(\RRc,\DDc,\EEc)$ iff
\begin{equation}
\EEc\leq E_{\st[L]}(P_{X},\DDc,\RRc)\teq\inf_{Q_{X}:R_{\st[L]}(Q_{X},\DDc)>\RRc}D(Q_{X}||P_{X}),\label{eq: Marton exponent}
\end{equation}
where $E_{\st[L]}(P_{X},\DDc,\RRc)$ is known as \emph{Marton's source
coding exponent}. It will be assumed that the required excess-distortion
exponent at the legitimate decoder is strictly positive and not larger
than Marton's exponent, i.e., $0<\EEc\leq E_{\st[L]}(P_{X},\DDc,\RRc)$. 
\item \emph{\uline{Partial ordering between distortion measures:}} The
distortion measure $d_{\st[E]}(\cdot,\cdot)$ will be termed more
\emph{lenient} than\emph{ $d_{\st[L]}(\cdot,\cdot)$,} if for every
$\mathbf{w}\in{\cal W}^{n}$, there exists $\mathbf{z}\in{\cal Z}^{n}$
such that 
\begin{equation}
\left\{ \mathbf{x}\in{\cal X}^{n}:d_{\st[L]}(\mathbf{x},\mathbf{w})\leq\mathsf{D}\right\} \subseteq\left\{ \mathbf{x}\in{\cal X}^{n}:d_{\st[E]}(\mathbf{x},\mathbf{z})\leq\mathsf{D}\right\} ,\label{eq: distortion ordering  - inclusion of distortion set}
\end{equation}
for every $\mathsf{D}\geq0$. This corresponds to a worst case assumption
regarding the distortion measure (and the reproduction alphabet ${\cal Z})$
used by the eavesdropper - it is at least not more demanding than
the distortion measure used by the legitimate decoder. In addition,
this also puts, in some sense, the distortion levels at the legitimate
decoder and at the eavesdropper decoder, on the same scale. Therefore,
it will be assumed that $\DD\geq\DDc$, namely, the distortion level
allowed by the eavesdropper is larger than the one allowed by the
legitimate decoder. It is also easily verified that this assumption
implies 
\[
R_{\st[E]}(Q_{X},\mathsf{D})\leq R_{\st[L]}(Q_{X},\mathsf{D})
\]
for every $\mathsf{D}>0$. 
\end{enumerate}
We denote by 
\begin{equation}
E_{e}^{*}(\DD)\teq\min_{Q_{X}}\left\{ D(Q_{X}||P_{X})+R_{\st[E]}(Q_{X},\DD)\right\} \label{eq: perfect secrecy exponent}
\end{equation}
the \emph{perfect-secrecy exponent.} Using standard method of types,
it can be shown that this is the maximal exiguous-distortion exponent
that can be achieved when the eavesdropper blindly estimates the source,
i.e. without using the cryptogram. Alternatively, as evident from
Theorem \ref{thm:exiguous distortion expoent}, this is the maximal
exponent for unlimited key rate. We are now ready to state our main
result.
\begin{thm}
\label{thm:exiguous distortion expoent}Let $\delta>0$ be given.
Then, there exists a sequence of codes ${\cal S}$ of fixed key rate
$\RR$, which satisfies a compression constraint \textup{$(\RRc+\delta,\DDc,\EEc)$
}and properties 1-5 above,  
\begin{equation}
{\cal E}_{d}^{-}({\cal S},\DD)\geq\min\left\{ \RR,E_{e}^{*}(\DD)\right\} -\delta\label{eq: main theorem achievable}
\end{equation}
for all $\DD\geq\DDc$. Conversely, for every sequence of codes ${\cal S}$
of average key rate $\E[r_{n}(\mathbf{x})]\leq\RR$ for all $n$,
which satisfies a compression constraint \textup{$(\RRc,\DDc,\EEc)$}\textup{\emph{
}}and properties 1-5 above,\textup{\emph{ }}
\begin{equation}
{\cal E}_{d}^{+}({\cal S},\DD)\leq\min\left\{ \RR,E_{e}^{*}(\DD)\right\} \label{eq: main theorem converse}
\end{equation}
for all $\DD\geq\DDc$.
\end{thm}
Section \ref{sec:Proof-of-Main} is devoted to the proof of Theorem
\ref{thm:exiguous distortion expoent}, and here we discuss its implications.
The main implication of this theorem is that the performance of lossy
compression and encryption are essentially decoupled. Note that in
Theorem \ref{thm:exiguous distortion expoent}, the exiguous-distortion
exponent of the eavesdropper is determined solely by the key rate
and the distortion level $\DD$ at the eavesdropper, and not by the
compression constraint $(\RRc,\DDc,\EEc)$ (as long as the assumptions
hold). Specifically, it holds for $\DDc=0$, which means that increasing
$\DDc$ does not increase $\DD$. In other words, reducing the amount
of information sent to the legitimate decoder cannot improve secrecy.
Nonetheless, on a positive note, as long as $\RR\leq E_{e}^{*}(\DD)$,
the maximal secrecy can be attained, for every $\DD\geq\DDc$, without
affecting the compression performance. In addition, note that in Theorem
\ref{thm:exiguous distortion expoent}, $\DD$ has a special stature:
A single sequence of codes ${\cal S}$ is \emph{universal} for all
$\DD\geq\DDc$. This enables the construction of secure rate-distortion
codes that are robust to the choice of $\DD$, which may be unspecified
when designing the system. 

As previously mentioned, the achievability part of Theorem \ref{thm:exiguous distortion expoent}
is proved using fixed rate codes. Since fixed rate codes clearly satisfy
the second assumption above, the maximal exiguous-distortion exponent
is fully characterized for fixed key rate coding. Furthermore, the
theorem shows that variable key rate codes, from the class of codes
which satisfy the above assumptions, offer no advantage over fixed
key rate codes in terms of exiguous-distortion exponent. This is in
contrast to similar problems (variable-rate channel coding with feedback
\cite{Burnashev_feeddback,channel_feedback_variable_rate}, variable-rate
Slepian-Wolf coding \cite{SW_paper_extended}), where the more lenient
average-rate constraint allows to increase the error exponent. It
should be mentioned that while the class of variable key rate codes
is restricted to satisfy uniform convergence in probability of the
conditional key rate (see the second assumption above), the important
class of \emph{type dependent variable key rate} codes satisfy this
assumption. In a type dependent variable key rate code, the key rate
$r_{n}(\mathbf{x})$ depends on $\mathbf{x}$ only via its type, namely,
$\hat{Q}_{\mathbf{x}}=\hat{Q}_{\mathbf{\tilde{\mathbf{x}}}}$ implies
$r_{n}(\mathbf{x})=r_{n}(\tilde{\mathbf{x}})=\rho(Q_{X})$ for some
\emph{key rate function} $\rho(\cdot):{\cal P}({\cal X})\to\mathbb{R}^{+}$.
Due to the symmetry of source blocks from the same type class, such
a key rate allocation is indeed plausible, and also practically motivated
due to its simplicity. Such codes trivially satisfy the convergence
requirement, and so the converse part of Theorem \ref{thm:exiguous distortion expoent}
is valid. 

Theorem \ref{thm:exiguous distortion expoent} essentially generalizes
\cite[Theorem 1]{secrecy_large_dev}. In \cite{secrecy_large_dev},
it was assumed that all alphabets are identical ${\cal X}={\cal W}={\cal Z}$,
and that $\DD=\DDc=0$. Thus, the legitimate decoder need to perfectly
reproduce the source block, and the eavesdropper performance is measured
by its probability of correct estimate, i.e. 
\[
p_{d}({\cal S}_{n},\DD)=\max_{\sigma_{n}\in\Sigma_{n}}\P(\mathbf{X}=\mathbf{Z}).
\]
Note also that for this specific case, the perfect-secrecy exponent
for this case is given by
\begin{align*}
E_{e}^{*}(\DD) & =\min_{Q_{X}}\left\{ D(Q_{X}||P_{X})+H(Q_{X})\right\} \\
 & =-\log\max_{x\in{\cal X}}P_{X}(x).
\end{align*}
Indeed, even without using the cryptogram, the eavesdropper can choose
$\mathbf{z}=(x^{*},\ldots,x^{*})$ where $x^{*}=\max_{x\in{\cal X}}P_{X}(x)$,
and achieve $E_{e}^{*}(\DD)$.

\section{Outline of the Proof of Theorem \ref{thm:exiguous distortion expoent}\label{sec:Outline-of-the proof}}

Since the proof of Theorem \ref{thm:exiguous distortion expoent}
is considerably involved, this section is devoted to an informal description
of the structure and the main ideas in this proof. Hopefully, this
will facilitate the reading of the formal proof, or at least give
the reader an idea of the main highlights. 

To begin, we observe, in Subsection \ref{sub:Type-Awarness}, that
the exiguous-distortion exponent remains unchanged even if the eavesdropper
is aware of the type of the source block $\hat{Q}_{\mathbf{x}}$.
This enables us to first, consider each type of the source separately,
and only then incorporate all types simultaneously, both in the achievability
and the converse parts. Next, in Subsection \ref{sub:Covering-A-Type},
we provide a technique which facilitates the construction of secure
rate-distortion codes, such that in view of the eavesdropper the cryptograms
are symmetric. The idea is to \emph{cover} a type class ${\cal T}_{n}(Q_{X})$
using an essentially minimal number of permutations of a constituent
set ${\cal D}_{n}\subseteq{\cal T}_{n}(Q_{X})$. To wit, if ${\cal D}_{n}\teq\left\{ \mathbf{x}(0),\ldots,\mathbf{x}(|{\cal D}_{n}|-1)\right\} $
then for any permutation $\pi$ over $\{1,\ldots,n\}$, we define
\begin{equation}
\pi({\cal D}_{n})\teq\left\{ \pi(\mathbf{x}(0)),\ldots,\pi(\mathbf{x}(|{\cal D}_{n}|-1))\right\} ,\label{eq: permutation of set defintion}
\end{equation}
and then find a set of permutations $\{\pi_{n,t}\}_{t=0}^{\kappa_{n}}$
such that
\begin{equation}
\bigcup_{t=0}^{\kappa_{n}}\pi_{n,t}({\cal D}_{n})={\cal T}_{n}(Q_{X}),\label{eq: permutations cover the entire type}
\end{equation}
where $\kappa_{n}$ is asymptotically close to its minimal value of
$\frac{\left|{\cal T}_{n}(Q_{X})\right|}{|{\cal D}_{n}|}$. For ordinary
rate-distortion, such covering lemma can be used to show the existence
of a good rate-distortion code (e.g. instead of \cite[Lemma 9.1]{csiszar2011information}).
Let us define, the \emph{D-cover }of $\mathbf{w}\in{\cal W}^{n}$
as 
\begin{equation}
\mathfrak{D}(\mathbf{w},Q_{X},\DDc)\teq\left\{ \mathbf{x}\in{\cal T}_{n}(Q_{X}):d_{\st[L]}(\mathbf{x},\mathbf{w})\leq\DDc\right\} .\label{eq: D-cover single codeword}
\end{equation}
If we set ${\cal D}_{n}=\mathfrak{D}(\mathbf{w},Q_{X},\DDc)$ and
find permutations $\{\pi_{n,t}\}_{t=0}^{\kappa_{n}}$ such that \eqref{eq: permutations cover the entire type}
holds, then the set $\hat{{\cal C}}_{n}\teq\{\pi_{n,t}(\mathbf{w})\}_{t=0}^{\kappa_{n}}$
is a rate-distortion code such that for every $\mathbf{x}\in{\cal T}_{n}(Q_{X})$
there exists $\mathbf{w}\in\hat{{\cal C}}_{n}$ such that $d_{\st[L]}(\mathbf{x},\mathbf{w})\leq\DDc$.
Such permutations can be found for all types of the source, and using
the method of types, it can be verified that Marton's source coding
exponent can be achieved by such a construction. For the construction
of secure rate-distortion codes, we will use permutations of more
complicated sets to cover the type.

The achievability part (lower bound) is proved in Subsection \ref{sub:Proof-of-Achievability}
using codes of fixed key rate $\RR$. Let us first focus on a single
type $Q_{X}$. For the legitimate decoder, a source block $\mathbf{x}\in{\cal T}_{n}(Q_{X})$
is reproduced by some $\mathbf{w}\in\overline{{\cal C}}_{n}\teq\left\{ \varphi_{n}(y,\mathbf{u}):y\in{\cal Y}_{n},\mathbf{u}\in\{0,1\}^{n\RR}\right\} $,
which satisfies $d_{\st[L]}(\mathbf{x},\mathbf{w})\le\DDc$, unless
no such $\mathbf{w}$ exists. The compression constraint $(\RRc,\DDc,\EEc)$
ensures that large-distortion reproduction occurs with an exponentially
decaying probability. The eavesdropper, on the other hand, reproduces
using only the cryptogram $y$. With a slight abuse of notation of
\eqref{eq: D-cover single codeword}, let us define, for a given the
D-cover of ${\cal C}_{n}\subseteq{\cal W}^{n}$ as 
\begin{equation}
\mathfrak{D}({\cal C}_{n},Q_{X},\DDc)\teq\bigcup_{\mathbf{w}\in{\cal C}_{n}}\mathfrak{D}(\mathbf{w},Q_{X},\DDc).\label{eq: D-cover codebook}
\end{equation}
When the eavesdropper observes $y$, it knows that the legitimate
decoder will reproduce $\mathbf{w}$ from the set ${\cal C}_{n}(y)=\left\{ \varphi_{n}(y,\mathbf{u}):\mathbf{u}\in\{0,1\}^{n\RR}\right\} $
of size $|{\cal C}_{n}(y)|=2^{n\RR}$. Furthermore, conditioning on
the cryptogram $y$ and the type $Q_{X}$, the source block $\mathbf{X}$
is distributed uniformly over $\mathfrak{D}({\cal C}_{n}(y),Q_{X},\DDc)$.
The proof of achievability is divided into three steps. In the first
step (Lemma \ref{lem: distortion packing exponent}), we demonstrate
the existence of a good and secure rate-distortion code conditioned
on a single cryptogram, in the second step, we extend this code for
an entire type class ${\cal T}_{n}(Q_{X})$ (Lemma \ref{lem: exponent lower bound, single type}),
and in the third step, we extend it to all types. 

In more detail, the first step of the proof (Lemma \ref{lem: distortion packing exponent})
shows, by a random selection mechanism, that there exists a set ${\cal C}_{n}^{*}$
of size $2^{n\RR}$ such that when $\mathbf{X}$ is distributed uniformly
over $\mathfrak{D}({\cal C}_{n}^{*},Q_{X},\DDc)$, the exiguous-distortion
probability of any eavesdropper is asymptotically not larger than
$2^{-n\cdot\min\{\RR,R_{\st[E]}(Q_{X},\DD)\}}$. Geometrically, this
implies that the \emph{D-covers} for $\mathbf{w}\in{\cal C}_{n}$
are distant from each other, under $d_{\st[E]}(\cdot,\cdot)$. Thus,
a secure rate-distortion code satisfying ${\cal C}_{n}(y)={\cal C}_{n}^{*}$
for some cryptogram $y$, will have a good conditional exiguous-distortion
probability given $y$. 

In the second step, we define the code for all $\mathbf{x}\in{\cal T}_{n}(Q_{X})$,
using a symmetry argument. Observe that the distortion measures of
both the legitimate and eavesdropper decoders are invariant to permutations
(see \eqref{eq: distortion definition leg} and \eqref{eq: distortion defintion eve}).
Thus, $\mathfrak{D}(\pi({\cal C}_{n}),Q_{X},\DDc)=\pi\left(\mathfrak{D}({\cal C}_{n},Q_{X},\DDc)\right)$,
and the exiguous-distortion probability for an eavesdropper when $\mathbf{X}$
is distributed uniformly over $\pi\left(\mathfrak{D}({\cal C}_{n},Q_{X},\DDc)\right)$
is the same as for $\mathfrak{D}({\cal C}_{n},Q_{X},\DDc)$. In Lemma
\ref{lem: exponent lower bound, single type}, we use a minimal number
of permutations (from Subsection \ref{sub:Covering-A-Type}) of a
good D-cover $\mathfrak{D}({\cal C}_{n}^{*},Q_{X},\DDc)$ to cover
${\cal T}_{n}(Q_{X})$, and then obtain a good secure rate-distortion
code for all ${\cal T}_{n}(Q_{X})$. There is a certain subtlety in
the proof of Lemma \ref{lem: exponent lower bound, single type}.
For an ordinary rate-distortion code, there might be more than a single
$\mathbf{w}\in\overline{{\cal C}}_{n}$ such that $d_{\st[L]}(\mathbf{x},\mathbf{w})\le\DDc$.
From the excess-distortion probability point of view, there is no
importance to which one of these $\{\mathbf{w}\}$ will reproduce
$\mathbf{x}$. However, this might result in $\mathbf{w}\in\overline{{\cal C}}_{n}$
for which only a small portion of $\mathfrak{D}(\mathbf{w},Q_{X},\DDc)$
is actually reproduced by $\mathbf{w}$ (as $\mathbf{x}\in\mathfrak{D}(\mathbf{w},Q_{X},\DDc)$
might be reproduced by some $\mathbf{w}'\in\overline{{\cal C}}_{n}$
which also satisfies $d_{\st[L]}(\mathbf{x},\overline{\mathbf{w}})\leq\DDc$),
which might be harmful for secrecy purposes. Indeed, the secure rate-distortion
code is constructed in Lemma \ref{lem: exponent lower bound, single type}
with the will that conditioned on any cryptogram $y$, the source
is distributed uniformly over $\mathfrak{D}({\cal C}_{n}^{*},Q_{X},\DDc)$.
But, since a source block must eventually be reproduced by a single
$\mathbf{w}$, then conditioned on some of the cryptograms $y$, the
source block will be distributed on a smaller set than $\mathfrak{D}({\cal C}_{n}^{*},Q_{X},\DDc)$.
For such cryptograms, the conditional exiguous-distortion probability
of the eavesdropper might be large. Lemma \ref{lem: exponent lower bound, single type}
shows that if the efficient covering described above is utilized,
then the total effect of such events is negligible. 

Until this stage, we have constructed a code for ${\cal T}_{n}(Q_{X})$
with appropriate conditional exiguous-distortion exponent. As we shall
see, in the construction of Lemma \ref{lem: distortion packing exponent}
and Lemma \ref{lem: exponent lower bound, single type}, the convergence
of probabilities to their asymptotic exponent is not necessarily uniform
(cf. Remark \ref{rem: uniform convergence is not assured}). In the
third step of the achievability proof, we prove that uniform convergence
is possible, using an elaborated construction, built from the previous
one. The idea is to consider a dense grid on the simplex ${\cal Q}({\cal X})$,
and construct a secure rate-distortion code, as in Lemma \ref{lem: exponent lower bound, single type},
for each of the types in the grid. Since the of number of types in
the grid is \emph{finite},\emph{ }then uniform convergence is assured
for types in the grid. If the type of the source block belongs to
the grid, then one of the constructed codes is used, according to
its type. Otherwise, the source block will be first modified, such
that the modified source block does have type within the grid, which
is not very far from the type of the original source block. The modified
source block will then be encoded using one of the codes of the grid,
and thus will have both low legitimate excess-distortion probability,
and large exiguous-distortion probability for the eavesdropper. It
will be shown that the overheads required for the legitimate decoder
to reproduce the original source block, rather than the modified source
block are negligible. 

In Subsection \ref{sub:Proof-of-Converse}, we prove the converse
part in two steps. Recall that in general, for any given type $Q_{X}\in{\cal P}({\cal X})$,
we have defined the average rate $\overline{R}({\cal S},Q_{X})$,
but we allow each source block $\mathbf{x}\in{\cal T}_{n}(Q_{X})$
to have a different key rate $r_{n}(\mathbf{x})\in[0,\log_{n}|{\cal X}|]$.
In addition, for a code satisfying the compression constraint $(\RRc,\DDc,\EEc)$,
and type $Q_{X}$ such that $D(Q_{X}||P_{X})\leq\EEc$, the legitimate
excess-distortion probability must decay to zero exponentially as
$2^{-n\left[\EEc-D(Q_{X}||P_{X})\right]}$ but does not need to be
strictly zero. In the first step of the proof of the converse, we
prove a lemma that shows that the optimal limit superior exiguous-distortion
exponent is not deteriorated, if we restrict $r_{n}(\mathbf{x})$
to be a constant within ${\cal T}_{n}(Q_{X})$, which is less than
$\overline{R}({\cal S},Q_{X})+\delta$, and also restrict the legitimate
excess-distortion probability to be exactly zero. It will be easier
to prove a converse for codes with such properties, as will be done
in the second step of the proof. In the second step, we assume the
structure of the code from the first step, and evaluate the performance
of an eavesdropper which adopts one of the following two simple strategies:
(1) It can guess the secret key bits, and then decode using these
bits just like the legitimate decoder. (2) It can ignore the cryptogram
altogether and choose an estimate $\mathbf{z}\in{\cal Z}^{n}$, based
on only $\hat{Q}_{\mathbf{x}}$. Clearly, in the first case, the probability
of success is $2^{-n\RR}$, and it is not difficult to show that the
exiguous-distortion probability for the second strategy is asymptotically
$2^{-nE_{e}^{*}(\DD)}$. This implies the upper bound \eqref{eq: main theorem converse}.
We remark that the asymptotic optimality of these two simple strategies
(sometimes called \emph{key-attack} and \emph{blind guessing}, respectively)
can also be found to some extent in related problems \cite{Henchman_ISIT,Neri_guessing2,Haroutunian_arxiv}. 

We conclude the outline of the proof with the following comments:
\begin{itemize}
\item \emph{\uline{Awareness of key-length:}} Since the number of possible
key-lengths is $n\log|{\cal X}|$, it can be compressed and fully
encrypted using negligible coding rate and key rate of $\frac{1}{n}\log(n\log|{\cal X}|)$
bits, and it can be assumed that the exiguous-distortion exponent
is not deteriorated if the eavesdropper is aware of the key-length
(as in Subsection \ref{sub:Type-Awarness}). Thus, in the converse
proof, we could have found the exiguous-distortion exponent conditioned
on both the type and the key-length, and then average over them. The
main obstacle in this approach is proving the second property (full
type covering) assured in Lemma \ref{lem: converse - fixed type key-rate}.
To show this property using the methods of Lemma \ref{lem: converse - fixed type key-rate},
would require showing that the subsets of the type classes of fixed
key-length, i.e., $\tilde{{\cal T}}_{n}(Q_{X},m)\teq{\cal T}_{n}(Q_{X})\cap\left\{ \mathbf{x}:k_{n}(\mathbf{x})=m\right\} $
for some $0\leq m\leq n\log|{\cal X}|$, can cover a type class by
essentially a minimal number of permutations, as in Lemma \ref{lem: Covering Lemma Awhlsede}
(Subsection \ref{sub:Covering-A-Type}). However, in turn, the proof
of Lemma \ref{lem: Covering Lemma Awhlsede} is based on the fact
that ${\cal T}_{n}(Q_{X})$ is invariant to permutations, which may
not hold for $\tilde{{\cal T}}_{n}(Q_{X},m)$. 
\item \emph{\uline{Full type covering:}}\emph{ }Let $Q_{X}\in{\cal P}({\cal X})$
be given such that $D(Q_{X}||P_{X})<\EEc$. The method of types and
the expression \eqref{eq: Marton exponent} reveal that to satisfy
the compression constraint $(\RRc,\DDc,\EEc)$, the following condition
should hold for any given $\{u_{i}\}_{i=1}^{\infty}$
\begin{equation}
\P\left[d_{\st[L]}(\mathbf{X},\varphi_{n}(f_{n}(\mathbf{X},\mathbf{u}),\mathbf{u}))>\DDc|\mathbf{X}\in{\cal T}_{n}(Q_{X})\right]\doteq2^{-n\left[\EEc-D(Q_{X}||P_{X})\right]}.\label{eq: not covering with positive exponent}
\end{equation}
For ordinary rate-distortion codes, it is well known%
\footnote{This can also be easily verified using Lemma \ref{lem: Covering Lemma Awhlsede}.%
} that if for a given $\epsilon\in(0,1)$ and for all $n$ sufficiently
large
\begin{equation}
\P\left[d_{\st[L]}(\mathbf{X},\mathbf{W})>\DDc\right]\leq1-\epsilon\label{eq: epsilon covering of a codeword}
\end{equation}
then there exists a rate-distortion code with almost the same rate,
such that 
\begin{equation}
\P\left[d_{\st[L]}(\mathbf{X},\mathbf{W})>\DDc\right]=0.\label{eq: zero covering of a codeword}
\end{equation}
Thus, to ensure an exponent constraint $\EEc$ for ordinary rate-distortion
codebook, the type classes of types which are `close' enough to $P_{X}$
(in the divergence sense) should be almost covered by the reproduction
set \eqref{eq: not covering with positive exponent}, but in fact,
can be \emph{fully} covered by the reproduction set \eqref{eq: zero covering of a codeword}.
Then, the minimal rate required to satisfy \eqref{eq: not covering with positive exponent}
is the same as the minimal rate to satisfy \eqref{eq: zero covering of a codeword},
and the compression rate cannot be decreased due to the softer requirement
in \eqref{eq: not covering with positive exponent}. By contrast,
in the presence of the eavesdropper, it might happen that the softer
requirement in \eqref{eq: not covering with positive exponent} can
lead to better exiguous-distortion exponent: Even if a type class
can be fully covered using the available coding rate, perhaps the
exiguous-distortion exponent can be improved if some of the source
blocks are reproduced with distortion larger than $\DDc$, but this
occurs with sufficiently small probability, as in \eqref{eq: not covering with positive exponent}.
Lemma \ref{lem: converse - fixed type key-rate} shows that this is
\emph{not} the case.
\item \emph{\uline{Compression constraint conditions:}}\emph{ }The conditions
required to satisfy the coding rate constraint \eqref{eq: code-rate constraint},
and the excess-distortion exponent constraint for the legitimate decoder
\eqref{eq: excess-distortion exponent constraint} can be weakened
without affecting Theorem \ref{thm:exiguous distortion expoent}.
First, \eqref{eq: code-rate constraint} can be weakened to 
\begin{equation}
\limsup_{n\to\infty}\frac{1}{n}H(Y)\leq\RRc,\label{eq: code-rate constraint weak}
\end{equation}
where $H(Y)$ is the entropy of the cryptogram. Second, the excess-distortion
exponent can be weakened to apply to the expectation constraint over
the key-bits $\{U_{i}\}_{i=1}^{\infty}$, rather than for every given
$\{u_{i}\}_{i=1}^{\infty}$, i.e. 
\begin{equation}
\liminf_{n\to\infty}-\frac{1}{n}\P\left[d_{\st[L]}(\mathbf{X},\varphi_{n}(f_{n}(\mathbf{X},\mathbf{U}),\mathbf{U}))\geq\DDc\right]\geq\EEc.\label{eq: excess-distortion exponent constraint weak}
\end{equation}
Obviously, since the achievability part is proved using the stronger
conditions \eqref{eq: code-rate constraint} and \eqref{eq: excess-distortion exponent constraint},
it also holds under the weaker conditions \eqref{eq: code-rate constraint weak}
and \eqref{eq: excess-distortion exponent constraint weak}. For the
converse, note that in Lemma \ref{lem: converse - fixed type key-rate}
and in the proof of the converse, the coding rate is essentially not
constrained. The excess-distortion exponent constraint is used in
the converse proof only in eq. \eqref{eq: usage of excess-distortion constraint},
which follows directly from the weaker condition \eqref{eq: excess-distortion exponent constraint weak}.
Therefore, the achievability part holds under the strong conditions,
and the converse part holds under the weak conditions. 
\item \emph{\uline{Legitimate excess-distortion exponent:}} As is evident
from Theorem \ref{thm:exiguous distortion expoent}, there is no improvement
in the exiguous-distortion exponent even if $\EEc$ vanishes (to wit,
the distortion $\DDc$ is achieved only on the average). Thus, the
excess-distortion exponent can be set to its maximal value of $E_{\st[L]}(P_{X},\DDc,\RRc)$,
as defined in \eqref{eq: Marton exponent}. 
\item \emph{\uline{Dependency on the source distribution:}} From the
proof of the achievability, it is evident that given $\hat{Q}_{\mathbf{x}}$,
the operation of the encoder, the legitimate decoder and the eavesdropper
decoder depend on $P_{X}$ only on whether $\RRc>R_{\st[L]}(\hat{Q}_{\mathbf{x}},\DDc)$
or not (equivalently, from the previous comment, whether $D(Q_{X}||P_{X})\leq\EEc$
or not). Since it can be assumed that $\hat{Q}_{\mathbf{x}}$ is known
to all parties, then prior knowledge of the source distribution $P_{X}$
is not required to either party. Hence, the secure rate-distortion
codes constructed are \emph{universal}. Of course, the exponents achieved
depend on $P_{X}$. 
\end{itemize}

\section{Proof of the Theorem \ref{thm:exiguous distortion expoent}\label{sec:Proof-of-Main}}

We remind the reader the \emph{reverse Markov inequality} \cite[Section 9.3, p. 159]{loeve},
which is a useful tool for the proof.
\begin{lem}
\label{lem: reverse Markov}Let $X$ be a positive random variable
which satisfies $\P(X\leq\alpha\E[X])=1$ for some $\alpha>1$. Then,
for any $\beta<1$,
\[
\P\left(X>\beta\E[X]\right)\geq\frac{1-\beta}{\alpha-\beta}.
\]

\end{lem}
The proof is based on the ordinary Markov inequality for the positive
random variable $\tilde{X}=\alpha\E[X]-X$.

\subsection{Type Awareness of the Eavesdropper\label{sub:Type-Awarness}}

Consider the following simple observation, which simplifies later
derivations: The largest achievable exiguous-distortion exponent is
not deteriorated if the eavesdropper is aware of the type of the source
block, in addition to the cryptogram. 
\begin{prop}
\label{prop: Type awarness does not help}For any $Q_{X}\in{\cal P}({\cal X})$
\[
{\cal E}_{d}^{-}({\cal S},\DD,Q_{X})=\liminf_{n\to\infty}\left\{ -\frac{1}{n}\max_{\sigma_{n}\in\tilde{\Sigma}_{n}}\log\P\left[d_{\st[E]}(\mathbf{X},\mathbf{Z})\leq\DD|\mathbf{X}\in{\cal T}_{n}(Q_{X})\right]\right\} .
\]
An analogous result holds for ${\cal E}_{d}^{+}({\cal S},\DD,Q_{X})$.\end{prop}
\begin{IEEEproof}
Since $\Sigma_{n}\subset\tilde{\Sigma}_{n}$ 
\begin{equation}
{\cal E}_{d}^{-}({\cal S},\DD,Q_{X})\geq\liminf_{n\to\infty}\left\{ -\frac{1}{n}\log\max_{\sigma_{n}\in\tilde{\Sigma}_{n}}\P\left[d_{\st[E]}(\mathbf{X},\mathbf{Z})\leq\DD|\mathbf{X}\in{\cal T}_{n}(Q_{X})\right]\right\} .\label{eq: type awareness conditional exponent}
\end{equation}
To show equality, let $\{\tilde{\sigma}_{n}^{*}\in\tilde{\Sigma}_{n}\}$
be the sequence of decoders which achieve the maximum in the right
hand side of \eqref{eq: type awareness conditional exponent}. Let
us define a sequence of decoders $\{\sigma{}_{n}\in\Sigma_{n}\}$
as follows. First, $\sigma_{n}$ produces a random guess $Q\in{\cal P}_{n}$
of the type of the source, with the uniform distribution over ${\cal P}_{n}$,
and second, it decodes 
\[
\sigma_{n}(y)=\tilde{\sigma}_{n}^{*}(y,Q).
\]
Given $Q_{X}\in{\cal P}$, the resulting conditional exiguous-distortion
probability is given by
\begin{align*}
 & \phantom{{}\geq{}}\P\left[d_{\st[E]}(\mathbf{X},\sigma_{n}(Y))\leq\DD|\mathbf{X}\in{\cal T}_{n}(Q_{X})\right]\\
 & \geq\P\left[d_{\st[E]}(\mathbf{X},\tilde{\sigma}_{n}^{*}(Y,Q))\leq\DD|Q=\hat{Q}_{\mathbf{x}},\mathbf{X}\in{\cal T}_{n}(Q_{X})\right]\cdot\P\left[Q=\hat{Q}_{\mathbf{x}}|\mathbf{X}\in{\cal T}_{n}(Q_{X})\right]\\
 & =\P\left[d_{\st[E]}(\mathbf{X},\tilde{\sigma}_{n}^{*}(Y,\hat{Q}_{\mathbf{x}}))\leq\DD|\mathbf{X}\in{\cal T}_{n}(Q_{X})\right]\cdot\frac{1}{|{\cal P}_{n}|}
\end{align*}
and as $|{\cal P}_{n}|\leq(n+1)^{|{\cal X}|}$, equality is achieved
in \eqref{eq: type awareness conditional exponent}.
\end{IEEEproof}

\subsection{Covering a Type Class via Permutations\label{sub:Covering-A-Type}}

In this subsection, we discuss the possibility to cover a type class
by means of permutations of a constituent subset. The fact that the
distortion measure of the eavesdropper is invariant to permutations
of both arguments hints on the usefulness of such a covering in the
construction of good secure rate-distortion codes. 

Given a type $Q_{X}\in{\cal P}({\cal X})$ and $\delta>0$, the method
of types implies that for $n>n_{0}(\delta,|{\cal X}|)$
\[
2^{n\left[H(Q_{X})-\delta\right]}\leq|{\cal T}_{n}(Q_{X})|\leq2^{nH(Q_{X})}.
\]
Now, consider the subset ${\cal D}_{n}\subset{\cal T}_{n}(Q_{X})$,
where the elements of ${\cal D}_{n}$ are distinct. We say that a
set of permutations $\{\pi_{n,t}\}_{t=0}^{\kappa_{n}}$ \emph{cover}
${\cal T}_{n}(Q_{X})$ if 
\[
\bigcup_{t=0}^{\kappa_{n}}\pi_{n,t}({\cal D}_{n})={\cal T}_{n}(Q_{X}),
\]
where $\pi_{n,t}({\cal D}_{n})$ means that the same permutation $\pi_{n,t}(\cdot)$
operates on all $\mathbf{x}\in{\cal D}_{n}$, as defined in \eqref{eq: permutation of set defintion}.
Let $\kappa_{n}^{*}$ be the minimal number of permutations of ${\cal D}_{n}$
required to cover ${\cal T}_{n}(Q_{X})$. By a simple counting argument,
we must have
\begin{equation}
\kappa_{n}^{*}\geq\frac{|{\cal T}_{n}(Q_{X})|}{|{\cal D}_{n}|}.\label{eq: lower bound on number of covering permutations}
\end{equation}
The following lemma guaranteed the existence of a cover which essentially
achieves the lower bound.
\begin{lem}[{\cite[Section 6, Covering Lemma 2]{ahlswede1979coloring}}]
\label{lem: Covering Lemma Awhlsede}For every ${\cal D}_{n}\subset{\cal T}_{n}(Q_{X})$,
$Q_{X}\in{\cal P}_{n}({\cal X})$ 
\[
\kappa_{n}^{*}\leq\frac{|{\cal T}_{n}(Q_{X})|}{|{\cal D}_{n}|}\cdot\log|{\cal T}_{n}(Q_{X})|.
\]

\end{lem}
The main application of this lemma is for a sequence of sets $\{{\cal D}_{n}\}_{n=1}^{\infty}$.
Let $n_{l}$ be the sequence of block-lengths such that ${\cal T}_{n_{l}}(Q_{X})$
is non-empty, and let ${\cal D}_{n_{l}}\subset{\cal T}_{n_{l}}(Q_{X})$
such that 
\[
|{\cal D}_{n_{l}}|\doteq2^{n_{l}\tilde{\RR}}.
\]
Then, Lemma \ref{lem: Covering Lemma Awhlsede} implies that for every
$\delta>0$ and $l\geq l_{0}(\delta,|{\cal X}|)$ both
\begin{align*}
\kappa_{n_{l}}^{*} & \geq\frac{2^{n_{l}\left[H(Q_{X})-\delta\right]}}{2^{n_{l}(\tilde{\RR}+\delta)}}\\
 & =2^{n_{l}\left[H(Q_{X})-\tilde{\RR}-2\delta\right]}
\end{align*}
from \eqref{eq: lower bound on number of covering permutations} and
\begin{align*}
\kappa_{n_{l}}^{*} & \leq\frac{2^{n_{l}H(Q_{X})}}{2^{n_{l}(\tilde{\RR}-\delta)}}n_{l}\left[H(Q_{X})+\delta\right]\\
 & \leq2^{n_{l}\left[H(Q_{X})-\tilde{\RR}+2\delta\right]}
\end{align*}
from Lemma \ref{lem: Covering Lemma Awhlsede}. Thus, the cover is
asymptotically efficient, and this implies that the permuted sets
cannot overlap too much. To further explore this property, let $\{\pi_{n_{l},t}\}_{t=0}^{\kappa_{n_{l}}^{*}}$
be the permutations constructed in Lemma \ref{lem: Covering Lemma Awhlsede}
for block-length $n_{l}$, and define the \emph{exclusive permutations
sets} as 
\begin{equation}
{\cal G}_{n_{l},t}\teq\pi_{n_{l},t}({\cal D}_{n_{l}})\backslash\left\{ \bigcup_{s=0}^{t-1}\pi_{n_{l},s}({\cal D}_{n_{l}})\right\} .\label{eq:exclusive permutations sets defintion}
\end{equation}
Note that ${\cal T}_{n_{l}}(Q_{X})$ is a disjoint union ${\cal G}_{n_{l},t}$,
and for any $\overline{\RR}<\tilde{\RR}$, consider the union of exclusive
permutations sets of small cardinality, namely 
\begin{equation}
{\cal H}(\overline{\RR})\teq\bigcup_{t:|{\cal G}_{n_{l},t}|\leq2^{n\overline{\RR}}}{\cal G}_{n_{l},t}.\label{eq: partial union of permutation sets}
\end{equation}
A simple aspect of the asymptotic efficiency of the covering is that
under the uniform distribution on the type class, the probability
that the source block belongs to a small exclusive permutations set
is also small. 
\begin{lem}
\label{lem: Efficient cover aspect 1}For any $\overline{\RR}\leq\tilde{\RR}$
\[
\P\left[\mathbf{X}\in{\cal H}(\overline{\RR})|\mathbf{X}\in{\cal T}_{n}(Q_{X})\right]\dotleq2^{-n(\tilde{\RR}-\overline{\RR})}
\]
\end{lem}
\begin{IEEEproof}
Let an arbitrary $\delta>0$ be given. For all $n$ sufficiently large,
if ${\cal T}_{n}(Q_{X})$ is empty then the statement of the lemma
is satisfied by convention. Otherwise, 
\begin{align*}
\P\left[\mathbf{X}\in{\cal H}(\overline{\RR})|\mathbf{X}\in{\cal T}_{n}(Q_{X})\right] & \leq\frac{\kappa_{n}^{*}\cdot e^{n\overline{\RR}}}{|{\cal T}_{n}(Q_{X})|}\\
 & \leq\frac{2^{n\left[H(Q_{X})-\tilde{\RR}+2\delta\right]}\cdot e^{n\overline{\RR}}}{2^{n\left[H(Q_{X})-\delta\right]}}\\
 & =2^{n(\overline{\RR}-\tilde{\RR}+3\delta)}.
\end{align*}

\end{IEEEproof}

\subsection{Proof of Achievability Part of Theorem \ref{thm:exiguous distortion expoent}
\label{sub:Proof-of-Achievability}}

We follow the three steps outlined in Section \ref{sec:Outline-of-the proof}.
In the first step of the proof, we focus on a single cryptogram, ${\cal C}_{n}(y)=\left\{ \varphi_{n}(y,\mathbf{u}):\mathbf{u}\in\{0,1\}^{n\RR}\right\} $,
which we generically denote by the set ${\cal {\cal C}}_{n}=\{\mathbf{w}(0),\ldots,\mathbf{w}(2^{n\RR}-1)\}\subset{\cal W}^{n}$.
We begin with some definitions and simple properties. For a given
$(\DDc,\DD)$ and $Q_{X}\in{\cal P}_{n}({\cal X})$, let $\tilde{\mathbf{X}}$
be uniformly distributed over $\mathfrak{D}({\cal C}_{n},Q_{X},\DDc)$
(defined in \eqref{eq: D-cover codebook}). The exiguous-distortion
probability for the set ${\cal C}_{n}$ is defined as%
\footnote{With a slight abuse of notation, we also use here the notation $p_{d}(\cdot)$. %
} 
\begin{equation}
p_{d}({\cal C}_{n},Q_{X},\DDc,\DD)\teq\max_{\mathbf{z}\in{\cal Z}^{n}}\P\left[d_{\st[E]}(\tilde{\mathbf{X}},\mathbf{z})\leq\DD\right].\label{eq: exigious distortion packing code probability}
\end{equation}
We have the following simple properties for $p_{d}({\cal C}_{n},Q_{X},\DDc,\DD)$. 
\begin{prop}
\label{prop: simple properties of distortion packing code}Let ${\cal C}_{n}\subset{\cal W}^{n}$
and $Q_{X}\in{\cal P}_{n}({\cal X})$ be given. Then:
\begin{enumerate}
\item \label{enu: permutation property}For every permutation $\pi$
\[
p_{d}({\cal C}_{n},Q_{X},\DDc,\DD)=p_{d}(\pi({\cal C}_{n}),Q_{X},\DDc,\DD),
\]
where $\pi({\cal C}_{n})$ is as defined in \eqref{eq: permutation of set defintion}.
\item \label{enu:subset property}Let $\overline{\mathbf{X}}$ be uniformly
distributed over ${\cal D}_{n}\subseteq\mathfrak{D}({\cal C}_{n},Q_{X},\DDc)$.
Then, 
\[
\max_{\mathbf{z}\in{\cal Z}^{n}}\P\left[d_{\st[E]}(\overline{\mathbf{X}},\mathbf{z})\leq\DD\right]\leq\frac{|\mathfrak{D}({\cal C}_{n},Q_{X},\DDc)|}{|{\cal D}_{n}|}\cdot p_{d}({\cal C}_{n},Q_{X},\DDc,\DD).
\]

\end{enumerate}
\end{prop}
\begin{IEEEproof}
~
\begin{enumerate}
\item Let $\mathbf{z}^{*}$ be the maximizer of \eqref{eq: exigious distortion packing code probability}.
Since $d_{\st[L]}(\mathbf{x},\mathbf{w})=d_{\st[L]}(\pi(\mathbf{x}),\pi(\mathbf{\mathbf{w}}))$
then $\mathfrak{D}(\pi({\cal C}_{n}),Q_{X},\DDc)=\pi\left(\mathfrak{D}({\cal C}_{n},Q_{X},\DDc)\right)$.
Since also $d_{\st[E]}(\mathbf{x},\mathbf{z})=d_{\st[E]}(\pi(\mathbf{x}),\pi(\mathbf{z}))$
then 
\begin{align*}
p_{d}\left[\pi({\cal C}_{n}),Q_{X},\DDc,\DD\right] & =\max_{\mathbf{z}\in{\cal Z}^{n}}\P\left[d_{\st[E]}(\pi(\tilde{\mathbf{X}}),\mathbf{z})\leq\DD\right]\\
 & \geq\P\left[d_{\st[E]}(\pi(\tilde{\mathbf{X}}),\pi(\mathbf{z}^{*}))\leq\DD\right]\\
 & =p_{d}({\cal C}_{n},Q_{X},\DDc,\DD),
\end{align*}
and the reverse inequality can be obtained similarly, by considering
the inverse permutation $\pi^{-1}$. 
\item For every $\mathbf{z}\in{\cal Z}^{n}$
\begin{align*}
\P\left[d_{\st[E]}(\overline{\mathbf{X}},\mathbf{z})\leq\DD\right] & =\frac{\left|\overline{\mathbf{x}}\in{\cal D}_{n}:d_{\st[E]}(\overline{\mathbf{x}},\mathbf{z})\leq\DD\right|}{|{\cal D}_{n}|}\\
 & \leq\frac{\left|\overline{\mathbf{x}}\in\mathfrak{D}({\cal C}_{n},Q_{X},\DDc):d_{\st[E]}(\overline{\mathbf{x}},\mathbf{z})\leq\DD\right|}{|{\cal D}_{n}|}\\
 & =\frac{|\mathfrak{D}({\cal C}_{n},Q_{X},\DDc)|}{|{\cal D}_{n}|}\cdot\frac{\left|\overline{\mathbf{x}}\in\mathfrak{D}({\cal C}_{n},Q_{X},\DDc):d_{\st[E]}(\overline{\mathbf{x}},\mathbf{z})\leq\DD\right|}{|\mathfrak{D}({\cal C}_{n},Q_{X},\DDc)|}\\
 & \leq\frac{|\mathfrak{D}({\cal C}_{n},Q_{X},\DDc)|}{|{\cal D}_{n}|}\cdot p_{d}({\cal C}_{n},Q_{X},\DDc,\DD).
\end{align*}

\end{enumerate}
\end{IEEEproof}
The next lemma is the first step in the proof, in which we prove the
existence of a good set ${\cal C}_{n}^{*}$ by a random selection.
\begin{lem}
\label{lem: distortion packing exponent}Let $\delta>0$ and $Q_{X}\in{\cal P}({\cal X})$
be given, and let $n_{l}$ be the sequence of block-lengths such that
${\cal T}_{n_{l}}(Q_{X})$ is non-empty. There exists a sequence of
sets ${\cal C}^{*}=\{{\cal C}_{n_{l}}^{*}\}$ of size $|{\cal C}_{n_{l}}^{*}|=2^{n_{l}\RR}$
such that for all $l$ sufficiently large 
\begin{equation}
\frac{1}{n_{l}}\log|\mathfrak{D}({\cal C}_{n_{l}}^{*},Q_{X},\DDc)|\geq H(Q_{X})+\RR-R_{\st[L]}(Q_{X},\DDc)-\delta,\label{eq: D-cover size}
\end{equation}
and 
\begin{equation}
-\frac{1}{n_{l}}\log\max_{\mathbf{z}\in{\cal Z}^{n_{l}}}\P\left[d_{\st[E]}(\tilde{\mathbf{X}},\mathbf{z})\leq\DD\right]\geq\min\left\{ \RR,R_{\st[E]}(Q_{X},\DD)\right\} -\delta,\label{eq: lower bound reliability function proof}
\end{equation}
for all $\DD\geq\DDc$, where $\tilde{\mathbf{X}}$ is distributed
uniformly over $\mathfrak{D}({\cal C}_{n}^{*},Q_{X},\DDc)$ .\end{lem}
\begin{IEEEproof}
Let $n$ be given such that ${\cal T}_{n}(Q_{X})$ is non-empty. Also,
let $\DD$ be given, choose any $Q_{W}\in{\cal P}_{n}({\cal W})$,
and consider an ensemble of randomly chosen sets ${\cal C}_{n}$,
where each member is selected independently at random, uniformly within
a type class ${\cal T}_{n}(Q_{W})$. By definition, for any given
${\cal C}_{n}$
\begin{equation}
p_{d}({\cal C}_{n},Q_{X},\DDc,\DD)=\frac{\max_{\mathbf{z}\in{\cal Z}^{n}}\left|\left\{ \mathbf{x}\in\mathfrak{D}({\cal C}_{n},Q_{X},\DDc):d_{\st[E]}(\mathbf{x},\mathbf{z})\leq\DD\right\} \right|}{|\mathfrak{D}({\cal C}_{n},Q_{X},\DDc)|}.\label{eq: random coding pd num/den}
\end{equation}
It should be noticed, that unlike the situation in standard random
coding bounds, here the denominator of \eqref{eq: random coding pd num/den}
is also a random variable. Nonetheless, we will show that there exists
a set ${\cal C}_{n}$ such that both the numerator and denominator
of \eqref{eq: random coding pd num/den} are close to their expected
values. To begin, let us analyze the expected value of the size of
the D-cover in the denominator of \eqref{eq: random coding pd num/den}.
We first consider the case $\RR\leq R_{\st[L]}(Q_{X},\DDc)$. For
a given ${\cal C}_{n}$ and $Q_{XW}$, define the \emph{type class
enumerator}
\[
N(Q_{XW}|\mathbf{x})\teq\left|\left\{ \mathbf{w}\in{\cal C}_{n}:\hat{Q}_{\mathbf{x}\mathbf{w}}=Q_{XW}\right\} \right|,
\]
and let
\[
E_{0}\teq H(Q_{X})+\RR-R_{\st[L]}(Q_{X},\DDc).
\]
Note that in the last equation the $X$-marginal ($W$-marginal) of
$Q$ is constrained to the given type $Q_{X}$ (respectively, $Q_{W}$).
For brevity, here and throughout the sequel, such constraints will
be omitted. Then,
\begin{align}
\E[|\mathfrak{D}({\cal C}_{n},Q_{X},\DDc)|] & =\E\left[\sum_{\mathbf{x}\in{\cal T}_{n}(Q_{X})}\I\left\{ \exists\mathbf{w}\in{\cal C}_{n}:d_{\st[L]}(\mathbf{x},\mathbf{w})\leq\DDc\right\} \right]\\
 & =\E\left[\sum_{\mathbf{x}\in{\cal T}_{n}(Q_{X})}\I\left\{ \bigcup_{Q_{XW}:\E_{Q}\left[d_{\st[L]}(X,W)\right]\leq\DDc}\left\{ N(Q_{XW}|\mathbf{x})\geq1\right\} \right\} \right]\\
 & \doteq\E\left[\sum_{\mathbf{x}\in{\cal T}_{n}(Q_{X})}\sum_{Q_{XW}:\E_{Q}\left[d_{\st[L]}(X,W)\right]\leq\DDc}\I\left\{ N(Q_{XW}|\mathbf{x})\geq1\right\} \right]\\
 & =\sum_{\mathbf{x}\in{\cal T}_{n}(Q_{X})}\sum_{Q_{XW}:\E_{Q}\left[d_{\st[L]}(X,W)\right]\leq\DDc}\P\left\{ N(Q_{XW}|\mathbf{x})\geq1\right\} \\
 & \trre[=,a]\sum_{\mathbf{x}\in{\cal T}_{n}(Q_{X})}\sum_{Q_{XW}:\E_{Q}\left[d_{\st[L]}(X,W)\right]\leq\DDc,I_{Q}(X;W)>\RR}\P\left\{ N(Q_{XW}|\mathbf{x})\geq1\right\} \\
 & \trre[\doteq,b]\sum_{\mathbf{x}\in{\cal T}_{n}(Q_{X})}\sum_{Q_{XW}:\E_{Q}\left[d_{\st[L]}(X,W)\right]\leq\DDc,I_{Q}(X;W)>\RR}2^{n\left[\RR-I_{Q}(X;W)\right]}\label{eq: typ enumerator aysmp 1}\\
 & \doteq2^{nH_{Q}(X)}\max_{Q_{XW}\in{\cal P}_{n}({\cal X}\times{\cal W}):\E_{Q}\left[d_{\st[L]}(X,W)\right]\leq\DDc,I_{Q}(X;W)>\RR}2^{n\left[\RR-I_{Q}(X;W)\right]}\\
 & \trre[=,c]\exp\left\{ n\cdot\left[H_{Q}(X)+\RR-\min_{Q_{XW}\in{\cal P}_{n}({\cal X}\times{\cal W}):\E_{Q}\left[d_{\st[L]}(X,W)\right]\leq\DDc}I_{Q}(X;W)\right]\right\} \\
 & \trre[=,d]2^{nE_{0}},\label{eq: expectation of den}
\end{align}
where in $(a)$ and $(c)$ we have used the assumption $\RR\leq R_{\st[L]}(Q_{X},\DDc)$,
and so, the set $\{Q_{XW}:\E_{Q}\left[d_{\st[L]}(X,W)\right]\leq\DDc,I_{Q}(X;W)\leq\RR\}$
is empty. In $(b)$, we have used the fact that $N(Q_{XW}|\mathbf{x})$
is a binomial random variable pertaining to $2^{n\RR}$ trials and
probability of success of exponential order $\exp\left[-nI_{Q}(X;W)\right]$.
Passage $(d)$ follows from the fact that ${\cal P}({\cal X}\times{\cal W})$
is dense in ${\cal Q}({\cal X}\times{\cal W})$ and $I_{Q}(X;W)$
is continuous. In addition, using the union bound, with probability
$1$, 
\begin{align}
|\mathfrak{D}({\cal C}_{n},Q_{X},\DDc)| & \leq\sum_{\mathbf{w}\in{\cal C}_{n}}\left|\left\{ \mathbf{x}\in{\cal T}_{n}(Q_{X}):d_{\st[L]}(\mathbf{x},\mathbf{w})\leq\DDc\right\} \right|\\
 & \dotleq2^{n\RR}\cdot\exp\left[n\cdot\max_{Q_{XW}\in{\cal P}_{n}({\cal X}\times{\cal W}):\E_{Q}\left[d_{\st[L]}(X,W)\right]\leq\DDc}H_{Q}(X|W)\right]\\
 & =2^{nE_{0}}.\label{eq: deterministic upper bound on den}
\end{align}
Next, we upper bound the numerator of \eqref{eq: random coding pd num/den}.
For a given ${\cal C}_{n}$ and $\mathbf{z}\in{\cal Z}^{n}$, define
now the type class enumerator
\[
N(Q_{ZW}|\mathbf{z})\teq\left|\left\{ \mathbf{w}\in{\cal C}_{n}:\hat{Q}_{\mathbf{z}\mathbf{w}}=Q_{ZW}\right\} \right|.
\]
Then, 
\begin{align*}
 & \hphantom{{}={}}\left|\left\{ \mathbf{x}\in\mathfrak{D}({\cal C}_{n},Q_{X},\DDc):d_{\st[E]}(\mathbf{x},\mathbf{z})\leq\DD\right\} \right|\\
 & =\left|\bigcup_{\mathbf{w}\in{\cal C}_{n}}\left\{ \mathbf{x}\in{\cal T}_{n}(Q_{X}):d_{\st[E]}(\mathbf{x},\mathbf{z})\leq\DD,d_{\st[L]}(\mathbf{x},\mathbf{w})\leq\DDc\right\} \right|\\
 & =\left|\bigcup_{Q_{ZW}}\bigcup_{\mathbf{w}\in{\cal T}_{n}(Q_{W|Z},\mathbf{z})\cap{\cal C}_{n}}\bigcup_{Q_{X|ZW}:\E_{Q}\left[d_{\st[E]}(X,Z)\right]\leq\DD,\E_{Q}\left[d_{\st[L]}(X,W)\right]\leq\DDc}\left\{ \mathbf{x}\in{\cal T}_{n}(Q_{X|ZW},\mathbf{z},\mathbf{w})\right\} \right|\\
 & \trre[\leq,a]\sum_{Q_{ZW}}\sum_{\mathbf{w}\in{\cal T}_{n}(Q_{W|Z},\mathbf{z})\cap{\cal C}_{n}}\sum_{Q_{X|ZW}:\E_{Q}\left[d_{\st[E]}(X,Z)\right]\leq\DD,\E_{Q}\left[d_{\st[L]}(X,W)\right]\leq\DDc}\left|\left\{ \mathbf{x}\in{\cal T}_{n}(Q_{X|ZW},\mathbf{z},\mathbf{w})\right\} \right|\\
 & \doteq\sum_{Q_{ZW}}\sum_{\mathbf{w}\in{\cal T}_{n}(Q_{W|Z},\mathbf{z})\cap{\cal C}_{n}}\sum_{Q_{X|ZW}:\E_{Q}\left[d_{\st[E]}(X,Z)\right]\leq\DD,\E_{Q}\left[d_{\st[L]}(X,W)\right]\leq\DDc}2^{nH_{Q}(X|ZW)}\\
 & \doteq\sum_{Q_{ZW}}\sum_{\mathbf{w}\in{\cal T}_{n}(Q_{W|Z},\mathbf{z})\cap{\cal C}_{n}}\max_{Q_{X|ZW}:\E_{Q}\left[d_{\st[E]}(X,Z)\right]\leq\DD,\E_{Q}\left[d_{\st[L]}(X,W)\right]\leq\DDc}2^{nH_{Q}(X|ZW)}\\
 & =\sum_{Q_{ZW}}N(Q_{ZW}|\mathbf{z})\max_{Q_{X|ZW}:\E_{Q}\left[d_{\st[E]}(X,Z)\right]\leq\DD,\E_{Q}\left[d_{\st[L]}(X,W)\right]\leq\DDc}2^{nH_{Q}(X|ZW)}\\
 & \doteq\max_{Q_{ZW}}\max_{Q_{X|ZW}:\E_{Q}\left[d_{\st[E]}(X,Z)\right]\leq\DD,\E_{Q}\left[d_{\st[L]}(X,W)\right]\leq\DDc}N(Q_{ZW}|\mathbf{z})2^{nH_{Q}(X|ZW)}\\
 & \doteq\sum_{Q_{XZW}:\E_{Q}\left[d_{\st[E]}(X,Z)\right]\leq\DD,\E_{Q}\left[d_{\st[L]}(X,W)\right]\leq\DDc}N(Q_{ZW}|\mathbf{z})2^{nH_{Q}(X|ZW)}
\end{align*}
where $(a)$ is the union bound, and in all the above equations, $Q_{XZW}\in{\cal P}_{n}({\cal X}\times{\cal Z}\times{\cal W})$.
Let 
\[
{\cal J}(\DDc,\DD)\teq\left\{ Q_{XZW}\in{\cal P}_{n}({\cal X}\times{\cal Z}\times{\cal W}):\E_{Q}\left[d_{\st[E]}(X,Z)\right]\leq\DD,\E_{Q}\left[d_{\st[L]}(X,W)\right]\leq\DDc\right\} .
\]
Taking expectation, and using the fact that $|{\cal P}_{n}({\cal X}\times{\cal Z}\times{\cal W})|\leq(n+1)^{|{\cal X}||{\cal Z}||{\cal W}|}$
i.e., increases with $n$ only polynomially, 
\begin{align}
 & \hphantom{{}\leq{}}\E\left[\max_{\mathbf{z}\in{\cal Z}^{n}}\left|\left\{ \mathbf{x}\in\mathfrak{D}({\cal C}_{n},Q_{X},\DDc):d_{\st[E]}(\mathbf{x},\mathbf{z})\leq\DD\right\} \right|\right]\\
 & \dotleq\E\left[\max_{\mathbf{z}\in{\cal Z}^{n}}\sum_{Q_{XZW}\in{\cal J}(\DDc,\DD)}N(Q_{ZW}|\mathbf{z})2^{nH_{Q}(X|ZW)}\right]\\
 & =\E\left[\lim_{\beta\to\infty}\left\{ \sum_{\mathbf{z}\in{\cal Z}^{n}}\left(\sum_{Q_{XZW}\in{\cal J}(\DDc,\DD)}N(Q_{ZW}|\mathbf{z})2^{nH_{Q}(X|ZW)}\right)^{\beta}\right\} ^{\nicefrac{1}{\beta}}\right]\\
 & \trre[=,a]\lim_{\beta\to\infty}\E\left[\left\{ \sum_{\mathbf{z}\in{\cal Z}^{n}}\left(\sum_{Q_{XZW}\in{\cal J}(\DDc,\DD)}N(Q_{ZW}|\mathbf{z})2^{nH_{Q}(X|ZW)}\right)^{\beta}\right\} ^{\nicefrac{1}{\beta}}\right]\\
 & \doteq\lim_{\beta\to\infty}\E\left[\left\{ \sum_{\mathbf{z}\in{\cal Z}^{n}}\left(\max_{Q_{XZW}\in{\cal J}(\DDc,\DD)}N(Q_{ZW}|\mathbf{z})2^{nH_{Q}(X|ZW)}\right)^{\beta}\right\} ^{\nicefrac{1}{\beta}}\right]\\
 & =\lim_{\beta\to\infty}\E\left[\left(\sum_{\mathbf{z}\in{\cal Z}^{n}}\max_{Q_{XZW}\in{\cal J}(\DDc,\DD)}N(Q_{ZW}|\mathbf{z})^{\beta}2^{n\beta H_{Q}(X|ZW)}\right)^{\nicefrac{1}{\beta}}\right]\\
 & \doteq\lim_{\beta\to\infty}\E\left[\left(\sum_{\mathbf{z}\in{\cal Z}^{n}}\sum_{Q_{XZW}\in{\cal J}(\DDc,\DD)}N(Q_{ZW}|\mathbf{z})^{\beta}2^{n\beta H_{Q}(X|ZW)}\right)^{\nicefrac{1}{\beta}}\right]\\
 & \trre[\leq,b]\lim_{\beta\to\infty}\left(\sum_{\mathbf{z}\in{\cal Z}^{n}}\sum_{Q_{XZW}\in{\cal J}(\DDc,\DD)}\E\left[N(Q_{ZW}|\mathbf{z})^{\beta}\right]2^{n\beta H_{Q}(X|ZW)}\right)^{\nicefrac{1}{\beta}}\\
 & =\lim_{\beta\to\infty}\biggl(\sum_{\mathbf{z}\in{\cal Z}^{n}}\sum_{Q_{XZW}\in{\cal J}(\DDc,\DD):I_{Q}(Z;W)\leq\RR}\E\left[N(Q_{ZW}|\mathbf{z})^{\beta}\right]2^{n\beta H_{Q}(X|ZW)}\nonumber \\
 & \hphantom{=\lim_{\beta\to\infty}\biggl(}+\sum_{\mathbf{z}\in{\cal Z}^{n}}\sum_{Q_{XZW}\in{\cal J}(\DDc,\DD):I_{Q}(Z;W)>\RR}\E\left[N(Q_{ZW}|\mathbf{z})^{\beta}\right]2^{n\beta H_{Q}(X|ZW)}\biggr)^{\nicefrac{1}{\beta}}\\
 & \trre[\doteq,c]\lim_{\beta\to\infty}\biggl(\sum_{\mathbf{z}\in{\cal Z}^{n}}\sum_{Q_{XZW}\in{\cal J}(\DDc,\DD):Q_{Z}=\hat{Q}_{\mathbf{z}},I_{Q}(Z;W)\leq\RR}2^{n\beta\left[\RR-I_{Q}(Z;W)\right]}2^{n\beta H_{Q}(X|ZW)}\nonumber \\
 & \hphantom{=\lim_{\beta\to\infty}\biggl(}+\sum_{\mathbf{z}\in{\cal Z}^{n}}\sum_{Q_{XZW}\in{\cal J}(\DDc,\DD):Q_{Z}=\hat{Q}_{\mathbf{z}},I_{Q}(Z;W)>\RR}2^{n\left[\RR-I_{Q}(Z;W)\right]}2^{n\beta H_{Q}(X|ZW)}\biggr)^{\nicefrac{1}{\beta}}\label{eq: type class enumerator asymp 2}\\
 & \doteq\lim_{\beta\to\infty}\biggl(\sum_{Q_{Z}}2^{nH_{Q}(Z)}\sum_{Q_{XW|Z}:\E_{Q}\left[d_{\st[E]}(X,Z)\right]\leq\DD,\E_{Q}\left[d_{\st[L]}(X,W)\right]\leq\DDc,I_{Q}(Z;W)\leq\RR}2^{n\beta\left[\RR-I_{Q}(Z;W)\right]}2^{n\beta H_{Q}(X|ZW)}\nonumber \\
 & \hphantom{=\lim_{\beta\to\infty}\biggl(}+\sum_{Q_{Z}}2^{nH_{Q}(Z)}\sum_{Q_{XW|Z}:\E_{Q}\left[d_{\st[E]}(X,Z)\right]\leq\DD,\E_{Q}\left[d_{\st[L]}(X,W)\right]\leq\DDc,I_{Q}(Z;W)>\RR}2^{n\left[\RR-I_{Q}(Z;W)\right]}2^{n\beta H_{Q}(X|ZW)}\biggr)^{\nicefrac{1}{\beta}}\\
 & \doteq\lim_{\beta\to\infty}\biggl(\max_{Q_{XZW}\in{\cal J}(\DDc,\DD):I_{Q}(Z;W)\leq\RR}2^{nH_{Q}(Z)}2^{n\beta\left[\RR-I_{Q}(Z;W)\right]}2^{n\beta H_{Q}(X|ZW)}\nonumber \\
 & \hphantom{=\lim_{\beta\to\infty}\biggl(}+\max_{Q_{XZW}\in{\cal J}(\DDc,\DD):I_{Q}(Z;W)>\RR}2^{nH_{Q}(Z)}2^{n\left[\RR-I_{Q}(Z;W)\right]}2^{n\beta H_{Q}(X|ZW)}\biggr)^{\nicefrac{1}{\beta}}\\
 & \doteq\lim_{\beta\to\infty}\biggl(\max\biggl\{\max_{Q_{XZW}\in{\cal J}(\DDc,\DD):I_{Q}(Z;W)\leq\RR}2^{nH_{Q}(Z)}2^{n\beta\left[\RR-I_{Q}(Z;W)\right]}2^{n\beta H_{Q}(X|ZW)},\nonumber \\
 & \hphantom{=\lim_{\beta\to\infty}\biggl(}\max_{Q_{XZW}\in{\cal J}(\DDc,\DD):I_{Q}(Z;W)>\RR}2^{nH_{Q}(Z)}2^{n\left[\RR-I_{Q}(Z;W)\right]}2^{n\beta H_{Q}(X|ZW)}\biggr\}\biggr)^{\nicefrac{1}{\beta}}\\
 & =\lim_{\beta\to\infty}\max\biggl\{\max_{Q_{XZW}\in{\cal J}(\DDc,\DD):I_{Q}(Z;W)\leq\RR}2^{n\frac{1}{\beta}H_{Q}(Z)}2^{n\left[\RR-I_{Q}(Z;W)\right]}2^{nH_{Q}(X|ZW)},\nonumber \\
 & \max_{Q_{XZW}\in{\cal J}(\DDc,\DD):I_{Q}(Z;W)>\RR}2^{n\frac{1}{\beta}H_{Q}(Z)}2^{n\frac{1}{\beta}\left[\RR-I_{Q}(Z;W)\right]}2^{nH_{Q}(X|ZW)}\biggr\}\\
 & =\max\biggl\{\max_{Q_{XZW}\in{\cal J}(\DDc,\DD):I_{Q}(Z;W)\leq\RR}2^{n\left[\RR-I_{Q}(Z;W)\right]}2^{nH_{Q}(X|ZW)},\nonumber \\
 & \hphantom{=\max\biggl\{}\max_{Q_{XZW}\in{\cal J}(\DDc,\DD):I_{Q}(Z;W)>\RR}2^{nH_{Q}(X|ZW)}\biggr\}\label{eq: expected value of num}
\end{align}
where $(a)$ is by the Lebesgue monotone convergence theorem \cite[Theorem 11.28]{rudin1964principles}
and the monotonicity of the argument inside the expectation operator
in $\beta$, and $(b)$ is by the Jensen inequality. In $(c)$, we
have used the analysis in \cite[Subsection 6.3]{Merhav09} of the
moments of $N(Q_{ZW}|\mathbf{z})$, which is a binomial random variable
with $2^{n\RR}$ trials and probability of success of the exponential
order of $\exp\left[-nI_{Q}(Z;W)\right]$. Also, note that in all
the above equations, $Q_{XZW}\in{\cal P}_{n}({\cal X}\times{\cal Z}\times{\cal W})$
but since ${\cal P}({\cal X}\times{\cal Z}\times{\cal W})$ is dense
in ${\cal Q}({\cal X}\times{\cal Z}\times{\cal W})$ and the arguments
of the maximization are continuous functions of $Q_{XZW}$, we can
change the maximization to be over ${\cal Q}({\cal X}\times{\cal Z}\times{\cal W})$.
Thus, 
\[
\E\left[\max_{\mathbf{z}\in{\cal Z}^{n}}\left|\left\{ \mathbf{x}\in\mathfrak{D}({\cal C}_{n},Q_{X},\DDc):d_{\st[E]}(\mathbf{x},\mathbf{z})\leq\DD\right\} \right|\right]\dotleq2^{nE_{1}(\DD)}
\]
where 
\[
E_{1}(\DD)\teq\max_{Q_{XZW}:\E_{Q}\left[d_{\st[E]}(X,Z)\right]\leq\DD,\E_{Q}\left[d_{\st[L]}(X,W)\right]\leq\DDc}\left\{ H_{Q}(X|ZW)+\left[\RR-I_{Q}(Z;W)\right]_{+}\right\} .
\]
Now, let $\delta>0$ be given. There exists $n_{0}(Q_{X})$ such that
for all $n\geq n_{0}(Q_{X})$, we have from \eqref{eq: expectation of den}
\begin{equation}
\E\left(|\mathfrak{D}({\cal C}_{n},Q_{X},\DDc)|\right)\geq2^{n(E_{0}-\frac{\delta}{2})},\label{eq: lower bound on expected value of D-cover}
\end{equation}
and from \eqref{eq: deterministic upper bound on den} 
\[
|\mathfrak{D}({\cal C}_{n},Q_{X},\DDc)|\leq2^{n(E_{0}+\frac{\delta}{2})}.
\]
Define, for the given ensemble of the random sets
\[
{\cal A}_{0}\teq\left\{ {\cal C}_{n}:|\mathfrak{D}({\cal C}_{n},Q_{X},\DDc)|>2^{-n\frac{\delta}{2}}\E[|\mathfrak{D}({\cal C}_{n},Q_{X},\DDc)|]\right\} .
\]
The reverse Markov lemma (Lemma \ref{lem: reverse Markov}) implies
\[
\P\left({\cal A}_{0}\right)\geq\frac{1-2^{-n\frac{\delta}{2}}}{2^{n\delta}-2^{-n\frac{\delta}{2}}}\geq2^{-2n\delta}
\]
where the second inequality is satisfied for all $n\geq n_{0}'$ for
some $n_{0}'\geq n_{0}(Q_{X})$. 

Now, note that we need to prove that a single set ${\cal C}_{n}^{*}$
satisfies \eqref{eq: lower bound reliability function proof} for
all $\DD\geq\DDc$. To show this, we consider a quantization of the
possible values of $\DD$. To this end, let an arbitrary $\eta>0$
be given, such that $J=\frac{R_{\st[E]}(Q_{X},\DDc)}{\eta}$ is integer,
and find $\DDbar$ sufficiently large such that%
\footnote{Note that if $d_{\st[E]}(x,z)<\infty$ for all $x\in{\cal X},z\in{\cal Z}$,
then $\lim_{\DD\to\infty}R_{\st[E]}(Q_{X},\DD)=0$.%
} 
\[
R_{\st[E]}(Q_{X},\DDbar)\leq\lim_{\DD\to\infty}R_{\st[E]}(Q_{X},\DD)+\eta.
\]
Let us quantize the interval $[R_{\st[E]}(Q_{X},\DDbar),R_{\st[E]}(Q_{X},\DDc)]$
to values $\{\RR(0),\ldots,\RR(J)\}$, where $\RR(j)=j\eta$ and let
$\DD(j)=R_{\st[E]}^{-1}(Q_{X},\RR(j))$, where $R_{\st[E]}^{-1}(Q_{X},\RR)$
is the inverse function of $R_{\st[E]}(Q_{X},\DD)$. By \eqref{eq: expected value of num},
there exists $n_{1}(j,Q_{X})$ such that for all $n\geq n_{1}(j,Q_{X})$
\[
\E\left[\max_{\mathbf{z}\in{\cal Z}^{n}}\left|\left\{ \mathbf{x}\in\mathfrak{D}({\cal C}_{n},Q_{X},\DDc):d_{\st[E]}(\mathbf{x},\mathbf{z})\leq\DD(j)\right\} \right|\right]\leq2^{n\left[E_{1}(\DD(j))+\delta\right]},
\]
where the expectation is over the random ensemble of sets ${\cal C}_{n}$.
By defining 
\[
{\cal A}_{1j}\teq\left\{ {\cal C}_{n}:\max_{\mathbf{z}\in{\cal Z}^{n}}\left|\left\{ \mathbf{x}\in\mathfrak{D}({\cal C}_{n},Q_{X},\DDc):d_{\st[E]}(\mathbf{x},\mathbf{z})\leq\DD(j)\right\} \right|\leq2^{n\left[E_{1}(\DD(j))+4\delta\right]}\right\} 
\]
the ordinary Markov lemma implies 
\begin{align*}
\P\left({\cal A}_{1j}\right) & \geq1-\frac{\E\left[\max_{\mathbf{z}\in{\cal Z}^{n}}\left|\left\{ \mathbf{x}\in\mathfrak{D}({\cal C}_{n},Q_{X},\DDc):d_{\st[E]}(\mathbf{x},\mathbf{z})\leq\DD(j)\right\} \right|\right]}{2^{n\left[E_{1}(\DD(j))+4\delta\right]}}\\
 & \geq1-2^{-3n\delta}.
\end{align*}
Defining ${\cal A}_{1}\teq\bigcap_{j=0}^{J}{\cal A}_{1j}$ we get
\begin{align*}
\P\left({\cal A}_{1}\right) & =\P\left(\bigcap_{j=0}^{J}{\cal A}_{1j}\right)\\
 & =1-\P\left(\bigcup_{j=0}^{J}{\cal A}_{1j}^{c}\right)\\
 & \geq1-\sum_{j=0}^{J}\P\left({\cal A}_{1j}^{c}\right)\\
 & \geq1-J\cdot2^{-3n\delta}.
\end{align*}
Thus, since $J$ does not depend on $n$, there exists $n_{1}'\geq\max_{0\leq j\leq J}n_{1}(j,Q_{X})$
such that for all $n\geq n_{1}'$ 
\begin{align*}
\P\left({\cal A}_{0}\cap{\cal A}_{1}\right) & =1-\P\left({\cal A}_{0}^{c}\cup{\cal A}_{1}^{c}\right)\\
 & \geq1-\P\left({\cal A}_{0}^{c}\right)-\P\left({\cal A}_{1}^{c}\right)\\
 & \geq1-(1-2^{-2n\delta})-J2^{-n\frac{5\delta}{2}}\\
 & =2^{-2n\delta}-J\cdot2^{-3n\delta}\\
 & >0.
\end{align*}
Therefore, for all sufficiently large $n>\max\{n_{0}',n_{1}'\}$,
there exists ${\cal C}_{n}\in{\cal A}_{0}\cap\{\bigcap_{j=0}^{J}{\cal A}_{1j}\}$,
i.e., ${\cal C}_{n}$ which satisfies both
\begin{equation}
|\mathfrak{D}({\cal C}_{n},Q_{X},\DDc)|>2^{-n\frac{\delta}{2}}\E[|\mathfrak{D}({\cal C}_{n},Q_{X},\DDc)|]\label{eq: D-cover size of good distortion packing code}
\end{equation}
and 
\[
\max_{\mathbf{z}\in{\cal Z}^{n}}\left|\left\{ \mathbf{x}\in\mathfrak{D}({\cal C}_{n},Q_{X},\DDc):d_{\st[E]}(\mathbf{x},\mathbf{z})\leq\DD(j)\right\} \right|\leq2^{4n\delta}2^{nE_{1}(\DD(j))}
\]
for all $0\leq j\leq J$. Thus we get 
\[
p_{d}\left[{\cal C}_{n},Q_{X},\DDc,\DD(j)\right]\leq\frac{2^{4n\delta}2^{nE_{1}(\DD(j))}}{2^{-n\frac{\delta}{2}}2^{n(E_{0}-n\frac{\delta}{2})}}=2^{5n\delta}\cdot2^{n\left[E_{1}(\DD(j))-E_{0}\right]}.
\]
If we now define $E(\DD)\teq E_{1}(\DD)-E_{0}$, then for any given
$Q_{W}\in{\cal P}_{n}({\cal W})$ 
\begin{equation}
\liminf_{n\to\infty}-\frac{1}{n}\log p_{d}\left[{\cal C}_{n},Q_{X},\DDc,\DD(j)\right]\geq E(\DD).\label{eq: lower bound with F' and F''}
\end{equation}
Now, choose let $Q_{W}$ be the $W$-marginal of $Q_{XW}$ which achieves
$R_{\st[L]}(Q_{X},\DDc)$. Then, 
\begin{align}
E(\DD) & \geq\min_{Q_{XZW}:\E_{Q}\left[d_{\st[E]}(X,Z)\right]\leq\DD,\E_{Q}\left[d_{\st[L]}(X,W)\right]\E_{Q}\left[d_{\st[L]}(X,W)\right]\leq\DDc,I_{Q}(Z;W)\leq\RR}\left\{ I_{Q}(Z;W)+I_{Q}(X;Z,W)\right\} \nonumber \\
 & \hphantom{\geq{}}-\min_{Q_{XW}:\E_{Q}\left[d_{\st[L]}(X,W)\right]\leq\DDc}I_{Q}(X;W)\label{eq: RCB F_tag}\\
 & \trre[\geq,a]\min_{Q_{XZW}:\E_{Q}\left[d_{\st[E]}(X,Z)\right]\leq\DD,\E_{Q}\left[d_{\st[L]}(X,W)\right]\leq\DDc,I_{Q}(Z;W)\leq\RR}\left\{ I_{Q}(Z;W)+I_{Q}(X;Z,W)-I_{Q}(X;W)\right\} \\
 & =\min_{Q_{XZW}:\E_{Q}\left[d_{\st[E]}(X,Z)\right]\leq\DD,\E_{Q}\left[d_{\st[L]}(X,W)\right]\leq\DDc,I_{Q}(Z;W)\leq\RR}\left\{ I_{Q}(Z;W)+I_{Q}(X;Z|W)\right\} \\
 & =\min_{Q_{XZW}:\E_{Q}\left[d_{\st[E]}(X,Z)\right]\leq\DD,\E_{Q}\left[d_{\st[L]}(X,W)\right]\leq\DDc,I_{Q}(Z;W)\leq\RR}I_{Q}(X,W;Z)\\
 & \trre[\geq,b]\min_{Q_{XZW}:\E_{Q}\left[d_{\st[E]}(X,Z)\right]\leq\DD}I_{Q}(X;Z)\\
 & =R_{\st[E]}(Q_{X},\DD)\label{eq: lower bound on F'}
\end{align}
where $(a)$ is by restricting $Q_{XW}$ to be the same in both minimizations
of \eqref{eq: RCB F_tag}, and $(b)$ is by the data processing property
of the mutual information. Similarly, 
\begin{align}
E(\DD) & \geq\RR+\min_{Q_{XZW}:\E_{Q}\left[d_{\st[E]}(X,Z)\right]\leq\DD,\E_{Q}\left[d_{\st[L]}(X,W)\right]\leq\DDc,I_{Q}(Z;W)>\RR}I_{Q}(X;Z,W)\nonumber \\
 & \hphantom{\geq{}}-\min_{Q_{XW}:\E_{Q}\left[d_{\st[L]}(X,W)\right]\leq\DDc}I_{Q}(X;W)\label{eq: lower bound on F''}\\
 & \geq\RR+\min_{Q_{XZW}:\E_{Q}\left[d_{\st[E]}(X,Z)\right]\leq\DD,\E_{Q}\left[d_{\st[L]}(X,W)\right]\leq\DDc,I_{Q}(Z;W)>\RR}I_{Q}(X;Z|W)\\
 & \geq\RR.\label{eq: lower bound on F'' -- R}
\end{align}
by restricting $Q_{XW}$ to be the same in both minimizations of \eqref{eq: lower bound on F''}. 

Therefore, \eqref{eq: lower bound with F' and F''}, \eqref{eq: lower bound on F'}
and \eqref{eq: lower bound on F''} imply that 
\[
\liminf_{n\to\infty}-\frac{1}{n}\log p_{d}\left[{\cal C}_{n},Q_{X},\DDc,\DD(j)\right]\geq\min\left\{ R_{\st[E]}(Q_{X},\DD(j)),\RR\right\} 
\]
for all $0\leq j\leq J$. By taking $\eta\downarrow0$, continuity
of $R_{\st[E]}(Q_{X},\DD)$ in $\DD$ provides the lower bound \eqref{eq: lower bound reliability function proof}
for all $\DD\geq\DDc$. Then, \eqref{eq: D-cover size} is obtained
from \eqref{eq: D-cover size of good distortion packing code} and
\eqref{eq: lower bound on expected value of D-cover}.

To complete the proof of the lemma, we consider the case of $\RR\geq R_{\st[L]}(Q_{X},\DDc)$.
Denote by $Q_{XW}^{(n)}$ a sequence of distributions such that $Q_{XW}^{(n)}\to Q_{XW}^{*}$
as $n\to\infty$, where $Q_{XW}^{*}$ achieves the rate-distortion
function $R_{\st[L]}(Q_{X},\DDc)$. For a given ${\cal C}_{n},$ let
$\tilde{{\cal C}}_{n}$ be a subset formed by the first $e^{nR_{\st[L]}(Q_{X},\DDc)}$
members of ${\cal C}_{n}$. The same analysis as before shows that
when randomly drawing a set ${\cal C}_{n}$ uniformly over the $W$-marginal
of $Q_{XW}^{(n)}$, there exists a sequence of sets $\{{\cal C}_{n}\}$
such that 
\[
|\mathfrak{D}(\tilde{{\cal C}}_{n},Q_{X},\DDc)|\geq2^{n(E_{0}-\delta)}\geq2^{n\left[H(Q_{X})-\delta\right]}.
\]
Then, for ${\cal C}_{n}$ 
\begin{align*}
p_{d}({\cal C}_{n},Q_{X},\DDc,\DD) & =\frac{\max_{\mathbf{z}\in{\cal Z}^{n}}\left|\left\{ \mathbf{x}\in\mathfrak{D}({\cal C}_{n},Q_{X},\DDc):d_{\st[E]}(\mathbf{x},\mathbf{z})\leq\DD\right\} \right|}{|\mathfrak{D}({\cal C}_{n},Q_{X},\DDc)|}\\
 & \leq\frac{\max_{\mathbf{z}\in{\cal Z}^{n}}\left|\left\{ \mathbf{x}\in\mathfrak{D}({\cal C}_{n},Q_{X},\DDc):d_{\st[E]}(\mathbf{x},\mathbf{z})\leq\DD\right\} \right|}{|\mathfrak{D}(\tilde{{\cal C}}_{n},Q_{X},\DDc)|}\\
 & \leq\frac{\max_{\mathbf{z}\in{\cal Z}^{n}}\left|\left\{ \mathbf{x}\in{\cal T}_{n}(Q_{X}):d_{\st[E]}(\mathbf{x},\mathbf{z})\leq\DD\right\} \right|}{|\mathfrak{D}(\tilde{{\cal C}}_{n},Q_{X},\DDc)|}\\
 & \leq\frac{\max_{\mathbf{z}\in{\cal Z}^{n}}\left|\left\{ \mathbf{x}\in{\cal T}_{n}(Q_{X}):d_{\st[E]}(\mathbf{x},\mathbf{z})\leq\DD\right\} \right|}{2^{n\left[H(Q_{X})-\delta\right]}}\\
 & =2^{-n\left[H(Q_{X})-\delta\right]}\max_{\mathbf{z}\in{\cal Z}^{n}}\sum_{Q_{X|Z}:\E_{Q}\left[d_{\st[E]}(X,Z)\right]\leq\DD}\left|{\cal T}_{n}(Q_{X|Z},\mathbf{z})\right|\\
 & \leq2^{-n\left[H(Q_{X})-\delta\right]}\max_{Q_{Z}}\sum_{Q_{X|Z}:\E_{Q}\left[d_{\st[E]}(X,Z)\right]\leq\DD}2^{nH_{Q}(X|Z)}\\
 & \doteq\exp\left(-n\left[H_{Q}(X)-\delta-\max_{Q_{XZ}:\E_{Q}\left[d_{\st[E]}(X,Z)\right]\leq\DD}H_{Q}(X|Z)\right]\right)\\
 & \leq2^{-n\left[R_{\st[E]}(Q_{X},\DD)-\delta\right]}
\end{align*}
and the proof of the lemma is complete, as $\delta$ is arbitrary.\end{IEEEproof}
\begin{rem}
\label{rem: uniform convergence is not assured}As mentioned in Section
\ref{sec:Outline-of-the proof}, to show achievability of an exiguous-distortion
exponent using the method of types, \emph{uniform} convergence of
$-\frac{1}{n}\log p_{d}({\cal C}_{n}^{*},Q_{X},\DDc,\DD)$ to the
exponent $\min\left\{ \RR,R_{\st[E]}(Q_{X},\DD)\right\} $ is required
(cf. eq. \eqref{eq: requirement of uniform convergence method of types}).
However, the proof of Lemma \ref{lem: distortion packing exponent}
is not sufficient to show this. Specifically, the convergence in the
asymptotic analysis of the type class enumerators, i.e. the relations
\[
\P\left\{ N(Q_{XW}|\mathbf{x})\geq1\right\} \doteq2^{n\left[\RR-I_{Q}(X;W)\right]}
\]
used in \eqref{eq: typ enumerator aysmp 1} and
\[
\E\left[N(Q_{ZW}|\mathbf{z})^{\beta}\right]\doteq\begin{cases}
2^{n\left[\RR-I_{Q}(Z;W)\right]}, & I_{Q}(Z;W)\leq\RR\\
2^{n\beta\left[\RR-I_{Q}(Z;W)\right]}, & I_{Q}(Z;W)>\RR
\end{cases}
\]
used in \eqref{eq: type class enumerator asymp 2}, are not uniform
in $Q_{X}$. 

We continue with the second step of the proof, which constructs from
the set ${\cal C}_{n}^{*}$ a secure rate-distortion code for all
$\mathbf{x}\in{\cal T}_{n}(Q_{X})$. The proof of the next lemma is
based on the permutations technique described in Subsection \ref{sub:Covering-A-Type}.\end{rem}
\begin{lem}
\label{lem: exponent lower bound, single type}For any given $Q_{X}\in{\cal P}({\cal X})\cap\interior{\cal Q}({\cal X})$
and $\delta>0$, there exists a sequence of secure rate-distortion
codes ${\cal S}^{*}$ of fixed key rate $\RR$ such that 
\begin{equation}
\lim_{n\to\infty}\frac{1}{n}\log|{\cal Y}_{n}|\leq R_{\st[L]}(Q_{X},\DDc)+\delta,\label{eq: achievability code rate constraint}
\end{equation}
and, 
\begin{equation}
\P\left[d_{\st[L]}(\mathbf{X},\varphi_{n}^{*}(f_{n}^{*}(\mathbf{X},\mathbf{u})))\geq\DDc|\mathbf{X}\in{\cal T}_{n}(Q_{X})\right]=0\label{eq: secrecy conditional exponent achievable perfect legitimate covering}
\end{equation}
for every $\mathbf{u}\in\{0,1\}^{n\RR}$, as well as

\begin{equation}
{\cal E}_{d}^{-}({\cal S}^{*},\DD,Q_{X})\geq\min\left\{ \RR,R_{\st[E]}(Q_{X},\DD)\right\} -\delta\label{eq: secrecy conditional error exponent bound inf}
\end{equation}
for all $\DD\geq\DDc$.\end{lem}
\begin{IEEEproof}
Assume that $Q_{X}\in\left[\interior{\cal Q}({\cal X})\right]\cap{\cal P}_{n_{0}}({\cal X})$
for some minimal $n_{0}\in\mathbb{N}$. Since the statements in the
lemma are only about conditional events given the type $Q_{X}$, it
is clear that the secure rate-distortion codes constructed ${\cal S}_{n}^{*}$,
may only encode $\mathbf{x}\in{\cal T}_{n}(Q_{X})$, and so only block-lengths
$n\bmod n_{0}=0$ should be considered, as otherwise ${\cal T}_{n}(Q_{X})$
is empty. 

Let ${\cal C}^{*}=\{{\cal C}_{n}^{*}\}$ be a sequence of sets of
size $2^{n\RR}$ constructed according to Lemma \ref{lem: distortion packing exponent}.
So for all $n$ sufficiently large 
\begin{equation}
p_{d}({\cal C}_{n}^{*},Q_{X},\DDc,\DD)\leq2^{-n\left[\min\left\{ \RR,R_{\st[E]}(Q_{X},\DD)\right\} -\delta\right]},\label{eq: exigious-distortion probability for set C*}
\end{equation}
and 
\begin{equation}
\left|\mathfrak{D}({\cal C}_{n}^{*},Q_{X},\DDc)\right|\geq2^{n(A-\delta)},\label{eq: size of good D-cover}
\end{equation}
where
\[
A\teq\min\left\{ H(Q_{X})+\RR-R_{\st[L]}(Q_{X},\DDc),H(Q_{X})\right\} .
\]
Now, let $\{\pi_{n,t}\}_{t=0}^{\kappa_{n}}$ be a set of permutations
constructed according to Lemma \ref{lem: Covering Lemma Awhlsede},
such that
\begin{equation}
\bigcup_{t=0}^{\kappa_{n}}\pi_{n,t}(\mathfrak{D}({\cal C}_{n}^{*},Q_{X},\DDc))={\cal T}_{n}(Q_{X}),\label{eq: covering of type using permutations of D-cover}
\end{equation}
where $\kappa_{n}\leq2^{n\left[H(Q_{X})-A+2\delta)\right]}$, and
let \emph{$\{{\cal G}_{n,t}\}$ }be the resulting exclusive permutation
sets, as defined in \eqref{eq:exclusive permutations sets defintion}.
We construct the following secure rate-distortion codes ${\cal S}_{n}^{*}=(f_{n}^{*},\varphi_{n}^{*})$
of fixed key rate $\RR$, which only encode $\mathbf{x}\in{\cal T}_{n}(Q_{X})$.
We utilize the covering of the type class ${\cal T}_{n}(Q_{X})$ by
permutations of a D-cover of the set ${\cal C}_{n}^{*}$ to encode
the source block in the following way. Assume that the elements of
${\cal C}_{n}^{*}$ are arbitrarily ordered, i.e. ${\cal C}_{n}^{*}=\{\mathbf{w}(0),\ldots,\mathbf{w}(2^{n\RR}-1)\}$.
For a given $\mathbf{x}\in{\cal T}_{n}(Q_{X})$, let 
\[
t^{*}(\mathbf{x})\teq\min\left\{ t:\mathbf{x}\in{\cal G}_{n,t}\right\} ,
\]
and 
\[
i^{*}(\mathbf{x})\teq\min\{i:\mathbf{w}(i)\in{\cal G}_{n,t^{*}(\mathbf{x})},d_{\st[L]}(\mathbf{x},\mathbf{w}(i))\leq\DDc\}
\]
The encoding is a concatenation of the following two parts $y=f_{n}^{*}(\mathbf{x},\mathbf{u})=(t_{y},i_{y})$:
\begin{itemize}
\item A description of the permutation set, defined as $t_{y}\teq\bin[t^{*}(\mathbf{x}),n(H(Q_{X})-A+2\delta)]$.
\item An encrypted description of the distortion covering codeword, defined
as $i_{y}\teq\bin[i^{*}(\mathbf{x}),n\RR]\oplus\mathbf{u}$. 
\end{itemize}
It is easily verified that given $\mathbf{u}$, the legitimate decoder
can reproduce $\mathbf{w}=\varphi_{n}(y,\mathbf{u})$ such that $d_{\st[L]}(\mathbf{x},\mathbf{w})\leq\DDc$,
for all $\mathbf{x}\in{\cal T}_{n}(Q_{X})$, and so \eqref{eq: secrecy conditional exponent achievable perfect legitimate covering}
is satisfied. Regarding the coding rate, note that 
\begin{align}
\frac{1}{n}\log|{\cal Y}_{n}| & =H(Q_{X})-A+2\delta+\RR\\
 & \leq R_{\st[L]}(Q_{X},\DDc)+3\delta\label{eq: achieavbility for type class code rate}
\end{align}
 for all $n$ sufficiently large, which results in \eqref{eq: achievability code rate constraint}. 

It remains to prove that for any eavesdropper $\sigma_{n}$, the conditional
exiguous-distortion exponent, given that $\mathbf{X}\in{\cal T}_{n}(Q_{X})$,
is larger than $\min\left\{ \RR,R_{\st[E]}(Q_{X},\DD)\right\} -\delta$.
From Proposition \ref{prop: Type awarness does not help}, it may
be assumed that the eavesdropper is aware of the type $Q_{X}$. Moreover,
given the cryptogram $Y=y$, the source block $\mathbf{X}$ is distributed
uniformly over ${\cal G}_{n,t_{y}}$, and independent of $i_{y}$.
Thus, the optimal eavesdropper has the same estimate for cryptograms
with the same $t_{y}$, and we may denote its estimate as $\mathbf{z}=\sigma_{n}(y)\teq\mathbf{z}(t_{y})$.
Since ${\cal G}_{n,0}=\mathfrak{D}({\cal C}_{n}^{*},Q_{X},\DDc)$,
then conditioned on the event $\{t^{*}(\mathbf{X})=0\}$, for any
$\mathbf{z}\in{\cal Z}^{n}$, Lemma \ref{lem: distortion packing exponent}
implies
\begin{align}
\P\left[d_{\st[E]}(\mathbf{X},\mathbf{z})\leq\DD|\mathbf{X}\in{\cal T}_{n}(Q_{X}),t^{*}(\mathbf{X})=0\right] & =\P\left[d_{\st[E]}(\mathbf{X},\mathbf{z})\leq\DD|\mathbf{X}\in{\cal G}_{n,0}\right]\\
 & \leq2^{-n\left[\min\left\{ \RR,R_{\st[E]}(Q_{X},\DD)\right\} -\delta\right]}
\end{align}
for all $n$ sufficiently large. It then follows that for $0<t\leq\kappa_{n}$,
\begin{align}
\P\left[d_{\st[E]}(\mathbf{X},\mathbf{z})\leq\DD|\mathbf{X}\in{\cal T}_{n}(Q_{X}),t^{*}(\mathbf{X})=t\right] & =\P\left[d_{\st[E]}(\mathbf{X},\mathbf{z})\leq\DD|\mathbf{X}\in{\cal G}_{n,t}\right]\nonumber \\
 & \trre[\leq,a]\frac{|{\cal G}_{n,0}|}{|{\cal G}_{n,t}|}\P\left[d_{\st[E]}(\mathbf{X},\mathbf{z})\leq\DD|\mathbf{X}\in{\cal G}_{n,0}\right]\nonumber \\
 & \leq\frac{|{\cal G}_{n,0}|}{|{\cal G}_{n,t}|}2^{-n\left(\min\left\{ \RR,R_{\st[E]}(Q_{X},\DD)\right\} -\delta\right)},\label{eq: ratio between G_0t and G_nt}
\end{align}
where $(a)$ follows from the fact that for any $0<t\leq\kappa_{n}$,
there exists a permutation $\pi$ such that $\pi\left({\cal G}_{n,t}\right)\subset{\cal G}_{n,0}=\mathfrak{D}({\cal C}_{n}^{*},Q_{X},\DDc)$
and Proposition \ref{prop: simple properties of distortion packing code}.
Thus, the exiguous-distortion probability conditioned on $t^{*}(\mathbf{X})=t$
can be larger than the same probability conditioned on $t^{*}(\mathbf{X})=0$,
but only up to a factor of $\frac{|{\cal G}_{n,0}|}{|{\cal G}_{n,t}|}$,
which is large if $|{\cal G}_{n,t}|$ is small. Next, we show that
the contribution to the exiguous-distortion probability of these small
sets does not impact its exponential behavior. To this end, for any
fixed $0<\eta<A+\delta$ such that $J=\frac{A+\delta}{\eta}$ is an
integer, let us quantize the interval $[0,A+\delta]$ to values $\{A_{0},\ldots,A_{J}\}$,
where $A_{j}=j\eta$. We will treat separately sets such that $2^{nA_{j}}\leq|{\cal G}_{n,t}|\leq2^{nA_{j+1}}$.
For all $n$ sufficiently large 
\begin{align*}
 & \hphantom{{}={}}\P\left[d_{\st[E]}(\mathbf{X},\mathbf{z})\leq\DD|\mathbf{X}\in{\cal T}_{n}(Q_{X})\right]\\
 & =\sum_{t=0}^{\kappa_{n}}\P\left[\mathbf{X}\in{\cal G}_{n,t}|\mathbf{X}\in{\cal T}_{n}(Q_{X})\right]\P\left[d_{\st[E]}(\mathbf{X},\mathbf{z}(t))\leq\DD|\mathbf{X}\in{\cal G}_{n,t},\mathbf{X}\in{\cal T}_{n}(Q_{X})\right]\\
 & =\sum_{j=0}^{J-1}\sum_{t:2^{nA_{j}}\leq|{\cal G}_{n,t}|\leq2^{nA_{j+1}}}\P\left[\mathbf{X}\in{\cal G}_{n,t}|\mathbf{X}\in{\cal T}_{n}(Q_{X})\right]\P\left[d_{\st[E]}(\mathbf{X},\mathbf{z}(t))\leq\DD|\mathbf{X}\in{\cal G}_{n,t},\mathbf{X}\in{\cal T}_{n}(Q_{X})\right]\\
 & \trre[\leq,a]\sum_{j=0}^{J-1}\sum_{t:2^{nA_{j}}\leq|{\cal G}_{n,t}|\leq2^{nA_{j+1}}}\P\left[\mathbf{X}\in{\cal G}_{n,t}|\mathbf{X}\in{\cal T}_{n}(Q_{X})\right]\frac{|{\cal G}_{n,0}|}{|{\cal G}_{n,t}|}2^{-n\left(\min\left\{ \RR,R_{\st[E]}(Q_{X},\DD)\right\} -\delta\right)}\\
 & \leq\sum_{j=0}^{J-1}\sum_{t:2^{nA_{j}}\leq|{\cal G}_{n,t}|\leq2^{nA_{j+1}}}\P\left[\mathbf{X}\in{\cal G}_{n,t}|\mathbf{X}\in{\cal T}_{n}(Q_{X})\right]\frac{2^{n(A+\delta)}}{2^{nA_{j}}}2^{-n\left(\min\left\{ \RR,R_{\st[E]}(Q_{X},\DD)\right\} -\delta\right)}\\
 & =\sum_{j=0}^{J-1}\frac{2^{n(A+\delta)}}{2^{nA_{j}}}2^{-n\left(\min\left\{ \RR,R_{\st[E]}(Q_{X},\DD)\right\} -\delta\right)}\sum_{t:2^{nA_{j}}\leq|{\cal G}_{n,t}|\leq2^{nA_{j+1}}}\P\left[\mathbf{X}\in{\cal G}_{n,t}|\mathbf{X}\in{\cal T}_{n}(Q_{X})\right]\\
 & \trre[\leq,b]\sum_{j=0}^{J-1}\frac{2^{n(A+\delta)}}{2^{nA_{j}}}2^{-n\left(\min\left\{ \RR,R_{\st[E]}(Q_{X},\DD)\right\} -\delta\right)}\P\left[\mathbf{X}\in{\cal H}(A_{j+1})|\mathbf{X}\in{\cal T}_{n}(Q_{X})\right]\\
 & \trre[\leq,c]\sum_{j=0}^{J-1}\frac{2^{n(A+\delta)}}{2^{nA_{j}}}2^{-n\left(\min\left\{ \RR,R_{\st[E]}(Q_{X},\DD)\right\} -\delta\right)}2^{-n(A-A_{j+1}-\delta)}\\
 & \leq J\cdot\max_{0\leq j\leq J-1}\frac{2^{n(A_{j+1}+2\delta)}}{2^{nA_{j}}}2^{-n\left(\min\left\{ \RR,R_{\st[E]}(Q_{X},\DD)\right\} -\delta\right)}\\
 & \leq2^{n(\eta+3\delta)}2^{-n\cdot\min\left\{ \RR,R_{\st[E]}(Q_{X},\DD)\right\} }\\
 & \trre[\leq,d]2^{n(\eta+4\delta)}2^{-n\cdot\min\left\{ \RR,R_{\st[E]}(Q_{X},\DD)\right\} }
\end{align*}
where $(a)$ is using \eqref{eq: ratio between G_0t and G_nt}, $(b)$
is using the definition in \eqref{eq: partial union of permutation sets},
$(c)$ is using Lemma \ref{lem: Efficient cover aspect 1}, and $(d)$
is since $J\doteq1$. The result follows by taking $\eta\downarrow0$.\end{IEEEproof}
\begin{rem}
\label{rem: on permutations ach. proof}Note that only the properties
\eqref{eq: exigious-distortion probability for set C*}-\eqref{eq: size of good D-cover}
of $\mathfrak{D}({\cal C}_{n}^{*},Q_{X},\DDc)$ were used in order
to prove Lemma \ref{lem: exponent lower bound, single type}. The
same proof of Lemma \ref{lem: exponent lower bound, single type}
can be used to show that if some other set ${\cal D}_{n}\subset\mathfrak{D}({\cal C}_{n}^{*},Q_{X},\DDc)$
satisfies similar properties, i.e. if for some $E>0$ 
\begin{equation}
\max_{\mathbf{z}\in{\cal Z}^{n}}\P\left[d_{\st[E]}(\tilde{\mathbf{X}},\mathbf{z})\leq\DD\right]\leq2^{-nE},\label{eq: exiguous-distortion probability for a subset of D-cover}
\end{equation}
where here $\tilde{\mathbf{X}}$ is distributed uniformly over ${\cal D}_{n}$,
and

\begin{equation}
\left|{\cal D}_{n}\right|\geq2^{n(A-\delta)}\label{eq: size of subset of D-cover}
\end{equation}
then a secure rate-distortion code can be constructed, with conditional
exiguous-distortion exponent $E$. In this case, the code is constructed
such that only source blocks in ${\cal D}_{n}$ are mapped to the
permutation index $t^{*}(\mathbf{x})=0$, but not source blocks from
$\mathfrak{D}({\cal C}_{n}^{*},Q_{X},\DDc)\backslash{\cal D}_{n}$.
In addition, if the coding rate is unconstrained, then the condition
\eqref{eq: size of subset of D-cover} is not required. This fact
will be utilized in the sequel in the proof of Lemma \ref{lem: converse - fixed type key-rate}. 
\end{rem}
In the third step of the achievability proof, we construct the secure
rate-distortion code for all types in ${\cal P}({\cal X})$. We will
need the following two lemmas.
\begin{lem}
\label{lem: hamming distance close types}Let $Q_{X},Q'_{X}\in{\cal P}_{n}({\cal X})$
and assume that%
\footnote{For two different types in ${\cal P}_{n}({\cal X})$, the minimal
variation distance is $\frac{2}{n}$.%
} $||Q_{X}-Q'_{X}||=\frac{2d^{*}}{n}$ where $d^{*}>0$. If $\mathbf{x}\in{\cal T}_{n}(Q_{X})$
then 
\[
\min_{\mathbf{x}'\in{\cal T}_{n}(Q'_{X})}d_{\st[H]}(\mathbf{x},\mathbf{x}')\leq d^{*}.
\]
\end{lem}
\begin{IEEEproof}
See the extended version of \cite[Lemma 20]{SW_paper_extended}.\end{IEEEproof}
\begin{lem}
\label{lem: truncation of types}Let $Q_{X}\in{\cal P}_{n}({\cal X})$
and $\mathbf{x}\in{\cal T}_{n}(Q_{X})$. For any given $1\leq k<n$
let $\mathbf{x}'=\mathbf{x}_{1}^{n-k}$. Then 
\[
||\hat{Q}_{\mathbf{x}}-\hat{Q}_{\mathbf{x}'}||<|{\cal X}|\cdot\frac{k}{n-k}.
\]
\end{lem}
\begin{IEEEproof}
See the extended version of \cite[Lemma 21]{SW_paper_extended}.
\end{IEEEproof}
We are now ready for the third and final step of the proof of the
achievability part of Theorem \ref{thm:exiguous distortion expoent}. 
\begin{IEEEproof}[Proof of achievability part of Theorem \ref{thm:exiguous distortion expoent}]
 Let $0<\epsilon<1$ be given, and find $n_{0}$ sufficiently large
such that for any $Q'_{X}\in{\cal P}({\cal X})$ there exists $Q_{X}\in{\cal P}_{n_{0}}({\cal X})\cap\interior{\cal Q}({\cal X})$
such that $||Q_{X}-Q'_{X}||\leq\frac{\epsilon}{2}$. We will term
\emph{${\cal P}_{n_{0}}({\cal X})\cap\interior{\cal Q}({\cal X})$
}as the\emph{ grid}. Also let $n_{1}=n_{0}\epsilon+2n_{0}|{\cal X}|$.
We construct the following sequence of secure rate-distortion codes
${\cal S}$ for all $n>\max\{n_{0},n_{1}\}$. We will use the following
definitions and constructions:
\begin{itemize}
\item Let $\tilde{n}=\left\lfloor \frac{n}{n_{0}}\right\rfloor \cdot n_{0}$. 
\item Enumerate the types of the source ${\cal P}_{n}({\cal X})$. 
\item Assume, w.l.o.g., that ${\cal X}=\{1,\ldots,|{\cal X}|\}$ and let
$\overline{{\cal X}}\teq\{0\}\cup{\cal X}$. 
\item Let 
\begin{equation}
{\cal B}_{\s[H]}^{n}(\epsilon)\teq\left\{ \mathbf{x}\in\overline{{\cal X}}^{n}:d_{\st[H]}(\mathbf{x},\mathbf{0})\leq\frac{n\epsilon}{2}\right\} ,\label{eq: Hamming ball definition}
\end{equation}
i.e., an Hamming ball of radius $\frac{n\epsilon}{2}$ and dimension
$n$. 
\item Construct the codes ${\cal S}_{\tilde{n},Q_{X}}^{*}=(f_{\tilde{n,}Q_{X}}^{*},\varphi_{\tilde{n,}Q_{X}}^{*})$
of key rate $\RR$ as in Lemma \ref{lem: exponent lower bound, single type},
for all $Q{}_{X}\in{\cal P}_{n_{0}}({\cal X})\cap\interior{\cal Q}({\cal X})$. 
\item For every given $Q_{X}\in{\cal P}_{n}({\cal X})$ find 
\[
\Phi_{\epsilon}(Q_{X})\teq\argmin_{Q'_{X}\in{\cal P}_{n_{0}}({\cal X})\cap\interior{\cal Q}({\cal X})}||Q_{X}-Q'_{X}||.
\]

\item For any given $\mathbf{x}\in{\cal X}^{n}$ and $\overline{\mathbf{x}}\in\overline{{\cal X}}^{n}$,
define the \emph{replacement operator} $\Psi:{\cal X}^{n}\times\overline{{\cal X}}^{n}\to\overline{{\cal X}}^{n}$
which for $\tilde{\mathbf{x}}=\Psi(\mathbf{x},\overline{\mathbf{x}})$
satisfies
\begin{equation}
\tilde{x}{}_{i}=\begin{cases}
x_{i}, & \overline{x}_{i}=0\\
\overline{x}_{i}, & \overline{x}_{i}\neq0
\end{cases}\label{eq: replacement operator}
\end{equation}

\item For a given $\mathbf{x}\in{\cal X}^{n}$, define the \emph{replacement
set} 
\begin{equation}
{\cal K}(\mathbf{x},\epsilon)\teq\left\{ \overline{\mathbf{x}}\in{\cal B}_{\s[H]}^{\tilde{n}}(\epsilon):\Psi(\mathbf{x}_{1}^{\tilde{n}},\overline{\mathbf{x}})\in{\cal T}_{\tilde{n}}(\Phi_{\epsilon}(\hat{Q}_{\mathbf{x}}))\right\} .\label{eq: replacment set}
\end{equation}
Note that the size of ${\cal K}(\mathbf{x},\epsilon)$ depends on
$\mathbf{x}$ only via its type $\hat{Q}_{\mathbf{x}}$. 
\end{itemize}
The above type enumeration and the codes constructed are revealed
to both the encoder and the decoder off-line. Before we provide the
details of the encoding and the legitimate decoding, we outline the
main ideas. Using the construction of Lemma \ref{lem: exponent lower bound, single type},
we construct secure rate distortion codes for each type in the\emph{
grid ${\cal P}_{n_{0}}({\cal X})\cap\interior{\cal Q}({\cal X})$.
}Since this grid has a \emph{finite} number of types, then for all
sufficiently large $n$, the normalized logarithm of the conditional
exiguous-distortion probability is close to the exponent \eqref{eq: secrecy conditional error exponent bound inf}
\emph{uniformly} over all types in the grid. As mentioned in the outline
of the proof in Section \ref{sec:Main-Result}, we will modify any
given source block so that it can be encoded using one of the codes
in the grid. In order to allow the legitimate decoder to be able to
reproduce with the desired distortion $\DDc$, the cryptogram will
be comprised of (at most) four parts, each one of them being encrypted
using key bits $\mathbf{u}^{(i)}$ for $1\leq i\leq4$. First, the
type of the source $\hat{Q}_{\mathbf{x}}$ is conveyed to the legitimate
decoder, and, in accordance with Proposition \ref{prop: Type awarness does not help},
the type information is not encrypted, and so $\mathbf{u}^{(1)}$
is the empty string. This type will be modified to the type $\Phi_{\epsilon}(\hat{Q}_{\mathbf{x}})$,
which is also known to the legitimate decoder and the eavesdropper.
Second, since if $n\bmod n_{0}\neq0$ then $\hat{Q}_{\mathbf{x}}$
may not belong to the grid, we first truncate the source block to
the length $\tilde{n}$. The truncated part $\mathbf{x}_{\tilde{n}+1}^{n}$
will be sent to the legitimate decoder losslessly, and fully encrypted
using $\mathbf{u}^{(2)}$. Third, we will modify $\mathbf{x}_{1}^{\tilde{n}}$
to the \emph{modified vector $\mathbf{\mathbf{v}}$}, such that $\hat{Q}_{\mathbf{v}}=\Phi_{\epsilon}(\hat{Q}_{\mathbf{x}})$.
This will be done by replacing a small number of the symbols of $\mathbf{x}$.
The symbols of $\mathbf{x}$ which were replaced in order to create
$\mathbf{v}$ will be sent to the legitimate decoder losslessly, and
fully encrypted using $\mathbf{u}^{(3)}$. Note, that there might
be more than one way to replace the symbols of $\mathbf{x}$, and
in fact, any $\overline{\mathbf{x}}\in{\cal K}(\mathbf{x},\epsilon)$
can be used for this purpose if we define $\mathbf{v}\teq\Psi(\mathbf{x}_{1}^{\tilde{n}},\overline{\mathbf{x}})$
using \eqref{eq: replacement operator} and \eqref{eq: replacment set}.
For the sake of the analysis, it will be convenient to choose a replacement
vector randomly from ${\cal K}(\mathbf{x},\epsilon)$. This will be
achieved using key bits $\overline{\mathbf{u}}$, which in this case,
function as common randomness rather than for encryption. Fourth,
the code $s_{\tilde{n},\Phi_{\epsilon}(\hat{Q}_{\mathbf{x}})}^{*}$
will be used to encode the modified vector $\mathbf{v}$ using the
key bits $\mathbf{u}^{(4)}$. As we will prove, the whole modification
procedure incurs a negligible cost on the compression and secrecy
performance, which we analyze after formally defining the encoder
and legitimate decoder.\\
\emph{Encoding:} Let $\mathbf{u}=(\mathbf{u}^{(1)},\mathbf{u}^{(2)},\mathbf{u}^{(3)},\mathbf{u}^{(4)},\overline{\mathbf{u}})$.
The following cryptogram parts are generated:
\begin{itemize}
\item Source block type: Find the type index $0\leq j^{*}\leq|{\cal P}_{n}({\cal X})|-1$
of the source block type in the enumeration of the types, and let
\[
y_{1}\teq\bin[j^{*},\log|{\cal P}_{n}({\cal X})|].
\]
Set $\mathbf{u}^{(1)}=\phi$, namely, the type information is not
encrypted, in accordance with Proposition \ref{prop: Type awarness does not help}.
\item Fully encrypted source block tail: 
\[
y_{2}\teq\bin[\mathbf{x}_{\tilde{n}+1}^{n},(n-\tilde{n})\log|{\cal X}|]\oplus\mathbf{u}^{(2)}
\]

\item Modification vector: Let $\overline{\mathbf{x}}$ be the $K_{\overline{\mathbf{u}}}$-th
vector in ${\cal K}(\mathbf{x},\epsilon)$, where $\overline{\mathbf{u}}$
is of length $\log|{\cal K}(\mathbf{x},\epsilon)|$ bits, and $K_{\overline{\mathbf{u}}}$
is integer corresponding to $\mathbf{u}$, i.e. 
\[
K_{\overline{\mathbf{u}}}\teq\sum_{l=1}^{\log|{\cal K}(\mathbf{x},\epsilon)|}\overline{\mathbf{u}}_{l}\cdot2^{(l-1)}+1.
\]
Also, let 
\begin{equation}
\mathbf{v}\teq\Psi(\mathbf{x}_{1}^{\tilde{n}},\overline{\mathbf{x}})\label{eq: modified vector def}
\end{equation}
and let $\mathbf{x}'''\in\overline{{\cal X}}^{n}$ where 
\[
x_{i}'''=\begin{cases}
0, & \overline{x}_{i}=0\\
x_{i}, & \overline{x}_{i}\neq0
\end{cases}.
\]
As clearly $\mathbf{x}'''\in{\cal B}_{\s[H]}^{\tilde{n}}(\epsilon)$,
let $i^{*}$ be the index of $\mathbf{x}'''$ in ${\cal B}_{\s[H]}^{\tilde{n}}(\epsilon)$
and 
\[
y_{3}\teq\bin[i^{*},\log|{\cal B}_{\s[H]}^{\tilde{n}}(\epsilon)|]\oplus\mathbf{u}^{(3)}.
\]

\item Cryptogram of modified vector: Let 
\[
y_{4}\teq s_{\tilde{n},\Phi_{\epsilon}(\hat{Q}_{\mathbf{x}})}^{*}(\mathbf{v},\mathbf{u}^{(4)})
\]
where $\mathbf{u}^{(4)}$ is of length $n\RR$ bits.
\end{itemize}
The encoding of the source block is separated into two cases, depending
on its type $\hat{Q}_{\mathbf{x}}$. If $\RRc<R_{\st[L]}(\hat{Q}_{\mathbf{x}},\DDc)$
then 
\[
y=f_{n}^{*}(\mathbf{x},\mathbf{u})=y_{1}.
\]
Otherwise, if $\RRc\geq R_{\st[L]}(\hat{Q}_{\mathbf{x}},\DDc)$ then
\begin{equation}
y=f_{n}^{*}(\mathbf{x},\mathbf{u})=(y_{1},y_{2},y_{3},y_{4}).\label{eq: cryptogram in 4 parts}
\end{equation}
To verify that such coding is possible, notice that from Lemma \ref{lem: truncation of types}
and the fact that $n>n_{1}$, we have
\[
||\hat{Q}_{\mathbf{x}_{1}^{\tilde{n}}}-\hat{Q}_{\mathbf{x}}||\leq\frac{\epsilon}{2}
\]
and by the triangle inequality 
\[
||\hat{Q}_{\mathbf{x}_{1}^{\tilde{n}}}-\hat{Q}_{\mathbf{v}}||\leq||\hat{Q}_{\mathbf{x}_{1}^{\tilde{n}}}-\hat{Q}_{\mathbf{x}}||+||\hat{Q}_{\mathbf{x}}-\hat{Q}_{\mathbf{v}}||\leq\frac{\epsilon}{2}+\frac{\epsilon}{2}=\epsilon.
\]
Thus, the definition \eqref{eq: Hamming ball definition}, and Lemma
\ref{lem: hamming distance close types} imply that\textbf{ ${\cal K}(\mathbf{x},\epsilon)$
}is indeed non-empty, and an appropriate $\overline{\mathbf{x}}$
can always be found.\\
\emph{Decoding by the legitimate decoder:} Upon observing $y=f_{n}^{*}(\mathbf{x},\mathbf{u})$:
\begin{itemize}
\item Recover the type $\hat{Q}_{\mathbf{x}}$ from $y_{1}$, and determine
$\Phi_{\epsilon}(\hat{Q}_{\mathbf{x}})$ and $|{\cal K}(\mathbf{x},\epsilon)|$.
\item If $\RRc<R_{\st[L]}(\hat{Q}_{\mathbf{x}},\DDc)$ then arbitrarily
choose a vector from $\tilde{\mathbf{w}}\in{\cal W}^{n}$, and reproduce
\[
\mathbf{w}\teq\varphi_{n}^{*}(y,\mathbf{u})=\tilde{\mathbf{w}}.
\]
Otherwise, if $\RRc\geq R_{\st[L]}(\hat{Q}_{\mathbf{x}},\DDc)$ then:

\begin{itemize}
\item Recover $\mathbf{x}_{\tilde{n}+1}^{n}$ from $y_{2}$ and $\mathbf{u}^{(2)}$.
Let $\mathbf{w}''\in{\cal W}^{n-\tilde{n}}$ be such that $d_{\st[L]}(\mathbf{x}_{\tilde{n}+1}^{n},\mathbf{w}'')=0$. 
\item Recover $\mathbf{x}'''$ from $y_{3}$ and $\mathbf{u}^{(3)}$, and
let $\mathbf{w}'''\in{\cal W}^{\tilde{n}}$ be such that $d_{\st[L]}(\mathbf{x}''',\mathbf{w}''')=0$. 
\item Reproduce $\mathbf{v}$ from $y_{4}$ and $\mathbf{u}^{(4)}$ as
\[
\mathbf{w}''''\teq\varphi_{\tilde{n,}\hat{Q}_{\mathbf{x}}}^{*}(y_{4},\mathbf{u}^{(4)})
\]

\item Reproduce the source block as 
\[
\mathbf{w}\teq\varphi_{n}^{*}(y,\mathbf{u})=(\Psi(\mathbf{w}'''',\mathbf{w}'''),\mathbf{w}'').
\]

\end{itemize}
\end{itemize}
Note that the decoder knows $|{\cal K}(\mathbf{x},\epsilon)|$ and
thus can compute the total length of $\mathbf{u}$. So, if multiple
source blocks are encoded in succession, the legitimate decoder can
stay synchronized with the encoder and use the correct key bits when
deciphering the message. 

For the sequence of codes ${\cal S}^{*}$ constructed, we need to
verify that the compression constraint is satisfied, and to find the
achievable exiguous-distortion exponent for any (type aware) eavesdropper,
as well as the key rate. First, consider the compression constraint.
For the rate, recall that the cryptogram is composed of at most four
parts \eqref{eq: cryptogram in 4 parts}. Let ${\cal Y}_{nj}$ be
the alphabet of the $j$-th part, for $1\leq j\leq4$, such that $|{\cal Y}_{n}|=\prod_{j=1}^{4}|{\cal Y}_{nj}|$.
We have,

\[
|{\cal Y}_{n1}|=|{\cal P}_{n}({\cal X})|\leq(n+1)^{|{\cal X}|},
\]
and 
\[
|{\cal Y}_{n2}|=|{\cal X}|^{n-\tilde{n}}.
\]
For ${\cal Y}_{n3}$,
\begin{align}
|{\cal Y}_{n3}|=\left|{\cal B}_{\s[H]}^{\tilde{n}}(\epsilon)\right| & =\sum_{k=0}^{\frac{\tilde{n}\epsilon}{2}}\binom{\tilde{n}}{k}|{\cal X}|^{k}\\
 & \leq\frac{\tilde{n}\epsilon}{2}\cdot\binom{\tilde{n}}{\left\lceil \frac{\tilde{n}\epsilon}{2}\right\rceil }|{\cal X}|^{\frac{\tilde{n}\epsilon}{2}}\\
 & \leq2^{\tilde{n}\left[h_{\st[B]}(\frac{\epsilon}{2})+\frac{\epsilon}{2}\log|{\cal X}|\right]}\\
 & \teq2^{\tilde{n}g(\epsilon)}\label{eq: bounding Hamming ball size}
\end{align}
where $g(\epsilon)$ was implicitly defined, and $g(\epsilon)\downarrow0$
as $\epsilon\downarrow0$. For ${\cal Y}_{n4}$, notice that the cryptogram
part $y_{4}$ is only used for types $Q_{X}$ which satisfy $\RRc\geq R_{\st[L]}(Q_{X},\DDc).$
Thus, 
\begin{align*}
|{\cal Y}_{n4}| & \leq\sum_{Q_{X}\in{\cal P}_{n}(Q_{X}):\RRc\geq R_{\st[L]}(Q_{X},\DDc)}2^{nR_{\st[L]}(Q_{X},\DDc)}\\
 & \leq|{\cal P}_{n}({\cal X})|\cdot2^{n\RRc}
\end{align*}
Therefore, for all $n$ sufficiently large
\begin{align*}
\limsup_{n\to\infty}\frac{1}{n}\log|{\cal Y}_{n}| & \leq\limsup_{n\to\infty}\sum_{j=1}^{4}\frac{1}{n}\log|{\cal Y}_{nj}|\\
 & \leq\RRc+g(\epsilon)+3\delta.
\end{align*}
Now, as the codes ${\cal S}_{\tilde{n},Q_{X}}^{*}$ are constructed
according to Lemma \ref{lem: exponent lower bound, single type},
it is easily verified that if $\RRc\geq R_{\st[L]}(\hat{Q}_{\mathbf{x}},\DDc)$
then for any $\mathbf{u}$ 
\[
d_{\st[L]}(\mathbf{x},\varphi_{n}^{*}(f_{n}^{*}(\mathbf{x},\mathbf{u}),\mathbf{u}))\leq\DDc
\]
(see \eqref{eq: secrecy conditional exponent achievable perfect legitimate covering}).
Thus, as $|{\cal P}_{n}({\cal X})|\leq(n+1)^{|{\cal X}|}$, for all
$n$ sufficiently large 
\begin{align}
 & \hphantom{{}={}}\P\left[d_{\st[L]}(\mathbf{X},\varphi_{n}^{*}(f_{n}^{*}(\mathbf{X},\mathbf{u}),\mathbf{u}))\geq\DDc\right]\\
 & =\sum_{Q_{X}\in{\cal P}_{n}({\cal X})}\P\left[\mathbf{X}\in{\cal T}_{n}(Q_{X})\right]\P\left[d_{\st[L]}(\mathbf{X},\varphi_{n}^{*}(f_{n}^{*}(\mathbf{X},\mathbf{u}),\mathbf{u}))\geq\DDc|\mathbf{X}\in{\cal T}_{n}(Q_{X})\right]\\
 & \leq\sum_{Q_{X}\in{\cal P}_{n}({\cal X}):\RRc<R_{\st[L]}(Q_{X},\DDc)}\P\left[\mathbf{X}\in{\cal T}_{n}(Q_{X})\right]\\
 & \le\sum_{Q_{X}\in{\cal P}_{n}({\cal X}):\RRc<R_{\st[L]}(Q_{X},\DDc)}2^{-nD(Q_{X}||P_{X})}\\
 & \leq2^{-n\left[E_{\st[L]}(P_{X},\DDc,\RRc)-\delta\right]}\\
 & \leq2^{-n\left(\EEc-\delta\right)}.\label{eq: achievability method of type legitimate decoder}
\end{align}
Second, let us analyze the exiguous-distortion exponent of ${\cal S}$
for an arbitrary eavesdropper. Let $\hat{\mathbf{v}}^{*}$ be the
eavesdropper which maximizes the exiguous-distortion probability for
the modified source block $\mathbf{v}$, given the cryptogram $y$.
Then, 
\begin{alignat}{1}
{\cal E}_{d}^{-}({\cal S},\DD) & \trre[=,a]\liminf_{n\to\infty}\min_{Q_{X}\in{\cal P}_{n}({\cal X})}\left\{ D\left(Q_{X}||P_{X}\right)-\frac{1}{n}\log\max_{\tilde{\sigma}_{n}\in\tilde{\Sigma}_{n}}\P\left[d_{\st[E]}(\mathbf{X},\mathbf{Z})\leq\DD|\mathbf{X}\in{\cal T}_{n}(Q_{X})\right]\right\} \\
 & \trre[\geq,b]\liminf_{n\to\infty}\min\Biggl\{\min_{Q_{X}\in{\cal P}_{n}({\cal X}):\RRc\geq R_{\st[L]}(Q_{X},\DDc)}\Biggl\{ D\left(Q_{X}||P_{X}\right)-\nonumber \\
 & \hphantom{\geq{}}\frac{1}{n}\log\left(|{\cal B}_{\s[H]}^{\tilde{n}}(\epsilon)|\P\left[d_{\st[E]}(\mathbf{V},\hat{\mathbf{V}}^{*})\leq\DD|\mathbf{V}\in{\cal T}_{\tilde{n}}(\Phi_{\epsilon}(Q_{X}))\right]\right)\Biggr\},\nonumber \\
 & \hphantom{\geq{}}\min_{Q_{X}\in{\cal P}_{n}({\cal X}):\RRc<R_{\st[L]}(Q_{X},\DDc)}\Biggl\{ D\left(Q_{X}||P_{X}\right)-\frac{1}{n}\log\max_{\tilde{\sigma}_{n}\in\tilde{\Sigma}_{n}}\P\left[d_{\st[E]}(\mathbf{X},\mathbf{Z})\leq\DD|\mathbf{X}\in{\cal T}_{n}(Q_{X})\right]\Biggr\}\Biggr\}\label{eq: achievabiliy erorr exponent modification-1}\\
 & \trre[\geq,c]\liminf_{n\to\infty}\min\Biggl\{\min_{Q_{X}\in{\cal P}_{n}({\cal X}):\RRc\geq R_{\st[L]}(Q_{X},\DDc)}\Biggl\{ D\left(Q_{X}||P_{X}\right)-\nonumber \\
 & \hphantom{\geq{}}\frac{1}{n}\log\left[|{\cal B}_{\s[H]}^{\tilde{n}}(\epsilon)|\P\left[d_{\st[E]}(\mathbf{V},\hat{\mathbf{V}}^{*})\leq\DD|\mathbf{V}\in{\cal T}_{\tilde{n}}(\Phi_{\epsilon}(Q_{X}))\right]\right]\Biggr\},\nonumber \\
 & \hphantom{\geq{}}\min_{Q_{X}\in{\cal P}_{n}({\cal X}):\RRc<R_{\st[L]}(Q_{X},\DDc)}\Biggl\{ D\left(Q_{X}||P_{X}\right)+R_{\st[E]}(Q_{X},\DD)-\delta\Biggr\}\Biggr\}\\
 & =\min\Biggl\{\liminf_{n\to\infty}\min_{Q_{X}\in{\cal P}_{n}({\cal X}):\RRc\geq R_{\st[L]}(Q_{X},\DDc)}\Biggl\{ D\left(Q_{X}||P_{X}\right)-\nonumber \\
 & \hphantom{\geq{}}\frac{1}{n}\log\left[|{\cal B}_{\s[H]}^{\tilde{n}}(\epsilon)|\P\left[d_{\st[E]}(\mathbf{V},\hat{\mathbf{V}}^{*})\leq\DD|\mathbf{V}\in{\cal T}_{\tilde{n}}(\Phi_{\epsilon}(Q_{X}))\right]\right]\Biggr\},\nonumber \\
 & \hphantom{\geq{}}\liminf_{n\to\infty}\min_{Q_{X}\in{\cal P}_{n}({\cal X}):\RRc<R_{\st[L]}(Q_{X},\DDc)}\Biggl\{ D\left(Q_{X}||P_{X}\right)+R_{\st[E]}(Q_{X},\DD)-\delta\Biggr\}\Biggr\},\label{eq: achievability - bounding the exgious distortion}
\end{alignat}
where the passages are explained as follows:
\begin{itemize}
\item Equality $(a)$ is standard method of types, (as, e.g., in \eqref{eq: achievability method of type legitimate decoder}).
Notice that the exiguous-distortion event $\{d_{\st[E]}(\mathbf{X},\mathbf{Z})\leq\DD\}$
in this equation is for the code ${\cal S}_{n}.$
\item Equality $(b)$ is verified by establishing the following properties:

\begin{itemize}
\item \emph{\uline{Property 1:}} Due to the permutation invariance of
type classes and Hamming spheres, given the event \textbf{$\mathbf{X}\in{\cal T}_{n}(Q_{X})$},
$\mathbf{V}$ is distributed uniformly over ${\cal T}_{\tilde{n}}(\Phi_{\epsilon}(Q_{X}))$.
Indeed, let $\mathbf{v}',\mathbf{v}''\in{\cal T}_{\tilde{n}}(\Phi_{\epsilon}(Q_{X}))$,
where $\mathbf{v}'=\pi(\mathbf{v}'')$ for some permutation $\pi$.
Then, if for some $\mathbf{x}\in{\cal T}_{n}(Q_{X})$ and $\overline{\mathbf{x}}\in{\cal K}(\mathbf{x},\epsilon)$
\[
\mathbf{v}'=\Psi(\mathbf{x}_{1}^{\tilde{n}},\overline{\mathbf{x}})
\]
then
\[
\mathbf{v}''=\Psi(\pi(\mathbf{x}_{1}^{\tilde{n}}),\pi(\overline{\mathbf{x}}))
\]
where $(\pi(\mathbf{x}_{1}^{\tilde{n}}),\mathbf{x}_{\tilde{n}+1}^{n})\in{\cal T}_{n}(Q_{X})$
and $\pi(\overline{\mathbf{x}})\in{\cal K}((\pi(\mathbf{x}_{1}^{\tilde{n}}),\mathbf{x}_{\tilde{n}+1}^{n}),\epsilon)$%
\footnote{Notice that ${\cal K}(\mathbf{x})$ depends on $\mathbf{x}$ only
via its first $\tilde{n}$ components.%
}. The property then follows from the fact that $|{\cal K}(\mathbf{x},\epsilon))|$
depends on $\mathbf{x}$ only via its type, which is identical for
both $\mathbf{x}$ and $(\pi(\mathbf{x}_{1}^{\tilde{n}}),\mathbf{x}_{\tilde{n}+1}^{n})$. 
\item \emph{\uline{Property 2:}} An eavesdropper for $\mathbf{v}$ is
aware of its type (as $\hat{Q}_{\mathbf{v}}=\Phi_{\epsilon}(\hat{Q}_{\mathbf{x}})$)%
\footnote{Which is in fact not even required, using Proposition \ref{prop: Type awarness does not help}.%
}, and the cryptogram $y_{2}$ is not relevant for its estimate. Also,
since $y_{3}$ is fully encrypted (pure random bits) then it is also
useless. Thus, an eavesdropper for $\mathbf{v}$ uses only the type
information in $y_{1}$ and $y_{4}$.
\item \emph{\uline{Property 3:}} Consider the case $\RRc\geq R_{\st[L]}(Q_{X},\DDc)$.
The source block $\mathbf{X}$ is distributed uniformly over ${\cal T}_{n}(Q_{X})$
and $\mathbf{V}$ is distributed uniformly over ${\cal T}_{\tilde{n}}(\Phi_{\epsilon}(Q_{X}))$.
Let $\hat{\mathbf{V}}^{*}$ be the eavesdropper which achieves the
maximal exiguous-distortion probability for $\mathbf{V}$, given $y_{4}$.
Then, for any eavesdropper decoder $\tilde{\sigma}_{n}$ which estimates
$\mathbf{z}$ 
\begin{equation}
\frac{1}{|{\cal B}_{\s[H]}^{\tilde{n}}(\epsilon)|}\P\left[d_{\st[E]}(\mathbf{X},\mathbf{Z})\leq\DD|\mathbf{X}\in{\cal T}_{n}(Q_{X})\right]\leq\P\left[d_{\st[E]}(\mathbf{V},\hat{\mathbf{V}}^{*})\leq\DD|\mathbf{V}\in{\cal T}_{\tilde{n}}(\Phi_{\epsilon}(Q_{X}))\right].\label{eq: Relation of guessing X and W}
\end{equation}
Indeed, since $\mathbf{X}_{\tilde{n}+1}^{n}$ is fully encrypted then
it is easy to verify that 
\begin{equation}
\P\left[d_{\st[E]}(\mathbf{X},\mathbf{Z})\leq\DD|\mathbf{X}\in{\cal T}_{n}(Q_{X})\right]\leq\P\left[d_{\st[E]}(\mathbf{X}_{1}^{\tilde{n}},\mathbf{Z}{}_{1}^{\tilde{n}})\leq\DD|\mathbf{X}\in{\cal T}_{n}(Q_{X})\right].\label{eq: distortion after truncation}
\end{equation}
Now, any eavesdropper $\mathbf{Z}_{1}^{\tilde{n}}$ for $\mathbf{X}_{1}^{\tilde{n}}$
can be transformed into an eavesdropper $\hat{\mathbf{V}}$ for $\mathbf{V}$,
by a uniformly distributed guess of $\overline{\mathbf{X}}$ over
${\cal B}_{\s[H]}^{\tilde{n}}(b)$ (see \eqref{eq: modified vector def})
and then setting
\[
\hat{\mathbf{v}}=\begin{cases}
\argmin_{z\in{\cal Z}}d_{\st[E]}(\overline{\mathbf{x}}_{i},z), & \overline{\mathbf{x}}_{i}\neq0\\
\mathbf{z}_{i}, & \overline{\mathbf{x}}_{i}=0
\end{cases}
\]
where by assumption, $\min_{z\in{\cal Z}}d_{\st[E]}(\overline{\mathbf{x}}_{i},z)=0$.
If the guess of $\mathbf{\overline{x}}$ is correct (according to
the relation \eqref{eq: modified vector def}) then 
\[
d_{\st[E]}(\mathbf{v},\hat{\mathbf{v}})\leq d_{\st[E]}(\mathbf{x},\mathbf{z}).
\]
Since this happens with probability larger than $\left[|{\cal B}_{\s[H]}^{\tilde{n}}(\epsilon)|\right]^{-1}$
, then \eqref{eq: distortion after truncation} implies \eqref{eq: Relation of guessing X and W}.
\end{itemize}

Equality $(b)$ then follows from the above considerations. 

\item Inequality $(c)$ is because in case $\RRc<R_{\st[L]}(Q_{X},\DDc)$
the eavesdropper has no knowledge beyond the type of the source block,
and so given such $y$, $\mathbf{x}$ is distributed uniformly over
${\cal T}_{n}(Q_{X})$. For any given $\mathbf{z}\in{\cal Z}^{n}$,
using standard method of types 
\begin{align*}
\P\left[d_{\st[E]}(\mathbf{X},\mathbf{z})\leq\DD|\mathbf{X}\in{\cal T}_{n}(Q_{X})\right] & =\sum_{\mathbf{x}\in{\cal T}_{n}(Q_{X}):d_{\st[E]}(\mathbf{x},\mathbf{z})\leq\DD}\frac{1}{\left|{\cal T}_{n}(Q_{X})\right|}\\
 & =\sum_{Q_{X|Z}:\E_{Q}\left[d_{\st[E]}(X,Z)\right]\leq\DD}\sum_{\mathbf{x}\in{\cal T}_{n}(Q_{X|Z},\mathbf{z})}\frac{1}{\left|{\cal T}_{n}(Q_{X})\right|}\\
 & =\frac{1}{\left|{\cal T}_{n}(Q_{X})\right|}\sum_{Q_{X|Z}:\E_{Q}\left[d_{\st[E]}(X,Z)\right]\leq\DD}\left|{\cal T}_{n}(Q_{X|Z},\mathbf{z})\right|\\
 & \doteq\exp\left\{ -n\cdot\min_{Q_{X|Z}:\E_{Q}\left[d_{\st[E]}(X,Z)\right]\leq\DD}\left[-H_{Q}(X|Z)+H(Q_{X})\right]\right\} 
\end{align*}
Then, \textbf{
\begin{equation}
\max_{\mathbf{z}}\P\left[d_{\st[E]}(\mathbf{X},\mathbf{z})\leq\DD|\mathbf{X}\in{\cal T}_{n}(Q_{X})\right]\leq2^{-n\left[R_{\st[E]}(Q_{X},\DD)-\delta\right]}.\label{eq: blind estimate exponent}
\end{equation}
}
\end{itemize}
Next, we further bound the first term in the minimization of \eqref{eq: achievability - bounding the exgious distortion}
as follows 
\begin{alignat}{1}
 & \hphantom{\geq{}}\liminf_{n\to\infty}\min_{Q_{X}\in{\cal P}_{n}({\cal X}):\RRc\geq R_{\st[L]}(Q_{X},\DDc)}\Biggl\{ D\left(Q_{X}||P_{X}\right)-\nonumber \\
 & \hphantom{\geq{}}\frac{1}{n}\log\left[|{\cal B}_{\s[H]}^{\tilde{n}}(\epsilon)|\P\left[d_{\st[E]}(\mathbf{V},\hat{\mathbf{V}}^{*})\leq\DD|\mathbf{V}\in{\cal T}_{\tilde{n}}(\Phi_{\epsilon}(Q_{X}))\right]\right]\Biggr\}\\
 & \trre[\geq,a]\liminf_{n\to\infty}\min_{Q_{X}\in{\cal P}_{n}({\cal X}):\RRc\geq R_{\st[L]}(Q_{X},\DDc)}\Biggl\{ D\left(Q_{X}||P_{X}\right)-\nonumber \\
 & \hphantom{\geq{}}\frac{1}{n}\log\P\left[d_{\st[E]}(\mathbf{V},\hat{\mathbf{V}}^{*})\leq\DD|\mathbf{V}\in{\cal T}_{\tilde{n}}(\Phi_{\epsilon}(Q_{X}))\right]-g(\epsilon)\Biggr\}\\
 & \trre[=,b]\liminf_{n\to\infty}\min_{Q_{X}\in{\cal P}_{n}({\cal X}):\RRc\geq R_{\st[L]}(Q_{X},\DDc)}\Biggl\{ D\left(Q_{X}||P_{X}\right)-\nonumber \\
 & \hphantom{\geq{}}\frac{1}{\tilde{n}}\log\P\left[d_{\st[E]}(\mathbf{V},\hat{\mathbf{V}}^{*})\leq\DD|\mathbf{V}\in{\cal T}_{\tilde{n}}(\Phi_{\epsilon}(Q_{X}))\right]-g(\epsilon)\Biggr\}\\
 & \trre[\geq,c]\liminf_{n\to\infty}\min_{Q_{X}\in{\cal P}_{n}({\cal X}):\RRc\geq R_{\st[L]}(Q_{X},\DDc)}\Biggl\{ D\left(Q_{X}||P_{X}\right)+\min\left\{ \RR,R_{\st[E]}(Q_{X},\DD)\right\} -\delta-g(\epsilon)\Biggr\}\label{eq: requirement of uniform convergence method of types}\\
 & \trre[\geq,d]\liminf_{n\to\infty}\min_{Q_{X}\in{\cal P}_{n}({\cal X}):\RRc\geq R_{\st[L]}(Q_{X},\DDc)}\Biggl\{ D\left(\Phi_{\epsilon}(Q_{X})||P_{X}\right)+\min\left\{ \RR,R_{\st[E]}(Q_{X},\DD)\right\} -\delta-\delta_{1}(\epsilon)-g(\epsilon)\Biggr\}\\
 & \trre[=,e]\liminf_{n\to\infty}\min_{Q_{X}\in{\cal P}_{n_{0}}({\cal X}):\RRc\geq R_{\st[L]}(Q_{X},\DDc)}\Biggl\{ D\left(\Phi_{\epsilon}(Q_{X})||P_{X}\right)+\min\left\{ \RR,R_{\st[E]}(Q_{X},\DD)\right\} -\delta-\delta_{1}(\epsilon)-g(\epsilon)\Biggr\}\\
 & \trre[=,f]\liminf_{n\to\infty}\min_{Q_{X}\in{\cal P}_{n_{0}}({\cal X}):\RRc\geq R_{\st[L]}(Q_{X},\DDc)}\Biggl\{ D\left(Q_{X}||P_{X}\right)+\min\left\{ \RR,R_{\st[E]}(Q_{X},\DD)\right\} -\delta-\delta_{1}(\epsilon)-g(\epsilon)\Biggr\}\\
 & =\min_{Q_{X}\in{\cal P}_{n_{0}}({\cal X}):\RRc\geq R_{\st[L]}(Q_{X},\DDc)}\Biggl\{ D\left(Q_{X}||P_{X}\right)+\min\left\{ \RR,R_{\st[E]}(Q_{X},\DD)\right\} -\delta-\delta_{1}(\epsilon)-g(\epsilon)\Biggr\},\\
 & \geq\liminf_{n\to\infty}\min_{Q_{X}\in{\cal P}_{n}({\cal X}):\RRc\geq R_{\st[L]}(Q_{X},\DDc)}\Biggl\{ D\left(Q_{X}||P_{X}\right)+\min\left\{ \RR,R_{\st[E]}(Q_{X},\DD)\right\} -\delta-\delta_{1}(\epsilon)-g(\epsilon)\Biggr\}\label{eq: achievability - bounding the exgious distortion less equal EEc}
\end{alignat}

\begin{itemize}
\item Inequality $(a)$ follows from the fact that since $0<\epsilon<1$,
for all $n$ sufficiently large $\left|{\cal B}_{\s[H]}^{\tilde{n}}(\epsilon)\right|\leq2^{\tilde{n}g(\epsilon)}$
as in \eqref{eq: bounding Hamming ball size}.
\item Equality $(b)$ is because $\frac{\tilde{n}}{n}\to1$ as $n\to\infty$. 
\item Inequality $(c)$ is because there exists $n_{2}$ sufficiently large,
such that for all $n>n_{2}$ the error probability of the any eavesdropper
decoder $\sigma_{\tilde{n},\Phi_{\epsilon}(Q_{X})}^{*}$ satisfies
\[
-\frac{1}{\tilde{n}}\log\P\left[\hat{\mathbf{V}}\neq\mathbf{V}|\mathbf{V}\in{\cal T}_{\tilde{n}}(\Phi_{\epsilon}(Q_{X}))\right]\geq\min\left\{ \RR,R_{\st[E]}(Q_{X},\DD)\right\} -\delta
\]
 \emph{uniformly }for all \emph{$Q_{X}\in{\cal P}_{n_{0}}(Q_{X})$}. 
\item Inequality $(d)$ is by defining 
\[
\delta_{1}(\epsilon)\teq\max_{Q_{X}}\left|D\left(\Phi_{\epsilon}(Q_{X})||P_{X}\right)-D(Q_{X}||P_{X})\right|.
\]
Note that since $D(Q_{X}||P_{X})$ is a continuous function of $Q_{X}$
in ${\cal Q}({\cal X})$ (as the support of $P_{X}$ is assumed to
be ${\cal X}$), it is also uniformly continuous. So, $\delta_{1}(\epsilon)\downarrow0$
as $\epsilon\downarrow0$.
\item Equalities $(e)$ and $(f)$ are because $\Phi_{\epsilon}(Q_{X})\in{\cal P}_{n_{0}}({\cal X})$
for all $Q_{X}\in{\cal P}_{n}({\cal X})$.
\end{itemize}
Substituting \eqref{eq: achievability - bounding the exgious distortion less equal EEc}
into \eqref{eq: achievability - bounding the exgious distortion},
and using the fact ${\cal P}_{n}({\cal X})\subset{\cal Q}({\cal X})$
we obtain
\begin{align}
{\cal E}_{d}^{-}({\cal S},\DD) & \geq\min\Biggl\{\min_{Q_{X}\in{\cal Q}({\cal X}):\RRc\geq R_{\st[L]}(Q_{X},\DDc)}\Biggl\{ D\left(Q_{X}||P_{X}\right)+\min\left\{ \RR,R_{\st[E]}(Q_{X},\DD)\right\} -g(\epsilon)-\delta_{1}(\epsilon),\nonumber \\
 & \hphantom{\geq{}}\min_{Q_{X}\in{\cal Q}({\cal X}):\RRc<R_{\st[L]}(Q_{X},\DDc)}\Biggl\{ D\left(Q_{X}||P_{X}\right)+R_{\st[E]}(Q_{X},\DD)\Biggr\}\Biggr\}-\delta\\
 & \geq\min\Biggl\{\min_{Q_{X}\in{\cal Q}({\cal X}):\RRc\geq R_{\st[L]}(Q_{X},\DDc)}\Biggl\{ D\left(Q_{X}||P_{X}\right)+\RR,\nonumber \\
 & \hphantom{\geq{}}\min_{Q_{X}\in{\cal P}({\cal X})}\Biggl\{ D\left(Q_{X}||P_{X}\right)+R_{\st[E]}(Q_{X},\DD)\Biggr\}\Biggr\}-\delta-\delta_{1}(\epsilon)-g(\epsilon)\\
 & \trre[\geq,a]\min\left\{ \RR,E_{e}^{*}(\DD)\right\} -\delta-\delta_{1}(\epsilon)-g(\epsilon)
\end{align}
where in $(a)$ we have used the definition in \eqref{eq: perfect secrecy exponent},
and the fact that the assumption $\EEc>0$ implies that $\RRc\geq R_{\st[L]}(P_{X},\DDc)$.

Next, we analyze the required key rate. If $\RRc<R_{\st[L]}(\hat{Q}_{\mathbf{x}},\DDc)$
then the required key rate is zero. Otherwise, if $\RRc\geq R_{\st[L]}(\hat{Q}_{\mathbf{x}},\DDc)$
then the total key rate required to encode $\mathbf{x}\in Q_{X}$
is given by
\[
\frac{1}{n}\left[(n-\tilde{n})\log|{\cal X}|+\log\left|{\cal K}(\mathbf{x},\epsilon)\right|+\log\left|{\cal B}_{\s[H]}^{\tilde{n}}(\epsilon)\right|+n\RR\right].
\]
Now, for all $n$ sufficiently large 
\[
\frac{1}{n}(n-\tilde{n})\log|{\cal X}|\leq\frac{n_{0}\log|{\cal X}|+1}{n}\leq\delta,
\]
\[
\frac{1}{n}\log\left|{\cal K}(\mathbf{x},\epsilon)\right|\leq\frac{1}{n}\log\left|{\cal B}_{\s[H]}^{\tilde{n}}(\epsilon)\right|\leq g(\epsilon),
\]
Thus, the required key rate is less than 
\[
\RR+2g(\epsilon)+2\delta.
\]
By taking $\epsilon\downarrow0$ we obtain $g(\epsilon)\downarrow0$
and $\delta_{1}(\epsilon)\downarrow0$, and so we obtain the achievability
part of Theorem \ref{thm:exiguous distortion expoent}.
\end{IEEEproof}

\subsection{Proof of Converse Part of Theorem \ref{thm:exiguous distortion expoent}
\label{sub:Proof-of-Converse}}

Following the outline of the converse, we begin with a lemma which
constructs from a given sequence of codes ${\cal S}$ a new sequence
${\cal S}^{*}$, with constant key rate, which is less than $\overline{R}({\cal S},Q_{X})+\delta$,
and a zero excess-distortion probability at the legitimate receiver.
\begin{lem}
\label{lem: converse - fixed type key-rate}Let ${\cal S}$ be an
arbitrary sequence of secure rate-distortion codes, which satisfies
a compression constraint $(\RRc,\DDc,\EEc)$. Also, let $Q_{X}\in{\cal P}({\cal X})$
be given such that $D(Q_{X}||P_{X})<\EEc$. Then, for every $\delta>0$,
there exists a sequence of secure rate-distortion codes ${\cal S}^{*}$
such that:
\begin{enumerate}
\item For all $n$ and all $\mathbf{x}\in{\cal T}_{n}(Q_{X})$, ${\cal S}_{n}^{*}$
has fixed key rate $r^{*}(\mathbf{x})=\RR^{*}$ where $\RR^{*}\leq\overline{R}({\cal S},Q_{X})+\delta$.
\item For all $n$ and $\{u_{i}\}_{i=1}^{\infty}$, ${\cal S}_{n}^{*}=(f_{n}^{*},\varphi_{n}^{*})$
satisfies 
\[
\P\left[d_{\st[L]}(\mathbf{X},\varphi_{n}^{*}(f_{n}^{*}(\mathbf{X},\mathbf{u}),\mathbf{u}))>\DDc|\mathbf{X}\in{\cal T}_{n}(Q_{X})\right]=0,
\]
and in addition, ${\cal S}^{*}$ satisfies a compression constraint
$(\RRc^{*},\DDc,\EEc)$ for $\RRc^{*}=\log|{\cal X}|$.
\item For every $\DD\geq\DDc$. 
\begin{equation}
{\cal E}_{d}^{+}({\cal S},\DD,Q_{X})\leq{\cal E}_{d}^{+}({\cal S}^{*},\DD,Q_{X})+\delta.\label{eq: converse - fixed rate lemma exiguous distortion}
\end{equation}

\end{enumerate}
\end{lem}
\begin{IEEEproof}
We will prove this lemma by modifying the sequence of codes ${\cal S}$
into the new sequence ${\cal S}^{*}$. Assume that $Q_{X}\in\interior{\cal Q}({\cal X}),$
and $Q_{X}\in{\cal P}_{n_{0}}({\cal X})$ for some minimal $n_{0}\in\mathbb{N}$.
Since the statements in the lemma are only about conditional events
given the type $Q_{X}$, it is clear that the new secure rate-distortion
codes constructed ${\cal S}_{n}^{*}$ need only be different from
${\cal S}_{n}$ for $\mathbf{x}\in{\cal T}_{n}(Q_{X})$, and so only
block-lengths $n\bmod n_{0}=0$ should be considered, as otherwise
${\cal T}_{n}(Q_{X})$ is empty. To wit, the limit $n\to\infty$ should
be read as limit $l\to\infty$ for $n=n_{0}l$, but this will not
be explicitly written, for the sake of brevity. Throughout the proof,
quantities that are related to ${\cal S}^{*}$ will be superscripted
by $*$. For brevity, we will denote the conditional key rate by $\overline{R}(Q_{X})$
and $\overline{R}^{*}(Q_{X})$ for ${\cal S}$ and ${\cal S}^{*}$,
respectively . 

Let $\delta>0$ be given. For any length $0\leq m\leq n\log|{\cal X}|$
and $y\in{\cal Y}_{n}$ define the \emph{ambiguity} \emph{sets for
a given key-length as}
\[
{\cal A}_{n}(y,m)\teq\left\{ \mathbf{x}\in{\cal T}_{n}(Q_{X}):k_{n}\mathbf{(x})=m,f_{n}(\mathbf{x},\mathbf{u})=y\mbox{ for some }\mathbf{u}\in\{0,1\}^{m}\right\} ,
\]
and with a slight abuse of notation define the \emph{ambiguity} \emph{set}%
\footnote{Called \emph{residue class} in the terminology of \cite{shannon1949communication}.%
} as 
\[
{\cal A}_{n}(y)\teq\bigcup_{m=0}^{n\log|{\cal X}|}{\cal A}_{n}(y,m).
\]
For any given $y$ and $\mathbf{x}\in{\cal A}_{n}(y)$, let us denote
the reproduction $\mathbf{w}(\mathbf{x},y)\teq\varphi(y,\mathbf{u})$,
where $\mathbf{u}$ satisfies $f_{n}(\mathbf{x},\mathbf{u})=y$, and
the \emph{ambiguity} \emph{set} \emph{without excess-distortion 
\[
{\cal D}_{n}(y)\teq\left\{ \mathbf{x}\in{\cal A}_{n}(y):d_{\st[L]}(\mathbf{x},\mathbf{w}(\mathbf{x},y))\leq\DDc\right\} .
\]
}Also, consider the \emph{modified ambiguity set} 
\[
{\cal A}_{n}^{*}(y)\teq\left\{ {\cal A}_{n}(y)\backslash\bigcup_{m=0}^{n(\overline{R}(Q_{X})-\delta)}{\cal A}_{n}(y,m)\backslash\bigcup_{m=n(\overline{R}(Q_{X})+\delta)}^{n\log|{\cal X}|}{\cal A}_{n}(y,m)\right\} \bigcap{\cal D}_{n}(y).
\]
For a given $y$, the eavesdropper knows that $\mathbf{x}\in{\cal A}_{n}(y)$
and chooses its estimate accordingly. However, conditioned on $y$,
the probability of $\mathbf{X}$ is not uniform over ${\cal A}_{n}(y)$,
since $k_{n}(\mathbf{x})$ is not the same for all $\mathbf{x}\in{\cal A}_{n}(y)$.
The proof of the lemma is divided into two steps and its outline is
as follows. In the first step, we will identify a sequence of cryptograms
$\{y_{n}^{*}\}$ which simultaneously satisfies the following properties: 
\begin{enumerate}
\item The conditional exiguous-distortion exponent of the eavesdropper when
$\mathbf{X}$ is distributed \emph{uniformly} over ${\cal A}_{n}^{*}(y_{n}^{*})$
is larger than the one for $\mathbf{X}$ distributed over ${\cal A}_{n}(y_{n}^{*})$
according to the distribution induced by ${\cal S}_{n}$.
\item The conditional exiguous-distortion exponent conditioned on $Y=y_{n}^{*}$
equals the same exponent without this conditioning. 
\end{enumerate}
In the second step of the proof, we utilize the set ${\cal A}_{n}^{*}(y_{n}^{*})$
to construct the new sequence of codes ${\cal S}^{*}$. This is done
by the same technique used in the achievability proof of Lemma \ref{lem: exponent lower bound, single type}
- by an efficient covering of the type class using permutations of
one good set ${\cal A}_{n}^{*}(y_{n}^{*})$. The two properties above
of $y_{n}^{*}$ will be used to show that the exiguous-distortion
exponent of ${\cal S}^{*}$ may be only slightly less than that of
${\cal S}$. 

We begin with the first step. For brevity, let us assume that $\mathbf{X}$
is distributed uniformly over the type class ${\cal T}_{n}(Q_{X})$,
and probabilities, expectations and entropies will be calculated w.r.t.
this probability distribution. So, we only consider $y$ such that
${\cal A}_{n}(y)$ is non-empty. If we let 
\begin{equation}
A(y)\teq\P\left[\overline{R}(Q_{X})-\delta\leq r_{n}(\mathbf{X})\leq\overline{R}(Q_{X})+\delta,d_{\st[L]}(\mathbf{X},\mathbf{W})\leq\DDc|Y=y\right]\label{eq: definition of A(y)}
\end{equation}
then for $n$ sufficiently large 
\begin{align}
\E\left[A(Y)\right] & =\P\left[\overline{R}(Q_{X})-\delta\leq r_{n}(\mathbf{X})\leq\overline{R}(Q_{X})+\delta,d_{\st[L]}(\mathbf{X},\mathbf{W})\leq\DDc\right]\nonumber \\
 & \geq\P\left[\overline{R}(Q_{X})-\delta\leq r_{n}(\mathbf{X})\leq\overline{R}(Q_{X})+\delta\right]-\P\left[d_{\st[L]}(\mathbf{X},\mathbf{W})>\DDc\right]\nonumber \\
 & \trre[\geq,a]\delta-\P\left[d_{\st[L]}(\mathbf{X},\mathbf{W})>\DDc\right]\nonumber \\
 & \trre[\geq,b]\delta-2^{-n\left[\EEc-D(Q_{X}||P_{X})-\delta\right]}\label{eq: usage of excess-distortion constraint}\\
 & \teq\frac{\delta}{2}\label{eq: lower bound on average of A(y)}
\end{align}
where $(a)$ is using the convergence in probability of $r_{n}(\mathbf{X})$
to $\overline{R}(Q_{X})$ (see \eqref{eq: uniform convergence in probability}),
and $(b)$ is since ${\cal S}$ satisfies a compression constraint
$(\RRc,\DDc,\EEc)$ and the assumption $D(Q_{X}||P_{X})<\EEc$. Defining
for any $0<\beta<1$ 
\[
{\cal V}_{n}^{(1)}\teq\left\{ y\in{\cal Y}_{n}:A(y)\geq\beta\cdot\frac{\delta}{2}\right\} ,
\]
then, since from the definition \eqref{eq: definition of A(y)} and
\eqref{eq: lower bound on average of A(y)} 
\[
0\leq\frac{A(y)}{\E\left[A(Y)\right]}\leq\frac{2}{\delta}
\]
for all $y\in{\cal Y}_{n}$, the reverse Markov inequity (Lemma \ref{lem: reverse Markov})
implies that 
\[
\P\left(Y\in{\cal V}_{n}^{(1)}\right)\geq\frac{1-\beta}{\frac{2}{\delta}-\beta}\teq\zeta(\delta,\beta),
\]
and choosing some $\beta^{*}<\min\{1,\frac{2}{\delta}\}$, we obtain
$\zeta^{*}(\delta)\teq\zeta(\delta,\beta^{*})>0$. Now, for $\gamma>1$,
let

\[
{\cal V}_{n}^{(2)}\teq\left\{ y\in{\cal Y}_{n}:\max_{\mathbf{z}}\P\left[d_{\st[E]}(\mathbf{X},\mathbf{z})\leq\DD|Y=y\right]<\gamma\cdot\max_{\tilde{\sigma}_{n}\in\tilde{\Sigma}_{n}}\P\left[d_{\st[E]}(\mathbf{X},\mathbf{Z})\leq\DD\right]\right\} .
\]
Then the Markov inequality implies 
\begin{alignat}{1}
\P(Y\not\in{\cal V}_{n}^{(2)}) & =\P\left[\max_{\mathbf{z}}\P\left[d_{\st[E]}(\mathbf{X},\mathbf{z})\leq\DD|Y\right]\geq\gamma\cdot\max_{\tilde{\sigma}_{n}\in\tilde{\Sigma}_{n}}\P\left[d_{\st[E]}(\mathbf{X},\mathbf{Z})\leq\DD\right]\right]\\
 & \trre[\leq,a]\frac{\E\left[\max_{\mathbf{z}}\P\left[d_{\st[E]}(\mathbf{X},\mathbf{z})\leq\DD|Y\right]\right]}{\gamma\cdot\max_{\tilde{\sigma}_{n}\in\tilde{\Sigma}_{n}}\P\left[d_{\st[E]}(\mathbf{X},\mathbf{Z})\leq\DD\right]}\\
 & =\frac{1}{\gamma}
\end{alignat}
where in $(a)$ is should be recalled that $\mathbf{z}$ is chosen
as a function of $Y$. Hence, by the union bound 
\begin{align*}
\P\left(Y\in{\cal V}_{n}^{(1)}\cap{\cal V}_{n}^{(2)}\right) & \geq1-\P\left(Y\not\in{\cal V}_{n}^{(1)}\right)-\P\left(Y\not\in{\cal V}_{n}^{(2)}\right)\\
 & \geq\zeta^{*}(\delta)-\frac{1}{\gamma}.
\end{align*}
Thus, for any given $\delta$, there exists $\gamma^{*}>1$ sufficiently
large (but independent of $n$) such that
\[
\P\left(Y\in{\cal V}_{n}^{(1)}\cap{\cal V}_{n}^{(2)}\right)>0.
\]
Therefore, there exists a sequence $\{y_{n}^{*}\}$ such that for
all $n$ sufficiently large, $y_{n}^{*}\in{\cal V}_{n}^{(1)}\cap{\cal V}_{n}^{(2)}.$ 

In the second step of the proof, we describe the construction of ${\cal S}_{n}^{*}$.
Note that by letting 
\[
{\cal U}^{*}\teq\left\{ \mathbf{u}:\exists\mathbf{x}\in{\cal A}_{n}^{*}(y_{n}^{*})\mbox{ such that }f_{n}(\mathbf{x},\mathbf{u})=y_{n}^{*}\right\} 
\]
and 
\[
{\cal C}_{n}^{*}\teq\left\{ \varphi_{n}(y_{n}^{*},\mathbf{u}):\mathbf{u}\in{\cal U}^{*}\right\} 
\]
we have that ${\cal A}_{n}^{*}(y_{n}^{*})\subseteq\mathfrak{D}({\cal C}_{n}^{*},Q_{X},\DDc)$.
Now, recall that in Lemma \ref{lem: exponent lower bound, single type}
of the achievability proof, we have utilized permutations of a D-cover
$\mathfrak{D}({\cal C}_{n}^{*},Q_{X},\DDc)$ (of a set ${\cal C}_{n}^{*}$)
which cover the type class ${\cal T}_{n}(Q_{X})$, to construct a
secure rate-distortion code. Following remark \ref{rem: on permutations ach. proof},
the set ${\cal A}_{n}^{*}(y_{n})$ can also be used as a constituent
set in the construction of a secure rate-distortion code, and the
conditional exiguous-distortion exponent equal to the exponent achieved
when the source block $\mathbf{X}$ is distributed uniformly over
${\cal A}_{n}^{*}(y_{n}^{*})$, as in \eqref{eq: exiguous-distortion probability for a subset of D-cover}.
Let us find the exponent achieved when $\mathbf{X}$ is distributed
uniformly over ${\cal A}_{n}^{*}(y_{n}^{*})$. To this end, denote

\[
{\cal M}(\delta)\teq\left[n\left(\overline{R}(Q_{X})-\delta\right),n\left(\overline{R}(Q_{X})+\delta\right)\right].
\]
and observe that for an arbitrary eavesdropper $\overline{\mathbf{z}}$,
and all $n$ sufficiently large, 
\begin{align*}
 & \hphantom{{}\hphantom{\geq{}}}\max_{\mathbf{z}}\P\left[d_{\st[E]}(\mathbf{X},\mathbf{z})\leq\DD|Y=y_{n}^{*}\right]\\
 & \geq\P\left[d_{\st[E]}(\mathbf{X},\overline{\mathbf{z}})\leq\DD|Y=y_{n}^{*}\right]\\
 & =\sum_{\mathbf{x}\in{\cal A}_{n}(y_{n}^{*}):d_{\st[E]}(\mathbf{x},\overline{\mathbf{z}})\leq\DD}\P\left[\mathbf{X}=\mathbf{x}|Y=y_{n}^{*}\right]\\
 & =\sum_{m=0}^{n\log|{\cal X}|}\sum_{\mathbf{x}\in{\cal A}_{n}(y_{n}^{*},m):d_{\st[E]}(\mathbf{x},\overline{\mathbf{z}})\leq\DD}\P\left[\mathbf{X}=\mathbf{x}|Y=y_{n}^{*}\right]\\
 & =\frac{\sum_{m=0}^{n\log|{\cal X}|}\sum_{\mathbf{x}\in{\cal A}_{n}(y_{n}^{*},m):d_{\st[E]}(\mathbf{x},\overline{\mathbf{z}})\leq\DD}\P\left(\mathbf{X}=\mathbf{x},Y=y_{n}^{*}\right)}{\P\left(Y=y_{n}^{*}\right)}\\
 & \geq\frac{\sum_{m\in{\cal M}(\delta)}\sum_{\mathbf{x}\in{\cal A}_{n}(y_{n}^{*},m):d_{\st[E]}(\mathbf{x},\overline{\mathbf{z}})\leq\DD}\P\left(\mathbf{X}=\mathbf{x},Y=y_{n}^{*}\right)}{\P\left(Y=y_{n}^{*}\right)}\\
 & \geq\frac{\sum_{m\in{\cal M}(\delta)}\sum_{\mathbf{x}\in{\cal A}_{n}(y_{n}^{*},m)\cap{\cal D}_{n}(y_{n}^{*}):d_{\st[E]}(\mathbf{x},\overline{\mathbf{z}})\leq\DD}\P\left(\mathbf{X}=\mathbf{x},Y=y_{n}^{*}\right)}{\P\left(Y=y_{n}^{*}\right)}\\
 & \trre[\geq,a]\beta\frac{\delta}{2}\cdot\frac{\sum_{m\in{\cal M}(\delta)}\sum_{\mathbf{x}\in{\cal A}_{n}(y_{n}^{*},m)\cap{\cal D}_{n}(y_{n}^{*}):d_{\st[E]}(\mathbf{x},\overline{\mathbf{z}})\leq\DD}\P\left(\mathbf{X}=\mathbf{x},Y=y_{n}^{*}\right)}{\P\left[\overline{R}(Q_{X})-\delta\leq r_{n}(\mathbf{X})\leq\overline{R}(Q_{X})+\delta,d_{\st[L]}(\mathbf{X},\mathbf{W})\leq\DDc,Y=y_{n}^{*}\right]}\\
 & =\beta\frac{\delta}{2}\cdot\frac{\sum_{m\in{\cal M}(\delta)}\sum_{\mathbf{x}\in{\cal A}_{n}(y_{n}^{*},m)\cap{\cal D}_{n}(y_{n}^{*}):d_{\st[E]}(\mathbf{x},\overline{\mathbf{z}})\leq\DD}\P\left(\mathbf{X}=\mathbf{x},Y=y_{n}^{*}\right)}{\sum_{m\in{\cal M}(\delta)}\sum_{\mathbf{x}\in{\cal A}_{n}(y_{n}^{*},m)\cap{\cal D}_{n}(y_{n}^{*})}\P\left(\mathbf{X}=\mathbf{x},Y=y_{n}^{*}\right)}\\
 & =\beta\frac{\delta}{2}\cdot\frac{\sum_{m\in{\cal M}(\delta)}\sum_{\mathbf{x}\in{\cal A}_{n}(y_{n}^{*},m)\cap{\cal D}_{n}(y_{n}^{*}):d_{\st[E]}(\mathbf{x},\overline{\mathbf{z}})\leq\DD}\P\left(\mathbf{X}=\mathbf{x},Y=y_{n}^{*}\right)}{\sum_{m\in{\cal M}(\delta)}\sum_{\mathbf{x}\in{\cal A}_{n}(y_{n}^{*},m)\cap{\cal D}_{n}(y_{n}^{*})}\P\left(\mathbf{X}=\mathbf{x},Y=y_{n}^{*}\right)}\\
 & =\beta\frac{\delta}{2}\cdot\frac{\sum_{m\in{\cal M}(\delta)}\sum_{\mathbf{x}\in{\cal A}_{n}(y_{n}^{*},m)\cap{\cal D}_{n}(y_{n}^{*}):d_{\st[E]}(\mathbf{x},\overline{\mathbf{z}})\leq\DD}\P\left(Y=y_{n}^{*}|\mathbf{X}=\mathbf{x}\right)}{\sum_{m\in{\cal M}(\delta)}\sum_{\mathbf{x}\in{\cal A}_{n}(y_{n}^{*},m)\cap{\cal D}_{n}(y_{n}^{*})}\P\left(Y=y_{n}^{*}|\mathbf{X}=\mathbf{x}\right)}\\
 & \trre[=,b]\beta\frac{\delta}{2}\cdot\frac{\sum_{m\in{\cal M}(\delta)}2^{-m}\cdot\left|\left\{ \mathbf{x}\in{\cal A}_{n}(y_{n}^{*},m)\cap{\cal D}_{n}(y_{n}^{*}):d_{\st[E]}(\mathbf{x},\overline{\mathbf{z}})\leq\DD\right\} \right|}{\sum_{m\in{\cal M}(\delta)}2^{-m}\cdot\left|{\cal A}_{n}(y_{n}^{*},m)\cap{\cal D}_{n}(y_{n}^{*})\right|}\\
 & \geq\beta\cdot\frac{\delta}{2}\frac{2^{-n\left(\overline{R}(Q_{X})+\delta\right)}\cdot\sum_{m\in{\cal M}(\delta)}\left|\left\{ \mathbf{x}\in{\cal A}_{n}(y_{n}^{*},m)\cap{\cal D}_{n}(y_{n}^{*}):d_{\st[E]}(\mathbf{x},\overline{\mathbf{z}})\leq\DD\right\} \right|}{2^{-n\left(\overline{R}(Q_{X})-\delta\right)}\cdot\sum_{m\in{\cal M}(\delta)}\left|{\cal A}_{n}(y_{n}^{*},m)\cap{\cal D}_{n}(y_{n}^{*})\right|}\\
 & =\beta\frac{\delta}{2}\cdot2^{-2n\delta}\frac{\sum_{m\in{\cal M}(\delta)}\left|\left\{ \mathbf{x}\in{\cal A}_{n}(y_{n}^{*},m)\cap{\cal D}_{n}(y_{n}^{*}):d_{\st[E]}(\mathbf{x},\overline{\mathbf{z}})\leq\DD\right\} \right|}{\sum_{m\in{\cal M}(\delta)}\left|{\cal A}_{n}(y_{n}^{*},m)\cap{\cal D}_{n}(y_{n}^{*})\right|}\\
 & \teq\beta\frac{\delta}{2}\cdot2^{-2n\delta}\P\left[d_{\st[E]}(\mathbf{X}^{*},\overline{\mathbf{z}})\leq\DD\right],
\end{align*}
where $(a)$ is because as $y_{n}^{*}\in{\cal V}_{n}^{(1)}$ implies
that 
\[
\frac{\P\left[\overline{R}(Q_{X})-\delta\leq r_{n}(\mathbf{X})\leq\overline{R}(Q_{X})+\delta,d_{\st[L]}(\mathbf{X},\mathbf{W})\leq\DDc,Y=y_{n}^{*}\right]}{\beta\frac{\delta}{2}}\geq\P(Y=y_{n}^{*}),
\]
and $(b)$ is because for admissible encoders and $\mathbf{x}\in{\cal A}_{n}(y_{n}^{*},m)$
\[
\P\left(Y=y_{n}^{*}|\mathbf{X}=\mathbf{x}\right)=2^{-m}.
\]
Thus, \textbf{
\begin{align*}
\limsup_{n\to\infty}-\frac{1}{n}\log\max_{\mathbf{z}}\P\left[d_{\st[E]}(\mathbf{X}^{*},\mathbf{z})\leq\DD\right] & \geq\limsup_{n\to\infty}-\frac{1}{n}\max_{\mathbf{z}}\log\P\left[d_{\st[E]}(\mathbf{X},\mathbf{z})\leq\DD|Y=y_{n}^{*}\right]-3\delta\\
 & \trre[=,a]{\cal E}_{d}^{+}({\cal S},\DD,Q_{X})-3\delta
\end{align*}
}where $(a)$ is because $y_{n}^{*}\in{\cal V}_{n}^{(2)}$. So, by
choosing $\delta$ sufficiently small, we can achieve \eqref{eq: converse - fixed rate lemma exiguous distortion}
by the permutation construction of Lemma \ref{lem: exponent lower bound, single type}. 

Finally, as the legitimate reconstruction $\mathbf{w}(\mathbf{x},y_{n}^{*})$
of any $\mathbf{x}\in{\cal A}_{n}^{*}(y_{n}^{*})$ satisfies $d_{\st[L]}(\mathbf{x},\mathbf{w}(\mathbf{x},y_{n}^{*}))\leq\DDc$,
the permutation construction assures this property for all $\mathbf{x}\in{\cal T}_{n}(Q_{X})$.
So, it is easy to verify that if ${\cal S}$ has excess-distortion
exponent $\EEc$ at distortion level $\DDc$, then ${\cal S}^{*}$
has an even larger exponent. As $\RRc^{*}=\log|{\cal X}|$, the compression
constraint $(\RRc^{*},\DDc,\EEc)$ is satisfied by ${\cal S}^{*}$.
\end{IEEEproof}
We are now ready for the second and final step of the proof of the
converse part of Theorem \ref{thm:exiguous distortion expoent}. 
\begin{IEEEproof}[Proof of converse part of Theorem \ref{thm:exiguous distortion expoent}]
 Let a sequence of secure rate-distortion codes ${\cal S}$ be given,
which satisfies the compression constraint $(\RRc,\DDc,\EEc)$, and
let $\delta>0$ be given. From Proposition \ref{prop: Type awarness does not help},
it may be assumed that the eavesdropper is aware of the type of the
source block $Q_{X}$. Moreover, from Lemma \ref{lem: converse - fixed type key-rate},
it may be assumed that ${\cal S}_{n}$ satisfies the three properties
in Lemma \ref{lem: converse - fixed type key-rate} for all $Q_{X}$
such that $D(Q_{X}||P_{X})<\EEc$. Specifically, the first property
implies that for some \emph{rate-function} $\rho:{\cal P}({\cal X})\to\mathbb{R}_{+}$
the code ${\cal S}_{n}$ has a fixed rate $r_{n}(\mathbf{x})=\rho(Q_{X})$
for all $\mathbf{x}\in{\cal T}_{n}(Q_{X})$, and $\rho(Q_{X})\leq\overline{R}({\cal S},Q_{X})+\delta$,
as long as $D(Q_{X}||P_{X})<\EEc$. 

Let us first focus on a type $Q_{X}$ that satisfies $D(Q_{X}||P_{X})<\EEc$,
and a specific (type-aware) eavesdropper for ${\cal S}_{n}$. The
eavesdropper first produces a guess $\hat{\mathbf{u}}$ of the key-bits
$\mathbf{u}$ (with a uniform probability over $\{0,1\}^{n\rho(Q_{X})}$,
and then decodes $\hat{\mathbf{w}}=\varphi_{n}(y,\hat{\mathbf{u}})$.
Since $d_{\st[E]}(\cdot,\cdot)$ is more lenient than $d_{\st[L]}(\cdot,\cdot)$,
and $\DD\geq\DDc$, there exists a $\hat{\mathbf{z}}\in{\cal Z}^{n}$
such that 
\begin{align*}
\left\{ \mathbf{x}\in{\cal X}^{n}:d_{\st[L]}(\mathbf{x},\hat{\mathbf{w}})\leq\DDc\right\}  & \subseteq\left\{ \mathbf{x}\in{\cal X}^{n}:d_{\st[E]}(\mathbf{x},\hat{\mathbf{z}})\leq\DDc\right\} \\
 & \subseteq\left\{ \mathbf{x}\in{\cal X}^{n}:d_{\st[E]}(\mathbf{x},\hat{\mathbf{z}})\leq\DD\right\} ,
\end{align*}
and so the final eavesdropper estimate is $\mathbf{z}=\hat{\mathbf{z}}$.
For any $n$, let us bound the resulting conditional exiguous-distortion
probability. 
\begin{align}
\P\left[d_{\st[E]}(\mathbf{X},\hat{\mathbf{Z}})\leq\DD|\mathbf{X}\in{\cal T}_{n}(Q_{X})\right] & \geq\P\left[\hat{\mathbf{U}}=\mathbf{U}|\mathbf{X}\in{\cal T}_{n}(Q_{X})\right]\times\nonumber \\
 & \hphantom{{}\geq{}}\P\left[d_{\st[E]}(\mathbf{X},\hat{\mathbf{Z}})\leq\DD|\mathbf{X}\in{\cal T}_{n}(Q_{X}),\hat{\mathbf{U}}=\mathbf{U}\right]\\
 & \geq2^{-n\rho(Q_{X})}\cdot\P\left[d_{\st[E]}(\mathbf{X},\hat{\mathbf{Z}})\leq\DD|\mathbf{X}\in{\cal T}_{n}(Q_{X}),\hat{\mathbf{U}}=\mathbf{U}\right]\\
 & \geq2^{-n\rho(Q_{X})}\cdot\P\left[d_{\st[L]}(\mathbf{X},\mathbf{W})\leq\DD|\mathbf{X}\in{\cal T}_{n}(Q_{X})\right]\\
 & \trre[=,a]2^{-n\rho(Q_{X})}\label{eq: converse lower bound R}
\end{align}
where $(a)$ is from the second property assured for ${\cal S}$ in
Lemma \ref{lem: converse - fixed type key-rate}. 

We now analyze the exiguous-distortion probability of ${\cal S}$.
Since $|{\cal P}_{n}({\cal X})|\leq(n+1)^{|{\cal X}|}$ 
\begin{alignat}{1}
p_{d}({\cal S}_{n},\DD) & =\sum_{Q_{X}\in{\cal P}_{n}({\cal X})}\P\left[\mathbf{X}\in{\cal T}_{n}(Q_{X})\right]\max_{\tilde{\sigma}_{n}\in\tilde{\Sigma}_{n}}\P\left[d_{\st[E]}(\mathbf{X},\mathbf{Z})\leq\DD|\mathbf{X}\in{\cal T}_{n}(Q_{X})\right]\\
 & \doteq\max_{Q_{X}\in{\cal P}_{n}({\cal X})}e^{-nD\left(Q_{X}||P_{X}\right)}\cdot\max_{\tilde{\sigma}_{n}\in\tilde{\Sigma}_{n}}\P\left[d_{\st[E]}(\mathbf{X},\mathbf{Z})\leq\DD|\mathbf{X}\in{\cal T}_{n}(Q_{X})\right]\\
 & =\exp\left(-n\cdot\min_{Q_{X}\in{\cal P}_{n}({\cal X})}\left\{ D\left(Q_{X}||P_{X}\right)-\vphantom{\frac{1}{n}\log\max_{\tilde{\sigma}_{n}\in\tilde{\Sigma}_{n}}\P\left[d_{\st[E]}(\mathbf{X},\mathbf{Z})\leq\DD|\mathbf{X}\in{\cal T}_{n}(Q_{X})\right]}\right.\right.\label{eq: converse method of type}\\
 & \hphantom{{}={}}\left.\left.\frac{1}{n}\log\max_{\tilde{\sigma}_{n}\in\tilde{\Sigma}_{n}}\P\left[d_{\st[E]}(\mathbf{X},\mathbf{Z})\leq\DD|\mathbf{X}\in{\cal T}_{n}(Q_{X})\right]\right\} \right)
\end{alignat}
Now, let $0<\epsilon<\EEc$ be given, and let $Q_{X}^{*}\in{\cal P}({\cal X})$
be such that 
\begin{multline}
D\left(Q_{X}^{*}||P_{X}\right)+\limsup_{n\to\infty}\left\{ -\frac{1}{n}\log\max_{\tilde{\sigma}_{n}\in\tilde{\Sigma}_{n}}\P\left[d_{\st[E]}(\mathbf{X},\mathbf{Z})\leq\DD|\mathbf{X}\in{\cal T}_{n}(Q_{X}^{*})\right]\right\} \leq\\
\inf_{Q_{X}\in{\cal P}({\cal X})}\left\{ D\left(Q_{X}||P_{X}\right)+\limsup_{n\to\infty}\left\{ -\frac{1}{n}\log\max_{\tilde{\sigma}_{n}\in\tilde{\Sigma}_{n}}\P\left[d_{\st[E]}(\mathbf{X},\mathbf{Z})\leq\DD|\mathbf{X}\in{\cal T}_{n}(Q_{X})\right]\right\} \right\} +\epsilon\label{eq: ineq1}
\end{multline}
and let $m_{0}$ be sufficiently large so that\textbf{
\begin{multline}
\sup_{n>m_{0}}\left\{ -\frac{1}{n}\log\max_{\tilde{\sigma}_{n}\in\tilde{\Sigma}_{n}}\P\left[d_{\st[E]}(\mathbf{X},\mathbf{Z})\leq\DD|\mathbf{X}\in{\cal T}_{n}(Q_{X}^{*})\right]\right\} \\
\leq\limsup_{n\to\infty}\left\{ -\frac{1}{n}\log\max_{\tilde{\sigma}_{n}\in\tilde{\Sigma}_{n}}\P\left[d_{\st[E]}(\mathbf{X},\mathbf{Z})\leq\DD|\mathbf{X}\in{\cal T}_{n}(Q_{X}^{*})\right]\right\} +\epsilon.\label{eq:ineq2}
\end{multline}
}Then, 
\begin{alignat}{1}
{\cal E}_{d}^{+}({\cal S},\DD) & =\limsup_{n\to\infty}\min_{Q_{X}\in{\cal P}_{n}({\cal X})}\left\{ D\left(Q_{X}||P_{X}\right)-\frac{1}{n}\log\max_{\tilde{\sigma}_{n}\in\tilde{\Sigma}_{n}}\P\left[d_{\st[E]}(\mathbf{X},\mathbf{Z})\leq\DD|\mathbf{X}\in{\cal T}_{n}(Q_{X})\right]\right\} \\
 & =\lim_{m\to\infty}\sup_{n\geq m}\min_{Q_{X}\in{\cal P}_{n}({\cal X})}\left\{ D\left(Q_{X}||P_{X}\right)-\frac{1}{n}\log\max_{\tilde{\sigma}_{n}\in\tilde{\Sigma}_{n}}\P\left[d_{\st[E]}(\mathbf{X},\mathbf{Z})\leq\DD|\mathbf{X}\in{\cal T}_{n}(Q_{X})\right]\right\} \\
 & \trre[=,a]\lim_{m\to\infty}\sup_{n\geq m}\inf_{Q_{X}\in{\cal P}({\cal X})}\left\{ D\left(Q_{X}||P_{X}\right)-\frac{1}{n}\log\max_{\tilde{\sigma}_{n}\in\tilde{\Sigma}_{n}}\P\left[d_{\st[E]}(\mathbf{X},\mathbf{Z})\leq\DD|\mathbf{X}\in{\cal T}_{n}(Q_{X})\right]\right\} \\
 & \leq\sup_{n\geq m_{0}}\inf_{Q_{X}\in{\cal P}({\cal X})}\left\{ D\left(Q_{X}||P_{X}\right)-\frac{1}{n}\log\max_{\tilde{\sigma}_{n}\in\tilde{\Sigma}_{n}}\P\left[d_{\st[E]}(\mathbf{X},\mathbf{Z})\leq\DD|\mathbf{X}\in{\cal T}_{n}(Q_{X})\right]\right\} \\
 & \leq\inf_{Q_{X}\in{\cal P}({\cal X})}\left\{ D\left(Q_{X}||P_{X}\right)+\sup_{n\geq m_{0}}\left\{ -\frac{1}{n}\log\max_{\tilde{\sigma}_{n}\in\tilde{\Sigma}_{n}}\P\left[d_{\st[E]}(\mathbf{X},\mathbf{Z})\leq\DD|\mathbf{X}\in{\cal T}_{n}(Q_{X})\right]\right\} \right\} \\
 & \leq\left\{ D\left(Q_{X}^{*}||P_{X}\right)+\sup_{n>m_{0}}\left\{ -\frac{1}{n}\log\max_{\tilde{\sigma}_{n}\in\tilde{\Sigma}_{n}}\P\left[d_{\st[E]}(\mathbf{X},\mathbf{Z})\leq\DD|\mathbf{X}\in{\cal T}_{n}(Q_{X}^{*})\right]\right\} \right\} \\
 & \trre[\leq,b]\inf_{Q_{X}\in{\cal P}({\cal X})}\left\{ D\left(Q_{X}||P_{X}\right)+\vphantom{\left\{ \limsup_{n\to\infty}\left\{ -\frac{1}{n}\log\max_{\tilde{\sigma}_{n}\in\tilde{\Sigma}_{n}}\P\left[d_{\st[E]}(\mathbf{X},\mathbf{Z})\leq\DD|\mathbf{X}\in{\cal T}_{n}(Q_{X})\right]\right\} \right\} }\right.\\
 & \hphantom{{}={}}\left.\limsup_{n\to\infty}\left\{ -\frac{1}{n}\log\max_{\tilde{\sigma}_{n}\in\tilde{\Sigma}_{n}}\P\left[d_{\st[E]}(\mathbf{X},\mathbf{Z})\leq\DD|\mathbf{X}\in{\cal T}_{n}(Q_{X})\right]\right\} \right\} +2\epsilon\\
 & =\inf_{Q_{X}\in{\cal P}({\cal X})}\left\{ D\left(Q_{X}||P_{X}\right)+{\cal E}_{d}^{+}({\cal S},\DD,Q_{X})\right\} +2\epsilon\\
 & \leq\inf_{Q_{X}\in{\cal P}({\cal X}):D(Q_{X}||P_{X})<\EEc}\left\{ D\left(Q_{X}||P_{X}\right)+{\cal E}_{d}^{+}({\cal S},\DD,Q_{X})\right\} +2\epsilon\\
 & \trre[\leq,c]\inf_{Q_{X}\in{\cal P}({\cal X}):D(Q_{X}||P_{X})<\EEc}\left\{ D\left(Q_{X}||P_{X}\right)+{\cal E}_{d}^{+}({\cal S},\DD,Q_{X})\right\} +2\epsilon+\delta\\
 & \trre[\leq,d]\inf_{Q_{X}\in{\cal P}({\cal X}):D(Q_{X}||P_{X})<\EEc}\left\{ D\left(Q_{X}||P_{X}\right)+\rho(Q_{X})\right\} +2\epsilon+\delta\\
 & \trre[\leq,e]\RR+2\epsilon+4\delta,\label{eq: bounding rate function}
\end{alignat}
where $(a)$ is because, by assumption, if ${\cal T}_{n}(Q_{X})$
is empty then $\P\left[d_{\st[E]}(\mathbf{X},\mathbf{Z})\leq\DD|\mathbf{X}\in{\cal T}_{n}(Q_{X})\right]=0$
, $(b)$ is from \eqref{eq: ineq1} and \eqref{eq:ineq2}, and $(c)$
is from the third property of ${\cal S}$ promised by Lemma \ref{lem: converse - fixed type key-rate}.
The passage $(d)$ follows from \eqref{eq: converse lower bound R},
and so it remains to prove $(e)$. To this end, recall that $\E[r_{n}(\mathbf{X})]\leq\RR$
for all $n$ was assumed. Define, for $0<\epsilon<\EEc$, the \emph{typical
set
\[
\tilde{{\cal T}}(P_{X},\epsilon)\teq\left\{ Q_{X}\in{\cal P}({\cal X}):D(Q_{X}||P_{X})\leq\epsilon\right\} ,
\]
}and with a slight abuse of notation, define $\tilde{{\cal T}}_{n}(P_{X},\epsilon)\teq\tilde{{\cal T}}(P_{X},\epsilon)\cap{\cal P}_{n}({\cal X})$.
Then, by the law of large numbers
\begin{equation}
\lim_{n\to\infty}\sum_{Q_{X}\in\tilde{{\cal T}}_{n}(P_{X},\epsilon)}\P\left[\mathbf{X}\in{\cal T}_{n}(Q_{X})\right]=1.\label{eq: typical set prob}
\end{equation}
Now, assume by contradiction, that for all $Q_{X}\in\tilde{{\cal T}}(P_{X},\epsilon)$
we have $\rho(Q_{X})\geq\RR+3\delta$. Since by construction $\rho(Q_{X})\leq\overline{R}({\cal S},Q_{X})+\delta$,
the uniform convergence of $\E[r_{n}(\mathbf{X})|\mathbf{X}\in{\cal T}_{n}(Q_{X})]$
to $\overline{R}({\cal S},Q_{X})$ (see \eqref{eq: uniform convergence in probability}
and the discussion that follows) implies that there exists $n_{0}$
such that for all $n>n_{0}$ 
\begin{align}
\E[r_{n}(\mathbf{X})|\mathbf{X}\in{\cal T}_{n}(Q_{X})] & \geq\overline{R}({\cal S},Q_{X})-\delta\nonumber \\
 & \geq\rho(Q_{X})-2\delta\nonumber \\
 & \geq\RR+\delta,\label{eq: lower bound on the key rate contradiction}
\end{align}
for all $Q_{X}\in\tilde{{\cal T}}_{n}(P_{X},\epsilon)$. So, from
\textbf{\eqref{eq: typical set prob}}, there exists $n_{1}$, such
that for all $n>n_{1}$ we have that $\P\left[\mathbf{X}\in\tilde{{\cal T}}_{n}(P_{X},\epsilon)\right]\geq\frac{1}{1+\nicefrac{\delta}{2\cdot\log|{\cal X}|}}$,
and then for all $n>\max\{n_{0},n_{1}\}$ 
\begin{align*}
\E\left[r_{n}(\mathbf{X})\right] & =\sum_{Q_{X}\in{\cal P}_{n}({\cal X})}\P\left[\mathbf{X}\in{\cal T}_{n}(Q_{X})\right]\cdot\E[r_{n}(\mathbf{X})|\mathbf{X}\in{\cal T}_{n}(Q_{X})]\\
 & \geq\sum_{Q_{X}\in\tilde{{\cal T}}_{n}(P_{X},\epsilon)}\P\left[\mathbf{X}\in{\cal T}_{n}(Q_{X})\right]\cdot\E[r_{n}(\mathbf{X})|\mathbf{X}\in{\cal T}_{n}(Q_{X})]\\
 & \geq\left(\min_{Q_{X}\in\tilde{{\cal T}}_{n}(P_{X},\epsilon)}\E[r_{n}(\mathbf{X})|\mathbf{X}\in{\cal T}_{n}(Q_{X})]\right)\cdot\sum_{Q_{X}\in\tilde{{\cal T}}_{n}(P_{X},\epsilon)}\P\left[\mathbf{X}\in{\cal T}_{n}(Q_{X})\right]\\
 & \trre[\geq,a]\left(\RR+\delta\right)\frac{1}{1+\nicefrac{\delta}{2\cdot\log|{\cal X}|}}\\
 & >\left(\RR+\delta\right)\frac{1}{1+\nicefrac{\delta}{\RR}}\\
 & =\RR,
\end{align*}
where $(a)$ follows from \eqref{eq: lower bound on the key rate contradiction}.
However, this is a contradiction to the fact that ${\cal S}_{n}$
satisfies $\E\left[r_{n}(\mathbf{X})\right]\leq\RR$ for all $n$.
Thus, there must exist $Q_{X}\in\tilde{{\cal T}}(P_{X},\epsilon)\subset\tilde{{\cal T}}(P_{X},\EEc)$
such that $\rho(Q_{X})<\RR+3\delta$, which directly leads to $(e)$
in \eqref{eq: bounding rate function}. Since $\epsilon>0$ and $\delta>0$
are arbitrary, the first term in the upper bound of \eqref{eq: main theorem converse}
is proved, i.e. ${\cal E}_{d}^{+}({\cal S},\DD)\leq\RR.$ 

To prove the second term in the upper bound of \eqref{eq: main theorem converse},
i.e. ${\cal E}_{d}^{+}({\cal S},\DD)\leq E_{e}^{*}(\DD)$, note that
the eavesdropper can always ignore the cryptogram and \emph{blindly}
choose its estimate $\mathbf{z}$ (based only on the type $Q_{X}$).
Thus, by similar arguments leading to \eqref{eq: blind estimate exponent},
it can be shown that\textbf{ }for all $n$ sufficiently large 
\[
{\cal E}_{d}^{+}({\cal S},\DD,Q_{X})\leq R_{\st[E]}(Q_{X},\DD).
\]
The method of types, as in \eqref{eq: converse method of type} and
the definition of $E_{e}^{*}(\DD)$ in \eqref{eq: perfect secrecy exponent},
complete the proof. 
\end{IEEEproof}
\bibliographystyle{plain}
\bibliography{Secrecy_Distortion}

\end{document}